\documentclass{aastex63}
\usepackage{amsmath}
\usepackage{ulem}
\usepackage{extarrows}

\received{2020 June 27}
\revised{2020 October 12}
\accepted{2020 October 21}
\submitjournal{...}

\shorttitle{Analytical Jet}
\shortauthors{Chen \& Zhang}
\graphicspath{{./}{figures/}}
\begin{document}

\title{Analytical Solution of Magnetically Dominated Astrophysical Jets and Winds: Jet Launching, Acceleration, and Collimation}

\correspondingauthor{Liang Chen}
\email{chenliang@shao.ac.cn}

\author[0000-0002-1908-0536]{Liang Chen}
\affiliation{Key Laboratory for Research in Galaxies and
Cosmology, Shanghai Astronomical Observatory, Chinese Academy of
Sciences, 80 Nandan Road, Shanghai 200030, China}

\author[0000-0002-9725-2524]{Bing Zhang}
\affiliation{Department of Physics and Astronomy, University of Nevada, Las Vegas, Las Vegas, NV 89154, USA}

%% Note that the \and command from previous versions of AASTeX is now
%% depreciated in this version as it is no longer necessary. AASTeX 
%% automatically takes care of all commas and "and"s between authors names.

%% AASTeX 6.3 has the new \collaboration and \nocollaboration commands to
%% provide the collaboration status of a group of authors. These commands 
%% can be used either before or after the list of corresponding authors. The
%% argument for \collaboration is the collaboration identifier. Authors are
%% encouraged to surround collaboration identifiers with ()s. The 
%% \nocollaboration command takes no argument and exists to indicate that
%% the nearby authors are not part of surrounding collaborations.

%% Mark off the abstract in the ``abstract'' environment. 

\begin{abstract}

We present an analytical solution of a highly magnetized jet/wind flow. The left side of the general force-free jet/wind equation (the ``pulsar'' equation) is separated into a rotating and a nonrotating term. The two equations with either term can be solved analytically, and the two solutions match each other very well. Therefore, we obtain a general approximate solution of a magnetically dominated jet/wind, which covers from the nonrelativistic to relativistic regimes, with the drift velocity well matching the cold plasma velocity. The acceleration of a jet includes three stages. (1) The jet flow is located within the Alfv\'{e}n critical surface (i.e. the light cylinder), has a nonrelativistic speed, and is dominated by toroidal motion.
(2) The jet is beyond the Alfv\'{e}n critical surface where the flow is dominated by poloidal motion and becomes relativistic. The total velocity in these two stages follows the same law $v\Gamma=\Omega R$. (3) The evolution law is replaced by $v\Gamma\approx1/\left(\theta\sqrt{2-\nu}\right)$, where $\theta$ is the half-opening angle of the jet and $0\leq\nu\leq2$ is a free parameter determined by the magnetic field configuration. This is because the earlier efficient acceleration finally breaks the causality connection between different parts in the jet, preventing a global solution. The jet has to carry local charges and currents to support an electromagnetic balance. This approximate solution is consistent with known theoretical results and numerical simulations, and it is more convenient to directly compare with observations. This theory may be used to constrain the spin of black holes in astrophysical jets.

\end{abstract}

%% Keywords should appear after the \end{abstract} command.
%% See the online documentation for the full list of available subject
%% keywords and the rules for their use.

%%\keywords{galaxies: jets - gamma-ray burst: general - magnetic fields - black hole physics - %%accretion, accretion disks}

%% From the front matter, we move on to the body of the paper.
%% Sections are demarcated by \section and \subsection, respectively.
%% Observe the use of the LaTeX \label
%% command after the \subsection to give a symbolic KEY to the
%% subsection for cross-referencing in a \ref command.
%% You can use LaTeX's \ref and \label commands to keep track of
%% cross-references to sections, equations, tables, and figures.
%% That way, if you change the order of any elements, LaTeX will
%% automatically renumber them.
%%
%% We recommend that authors also use the natbib \citep
%% and \citet commands to identify citations.  The citations are
%% tied to the reference list via symbolic KEYs. The KEY corresponds
%% to the KEY in the \bibitem in the reference list below.

\section{Introduction}
\label{sec:Introduction}

Astrophysical jets/winds are a very common astronomical phenomenon. They have been observed in different types of sources, such as active galactic nuclei \citep[AGNs;][]{1995PASP..107..803U, 2019ARA&A..57..467B}, $\gamma$-ray bursts \citep[GRBs;][]{2004RvMP...76.1143P, 2006RPPh...69.2259M,2015PhR...561....1K, 2018pgrb.book.....Z}, tidal disruption events \citep[TDEs;][]{2011Natur.476..421B, 2011Natur.476..425Z}, and X-ray binaries \citep[][]{1999ARA&A..37..409M}. Observationally, a collimated jet can be accelerated to a highly relativistic speed \citep[e.g.,][]{2001ApJ...555..540L, 2014Natur.515..376G, 2018ApJS..235...39C, 2019ARA&A..57..467B}, and its scale ranges many orders of magnitude during its propagation outward \citep[][]{1995PASP..107..803U, 2012ApJ...745L..28A, 2019ARA&A..57..467B}. Jets play an important role in many astrophysical phenomena, such as galaxy evolution \citep[through feedback; see, e.g.,][]{2012ARA&A..50..455F}, black hole (BH) growth \citep[through shedding the angular momentum to facilitate accretion; see, e.g.,][]{1982MNRAS.199..883B, 2004MNRAS.351..169M} and particle acceleration \citep[related to ultra/very-high-energy cosmic rays and neutrinos; see, e.g.,][]{2018Sci...361..147I, 2019PhR...801....1A, 2019ARA&A..57..467B, 2019ApJ...871...81X}. They are also the laboratory to study electrodynamical processes in curved spacetime \citep[e.g., extracting BH rotating energy, see][]{1977MNRAS.179..433B, 1982MNRAS.198..339T, 1982MNRAS.198..345M}.

The observations of astrophysical jets, especially AGN jets, have accumulated a huge amount of data and present a variety of observational phenomena. Very-long-baseline interferometry (VLBI) observations of AGN jets have revealed a feature of collimation and acceleration during the propagation of the jet from the core \citep[see, e.g.,][and references therein]{2015ApJ...798..134H}. Because of its proximity to  Earth, the M87 jet has been extensively observed, with a large amount of observational data showing that the jet is continuously accelerated from nonrelativistic to relativistic speed and collimated up to a distance over $\sim300$ pc \citep[see, e.g.,][]{2007ApJ...668L..27K, 2014ApJ...781L...2A, 2017PASJ...69...71H}. This source presents the typical feature of a magnetically dominated jet \citep{2001Sci...291...84M, 2008Natur.452..966M}. Some AGN jets present a limb-brightening phenomenon \citep[i.e. the edge of the jet is brighter than its spine, for example, the M87 jet;][]{2007ApJ...668L..27K, 2007ApJ...660..200L, 2018ApJ...855..128W}. The radio emission from AGN jets is usually polarized, and its Faraday rotation measure (RM) sometimes presents a systematic gradient with respect to the jet axis and also along the jet axis \citep[e.g.,][]{2002PASJ...54L..39A, 2004MNRAS.351L..89G, 2012AJ....144..105H, 2019ApJ...871..257P}. The polarization angle of AGN optical emission presents a smoothly rotating feature by a large angle during the outburst \citep[e.g., over $\sim 720^{\circ}$ for PKS 1510-089;][]{2008Natur.452..966M, 2010ApJ...710L.126M, 2010Natur.463..919A}. The $\gamma$-ray emission from some AGN jets presents a periodic variability with a timescale of months to years \citep[e.g.,][]{2015ApJ...813L..41A, 2018NatCo...9.4599Z}. Also, the innermost position angle of some AGN jets shows evidence of an oscillatory behavior \citep[e.g.,][]{2013AJ....146..120L, 2018ApJ...855..128W}.

The long-standing issue about how a jet is launched and which mechanism determines its collimation and acceleration is one of the most fundamental questions in astrophysics. Theoretically, one needs to obtain a self-consistent global jet model that connects the central accreting system to the large distance where the observed radiation is produced. It is generally believed that jets are a magnetic phenomenon. This motivates the investigation of  magnetohydrodynamic (MHD) models for jets. Through releasing the gravitational energy of an accretion disk \citep[AD; see][for the Blandford-Payne (BP) process]{1982MNRAS.199..883B} or extracting the rotational energy of the central compact object \citep[CO; see][for BH case and the Blandford-Znajek (BZ) process]{1977MNRAS.179..433B, 2014Natur.510..126Z}, the central accreting system can launch an electromagnetically dominated jet\footnote{Even a turbulent AD can form a simple power-law profile of the height-integrated toroidal current, and therefore can further launch a nearly force-free, stationary, collimated and ordered Poynting-flux-dominated jet in the polar region as shown in some MHD simulations \citep[e.g.,][]{2006ApJ...641..103H, 2006MNRAS.368.1561M, 2007MNRAS.375..513M}.} \citep[see e.g.,][and references therein]{2012MNRAS.423.3083M}, of which the electromagnetic stress provides the principal torque acting on the AD/CO, and most of the energy is liberated electromagnetically. Generally speaking, an ordered rotating magnetic field is the indispensable ingredient to globally launch a collimated relativistic jet.

The MHD equations governing jet physics are highly nonlinear, and therefore generally need to be solved numerically \citep[see, e.g.,][and references therein]{2003ApJ...596.1080V, 2012MNRAS.423.3083M, 2019MNRAS.490.2200C}. Compared with the complex numerical solutions, analytical solutions are only achieved for some special cases. Yet, such analytical studies may provide simple scalings, which could offer a qualitative understanding of the basic properties of relativistic MHD flows. Until now, a great deal of analytical work has been done on MHD outflows \citep[see, e.g.,][and references therein]{1976MNRAS.176..465B, 1994ApJ...426..269B, 2003ApJ...596.1080V, 2003ApJ...596.1104V, 2006MNRAS.367..375B, 2007MNRAS.375..548N, 2009ApJ...698.1570L}. If the pressure and internal energy are negligible, the plasma only contributes to the dynamics. One can then make a cold gas assumption to simplify the MHD equations. If the gas inertia can be further ignored in a highly magnetized flow, the plasma plays a dynamically negligible role and the electromagnetic field can be solved self-consistently. In this case, the force-free condition becomes a good approximation \citep[see, e.g.,][and references therein]{1977MNRAS.179..433B, 1996ASIC..477..249S, 2010LNP...794..233S, 2006MNRAS.368.1561M, 2007MNRAS.375..548N}.

In the case of a highly magnetized flow, a magnetic stream function ($\Psi$) is usually employed to measure the poloidal magnetic field in an axisymmetric system, which is conserved along a magnetic field line \citep[e.g.,][]{2007MNRAS.375..548N}. To solve MHD equations, a self-similar distribution of the magnetic stream function (e.g., $\Psi\propto R_{0}^{\nu}$ holding on the foot-point of magnetic field lines threading the AD plane) is often used to separate variables. Under a further assumption of ``flat rotation" on the AD plane \citep[i.e. an angular velocity $\Omega\propto R_{0}^{-1}$, not a Keplerian profile; see, e.g.,][]{1992ApJ...394..459L, 2007MNRAS.375..548N}, one can derive a globally analytical solution for the following cases: (1) the $\nu=0$ monopole solution \citep[this case does not need the ``flat rotation" assumption; see][]{1973ApJ...180L.133M}, (2) the $\nu=1$ parabolic solution \citep{1976MNRAS.176..465B, 2007MNRAS.375..548N}, and (3) the $\nu=3/4$ approximate solution of \cite{1982MNRAS.199..883B}. In general, numerical methods are necessarily needed to solve the equation for an arbitrary $\nu$ value, and an asymptotic behavior may be obtained in the limit of the highly relativistic case \citep[e.g.,][]{2003ApJ...596.1080V, 2004MNRAS.347..587B, 2006MNRAS.367..375B, 2007MNRAS.375..548N, 2008MNRAS.388..551T, 2009MNRAS.394.1182K, 2020ApJ...892...37P}. In another aspect, some MHD simulations explore some parameter spaces and find that the poloidal configuration of magnetic fields changes little from the nonrotation to the rotation cases \citep[e.g.,][]{2008MNRAS.388..551T}. In the case of a highly magnetized flow, given a magnetic field configuration, one can derive the velocity profile during the outward propagation of a flow. It has been found that the bulk Lorentz factor of the flow has an asymptotic power-law dependence on the distance from the central CO/AD in the highly relativistic regime \citep[e.g.,][]{2007MNRAS.375..548N, 2008MNRAS.388..551T}. However, the question regarding how the flow is accelerated from the nonrelativistic case to the relativistic case only can be studied numerically \citep[e.g.,][]{2008MNRAS.388..551T, 2012MNRAS.423.3083M, 2018ApJ...868..146N}.

The status of the field may be summarized as follows. (1) Numerical simulations become the reliable and popular method to explore the underlying physics of magnetic field configuration, jet launching, collimating, and acceleration (e.g. \citealt{2019MNRAS.490.2200C} recently simulated a jet spanning over five orders of magnitude in distance). On the other hand, due to its complexity, some simulation results are not very intuitive for understanding the physical details.
It is also challenging to directly test theoretical results against observations, although some efforts have been made recently \citep[see, e.g.,][and references therein]{2012MNRAS.421.1517D, 2016A&A...586A..38M, 2018MNRAS.473.4417C, 2019A&A...632A...2D, 2019MNRAS.486.2873C, 2019ApJ...875L...5E, 2020PASJ...72...32T}. (2) As a complementary method to numerical simulations, an analytical self-consistent model can present a better understanding of jet physics and may be easier to confront with observations directly. However, the analytic models proposed so far do not cover a large-enough parameter space and cannot describe the transition regime from the nonrelativistic to the relativistic regime. (3) An increasing amount of observational data have been collected, but direct testing of models against these observational data is still lacking \citep[][]{2010LNP...794..233S}.

To achieve a comparison between observations and theory, a global relativistic jet solution is needed, which should satisfy the following conditions: (1) The solution should be physically (mathematically) reasonable. (2) The solution can describe the transition from the nonrelativistic regime to the relativistic regime. (3) The solution for an ensemble of jets should have features and trends  consistent with the previous theoretical results (analytic/semianalytic, numerical simulation results) and observations. (4) The expression of the solution should be explicitly analytical and comprehensive, so that it can be easily used for further developments (e.g., adding radiative processes) and comparisons with observations. (5) The solution can be an approximation (i.e. an approximately quantitative description of a global jet), but it should be accurate enough in the collimated jet region.

In this paper, we study the problem of a magnetically dominated jet/wind under the force-free condition, and we find an approximately analytical solution to the magnetic stream function that meets the above requirements. Based on this solution, the jet properties (velocity, current, charge, and so on) can be further explored in detail. In Section \ref{sec:basic_equation}, starting from the first principles, we present the basic MHD equations to describe a highly magnetized jet flow. An approximate solution is provided in Section \ref{sec:approximate_solution} in the case of negligible gravity. Based on the approximate solution, the electromagnetic field configuration is presented in Section \ref{sec:magnetic_configuration}, and a detailed flow velocity profile is obtained and drawn in Section \ref{sec:jet_acceleration}. The questions about how much current, charge, and power a jet carries are discussed in Section \ref{sec:charge_current_Power}. Under the force-free condition, in principle, one cannot derive plasma fluid acceleration becuase of the omission of inertia. We discuss the flow dynamics and how a cold plasma velocity is related to the electromagnetic field drift velocity in Section \ref{sec:jet_dynamics} (and also the jet flow density). In general, we have ignored the effect of general relativity (GR) to derive the solutions. In Section \ref{BHjet}, we apply the approximate solution on a BH system with the gravity effect explicitly included near the BH.  As a boundary condition, the CO/AD offers gas, charges, and currents to support the electromagnetic field in the jet flow, which is discussed in Section \ref{Bp_AD}. A further note on jet stability is presented in Section \ref{sec:stability}, which is followed by a summary Section in \ref{sec:summary}. For simplicity, we employ the natural unit system throughout the paper, using the light speed $c=1$ and the gravitational constant $G=1$. Some important formulae are also presented in Gaussian units in Appendix \ref{app:Gaussianformula}.

\section{Basic Equations}
\label{sec:basic_equation}

We start with a brief derivation of the well-known ``pulsar equation'' following, for example, \cite{2007MNRAS.375..548N}. We consider a steady-state ($\partial/\partial t = 0$) jet flow with an infinite conductivity,
\begin{equation}
\mathbf{E}+\mathbf{v}\times\mathbf{B}
%= \mathbf{E}+(\mathbf{\Omega}\times \mathbf{r}) \times\mathbf{B}
= 0,
%%%%%%%%%%% E\propto vB/c
\label{ideal_MHD}
\end{equation}
which means that the electric field vanishes in the plasma fluid comoving frame\footnote{The perfect/ideal MHD condition corresponds to flux freezing with a high magnetic Reynold's number.}.

A magnetic field has no divergence and therefore can be expressed as $\mathbf{B}=\nabla\times\mathbf{A}$, where $\mathbf{A}$ is the vector potential. In axisymmetric coordinates, one can define a magnetic stream function $\Psi=r\sin\theta A_{\phi}$ and $\Phi=r\sin\theta B_{\phi}$ for convenience.
%\footnote{Both spherical (the unit vectors $r,\theta,\phi$) and cylindrical coordinates (the unit vectors $R,\phi,z$) are used throughout the paper.}.
Therefore, $\mathbf{B}$ can be expressed as
\begin{equation}
\begin{aligned}
\mathbf{B}=\frac{1}{r^{2}\sin\theta}\frac{\partial\Psi}{\partial\theta}\mathbf{\hat{r}}-\frac{1}{r\sin\theta}\frac{\partial\Psi}{\partial r}\hat{\mathbf{\theta}}+\frac{\Phi}{r\sin\theta}\mathbf{\hat{\phi}}
=-\frac{1}{R}\frac{\partial\Psi}{\partial z}\mathbf{\hat{R}}+\frac{\Phi}{R}\mathbf{\hat{\phi}}+\frac{1}{R}\frac{\partial\Psi}{\partial R}\hat{\mathbf{z}}
\label{Bmag_Psi}
\end{aligned}
\end{equation}
in spherical ($r,\theta,\phi$) and cylindrical coordinates ($R,\phi,z$), respectively.
It is easy to prove that $\mathbf{B}\cdot\nabla\Psi=0$, suggesting that $\Psi$ is conserved along a magnetic field line. This also suggests conservation of the magnetic flux enclosed within radius $R$, i.e.
\begin{equation}
F_{B}=\iint\mathbf{B}\cdot d\mathbf{S}=2\pi\Psi = {\rm const}.
\label{magneticflux}
\end{equation}
This suggests that the interior of a jet has smaller $\Psi$ values. For an axisymmetric system,  rotation ensures that a magnetic field line stays in a magnetic stream surface $\Psi=$ const, which is also where the frozen plasma fluid streams. Therefore, the global structure of the magnetic field configuration can evolve in a self-similar form.

The frozen-in condition also implies that the magnetic stream surface would be equipotential ($\mathbf{E}\cdot\mathbf{B}=0$, see Equation \ref{ideal_MHD}). Therefore, $\mathbf{E}$ can be written in the form
\begin{equation}
\mathbf{E}=-\Omega\nabla\Psi=-\Omega r\sin\theta\hat{\phi}\times\mathbf{B},
%%%%%%%%% E\propto\frac{\Omega RB_{p}}{c}
\label{E_general}
\end{equation}
where $\Omega$ is the angular velocity of the magnetic field line, which in principle is not necessarily equal to any angular velocity of the fluid matter. One can imagine that the angular velocity of the entire field line all the way up to infinity is determined by the rotation of the CO/AD at the foot-point, which implies that $\Omega$ would be conserved along a magnetic field line. This is indeed the case in a steady state, where the electric field is noncurling:
\begin{equation}
0=\nabla\times\mathbf{E}=-r\sin\theta\hat{\phi}\left(\mathbf{B}\cdot\nabla\Omega\right).
\label{E_curl}
\end{equation}
This implies that $\Omega$ is conserved along a magnetic field line and therefore is only a function of $\Psi$. Since the electric field is perpendicular to the magnetic stream surface, the electric force would be crucial to maintaining the force balance between different surfaces (see Section \ref{sec:Lorentz_Force} for a discussion of force balance).

Generally speaking, the motion of a cold gas with negligible gravity can be written as
\begin{equation}
\rho\left(\mathbf{u}\cdot\bigtriangledown\right)\mathbf{u}
=\rho_{\rm e}\mathbf{E}+\mathbf{j}\times\mathbf{B},
%%%%%%%% \rho_{\rm e}E \propto jB/c
\label{motion_Eq}
\end{equation}
where $\rho_{\rm e}$ refers to the charge density, $\mathbf{j}$ the current density, $\mathbf{u}=\Gamma\mathbf{v}$ the four-velocity (the spatial part), and $\rho$ the plasma proper density. If a plasma is sufficiently rarefied so that it does not exert significant force on the magnetic field (i.e. the inertial term can be neglected) - but it is still sufficiently dense to support charges and currents to maintain the magnetic field, one then has the so-called force-free approximation\footnote{This is a reasonable approximation for highly magnetized flows \citep[e.g.,][]{1977MNRAS.179..433B, 2006MNRAS.368.1561M}. In terms of the standard magnetization parameter $\sigma$, we assume $\sigma\gg1$ \citep[see discussion in Section \ref{sec:dynamics_velocity} and][]{1969ApJ...158..727M, 1970ApJ...160..971G}.}, which is reduced from Equation \ref{motion_Eq} as
\begin{equation}
%%%%%%%%% rho E+j B/c
\rho_{\rm e}\mathbf{E}+\mathbf{j}\times\mathbf{B}=0,
\label{force_free}
\end{equation}
where the electric current density follows
\begin{eqnarray}
4\pi\mathbf{j} & = & \nabla\times\mathbf{B}=-\frac{1}{R}\frac{\partial\Phi}{\partial z}\mathbf{\hat{R}}-\frac{1}{R}\left(\frac{\partial^{2}\Psi}{\partial z^{2}}+\frac{\partial^{2}\Psi}{\partial R^{2}}-\frac{1}{R}\frac{\partial\Psi}{\partial R}\right)\hat{\phi}+\frac{1}{R}\frac{\partial\Phi}{\partial R}\mathbf{\hat{z}}, \nonumber \\
 & = & \frac{1}{r^{2}\sin\theta}\frac{\partial\Phi}{\partial \theta}\mathbf{\hat{r}}-\frac{1}{r\sin\theta}\frac{\partial\Phi}{\partial r}\mathbf{\hat{\theta}}-\frac{1}{r\sin\theta}\left(\frac{\partial^{2}\Psi}{\partial r^{2}}+\frac{1}{r^{2}}\frac{\partial^{2}\Psi}{\partial\theta^{2}}
-\frac{\cot\theta}{r^{2}}\frac{\partial\Psi}{\partial\theta}\right)\hat{\phi}.
\label{J_current}
\end{eqnarray}
Because the $\hat{\phi}$ component of $\mathbf{E}$ vanishes, one expects that the poloidal components $\mathbf{j}_{p}$ and $\mathbf{B}_{p}$ would be parallel to each other ($\mathbf{j}_{p}=\Phi'\mathbf{B}_{p}/4\pi$ with $\Phi'\equiv d\Phi/d\Psi$), which implies that the electric current would also flow in the magnetic stream surface. One immediately has $\mathbf{B}\cdot\nabla\Phi=0$, which implies that $\Phi$ is also conserved along the magnetic field line\footnote{$\Phi$ conserves approximately in the limit of a highly magnetized jet; see discussion in Section \ref{sec:jet_dynamics}.} and therefore is only a function of $\Psi$. This indicates that the current enclosed within $R$ is also conserved:
\begin{equation}
\begin{aligned}
J=\iint\mathbf{j}\cdot d\mathbf{S}=\frac{\Phi}{2} = {\rm const}.
\label{JzJr}
\end{aligned}
\end{equation}

Substituting relations $\rho_{\rm e}=\left(\nabla\cdot\mathbf{E}\right)/4\pi$, $\mathbf{j}=\left(\nabla\times\mathbf{B}\right)/4\pi$, and Equations \ref{Bmag_Psi} and \ref{E_general} into Equation \ref{force_free}, one finally gets the ``pulsar equation" \citep[i.e. the cross-field equation; see, e.g.,][]{1986ApJS...62....1L, 1974MNRAS.167..457O, 2007MNRAS.375..548N}\footnote{The $\hat{\phi}$ component of Equation \ref{force_free} is automatically satisfied, while the $\hat{r}$ and $\hat{\theta}$ components yield the same equation.},
\begin{eqnarray}
& \frac{\partial^{2}\Psi}{\partial r^{2}}+\frac{1}{r^{2}}\frac{\partial^{2}\Psi}{\partial\theta^{2}}
-\frac{\cot\theta}{r^{2}}\frac{\partial\Psi}{\partial\theta} \nonumber \\
& +\Phi'\Phi-\left\{\frac{\Omega'}{\Omega}\left[\left(\frac{\partial\Psi}{\partial r}\right)^{2}+\left(\frac{1}{r}\frac{\partial\Psi}{\partial\theta}\right)^{2}\right]
+\frac{\partial^{2}\Psi}{\partial r^{2}}+\frac{2}{r}\frac{\partial\Psi}{\partial r}+\frac{1}{r^{2}}\frac{\partial^{2}\Psi}{\partial\theta^{2}}+\frac{\cot\theta}{r^{2}}
\frac{\partial\Psi}{\partial\theta}\right\}\left(\Omega r\sin\theta\right)^{2}=0,
\label{DE_Psi}
\end{eqnarray}
with $\Omega'\equiv d\Omega/d\Psi$, which can also be reduced from the Grad-Shafranov equation in the limit of force-free conditions \citep[see, e.g.,][]{2009ApJ...698.1570L}. The first line in Equation \ref{DE_Psi} corresponds to the nonrotation term, while the second line is related to the rotation-induced term. This partial differential equation is clearly singular\footnote{The coefficients of the highest derivatives ($\frac{\partial^{2}\Psi}{\partial r^{2}}$ and $\frac{\partial^{2}\Psi}{\partial\theta^{2}}$) go to zero at the singularity.} at $\left(1-\Omega^{2}r^{2}\sin^{2}\theta\right)\rightarrow0$, which
corresponds to the Alfv\'{e}n critical surface (ACS) where the rotation velocity of the magnetic field lines reaches the speed of light (the so-called ``light cylinder'' in the case of pulsars). Both $\Psi$ and $\Omega$ are conserved along magnetic field lines, and they can be determined by the properties of the CO/AD at the foot-point, which can be taken as boundary conditions. The enclosed $z$-direction magnetic flux would vanish when approaching the polar axis, which implies $\Psi\left(r,\theta=0\right)=0$. On the AD plane or the CO ``surface," the value of $\Psi$ is finite. Given proper boundary conditions, the above 2D equation can be solved in principle, at least numerically. On the other hand, solving this equation analytically may give simple analytical scalings, which could offer a qualitative understanding of the basic properties of relativistic MHD flows. Many analytical or semianalytical studies have done  this, but only for very specific cases \citep[see Section \ref{sec:Introduction} and e.g.,][and references therein]{1976MNRAS.176..465B, 1977MNRAS.179..433B, 1982MNRAS.199..883B, 1995ApJ...446...67C, 2007MNRAS.375..548N}. Comparing with previous results, in this paper we consider whether there is a more general analytic solution for Equation \ref{DE_Psi}, even an approximate one. To do this, the plasma fluid is assumed to be flowing outward from an accreting system consisting of a spherical CO surrounded by an infinitely thin AD \citep[i.e., the foot-points of magnetic field lines; e.g.,][]{2007MNRAS.375..548N, 2010ApJ...711...50T}. In principle, at large distances from the central engine, the external environment inevitably plays a role in balancing the Lorentz force in the jet flow \citep[e.g.,][]{2009ApJ...698.1570L}. From this perspective, the force-free approach can only apply where the magnetic pressure dominates over the external pressure \citep[e.g.,][]{2006MNRAS.368.1561M}. The jet/wind (our region of interest) would have a finite magnetic flux. The boundary condition at the equator should be taken to balance with the external magnetic or gas pressure \citep[e.g.,][]{2000AstL...26..208B, 2009MNRAS.397.1486B, 2009ApJ...698.1570L}. For the problem tackled here, following \citet{2007MNRAS.375..548N}, the boundary is treated with externally supplied parameters (for example $\nu$ and $\lambda$, see below), which would be determined by disk/external boundary properties.

\section{An Approximate Solution}
\label{sec:approximate_solution}

In the region of $\Omega r\sin\theta\gg1$ (it will be shown in Section \ref{sec:total_velocity} that this corresponds to the relativistic case), the second line of Equation \ref{DE_Psi} dominates:
\begin{equation}
\begin{aligned}
\frac{\Phi'\Phi}{\Omega^{2} r^{2}\sin^{2}\theta}-\left\{\frac{\Omega'}{\Omega}\left[\left(\frac{\partial\Psi}{\partial r}\right)^{2}+\left(\frac{1}{r}\frac{\partial\Psi}{\partial\theta}\right)^{2}\right]
+\frac{\partial^{2}\Psi}{\partial r^{2}}+\frac{2}{r}\frac{\partial\Psi}{\partial r}+\frac{1}{r^{2}}\frac{\partial^{2}\Psi}{\partial\theta^{2}}+\frac{\cot\theta}{r^{2}}
\frac{\partial\Psi}{\partial\theta}\right\}=0.
\label{DE_F2_Psi}
\end{aligned}
\end{equation}
We call this the rotation term equation. Conversely, in the region of $\Omega r\sin\theta\ll1$ (corresponding to the nonrelativistic case), the first line of Equation \ref{DE_Psi} dominates:
\begin{equation}
\begin{aligned}
\frac{\partial^{2}\Psi}{\partial r^{2}}+\frac{1}{r^{2}}\frac{\partial^{2}\Psi}{\partial\theta^{2}}
-\frac{\cot\theta}{r^{2}}\frac{\partial\Psi}{\partial\theta}=0.\\
\label{DE_F1_Psi}
\end{aligned}
\end{equation}
We call this the nonrotation term equation. In fact, in the nonrelativistic region (Section \ref{sec:total_velocity}), whether the rotation term can be ignored depends on whether the rotation velocity of electromagnetic field ($\Omega r\sin\theta$) is relativistic or not: if the rotation velocity is close to the speed of light, one cannot ignore the rotation term, whereas in the nonrelativistic case, one can safely ignore the rotation term. It is interesting that the solution of the the rotation term, Equation \ref{DE_F2_Psi}, can approximately satisfy the nonrotation term, Equation \ref{DE_F1_Psi} (under certain conditions; see below), in the regimes either relativistic or nonrelativistic. Therefore, the solution of Equation \ref{DE_F2_Psi} (or \ref{DE_F1_Psi}) can always approximately satisfy Equation \ref{DE_Psi}. We prove this in the following.

Even though $\Omega$ is only a function of $\Psi$, the physical connection between the two at the foot-point is not very obvious\footnote{They may be constrained by proper boundary conditions \citep[e.g.,][]{2013ApJ...765..113C}.}. Likely most researchers,  here we assume that they generally follow the ansatz
\begin{equation}
\begin{aligned}
\Omega=\alpha\Psi^{\lambda}.
\label{Omega}
\end{aligned}
\end{equation}
In the region approaching the polar axis, one expects that the magnetic flux vanishes and thus $\Psi\rightarrow0$. Therefore, in  the region near the polar axis, mathematically $\lambda\geq0$ is required to guarantee a finite value of $\Omega$ (notice that $\lambda > 0$ is likely unphysical). Magnetic field lines just around the polar axis would connect to the central CO, which is expected to rotate with a roughly constant angular velocity\footnote{In the case of threading a BH, magnetic field lines at different polar angles may not rotate with exactly the same frequency; see discussion in Section \ref{BHjet}.} (see details below), implying $\lambda=0$ for the case of magnetic field lines threading the CO. In the region where magnetic field lines are threading the AD, one usually expects $\lambda<0$. The choice of $\Phi$, which determines the toroidal magnetic field, is important \citep[e.g.,][]{1990ApJ...350..732S, 1987A&A...184..341C}. Physically, it is the rotation that develops a toroidal magnetic field and a poloidal electric field in the lab frame (or the inertial frame), which seems to imply that $\Phi$ would not be chosen arbitrarily and would be self-consistently determined by the MHD equations, given the $\Psi$ and $\Omega$ specified \citep[e.g.,][]{2005A&A...433..619B, 2013ApJ...765..113C, 2019ApJ...880...93H, 2020ApJ...894...45H}. To solve the rotation term, Equation \ref{DE_F2_Psi}, one can see that in the case of
\begin{equation}
\begin{aligned}
\Phi=-\beta\Omega\Psi=-\beta\alpha\Psi^{\lambda+1},
\label{Phi}
\end{aligned}
\end{equation}
the function $\Psi$ can be variable-separated,
\begin{equation}
\begin{aligned}
\Psi=H_{\rm r}\left(r\right)T_{\rm r}\left(\theta\right),
\label{Psi}
\end{aligned}
\end{equation}
where the subscript `r' denotes `rotation.' The negative sign in Equation \ref{Phi} exists to guarantee that there is always a swept-back magnetic field line with respect to the direction of rotation. This relation implies that the strength of the toroidal magnetic field would be proportional to the angular velocity and the enclosed magnetic flux, which can be reasonably understood because it is the rotation of the poloidal magnetic field (related to the magnetic flux) that produces the toroidal magnetic field. In terms of Equation \ref{Psi}, Equation \ref{DE_F2_Psi} can be expressed as
\begin{equation}
\begin{aligned}
\lambda r^{2}\left(\frac{H_{\rm r}'}{H_{\rm r}}\right)^{2}+r^{2}\frac{H_{\rm r}''}{H_{\rm r}}+2r\frac{H_{\rm r}'}{H_{\rm r}} =\frac{\beta^{2}\left(1+\lambda\right)}{\sin^{2}\theta}
-\lambda\left(\frac{T_{\rm r}'}{T_{\rm r}}\right)^{2}-\frac{T_{\rm r}''}{T_{\rm r}}-\cot\theta\frac{T_{\rm r}'}{T_{\rm r}},
\label{DE_F2_Psi_sep}
\end{aligned}
\end{equation}
where $H_{\rm r}'=dH_{\rm r}/dr$, $H_{\rm r}''=d^{2}H_{\rm r}/dr^{2}$, $T_{\rm r}''=d^{2}T_{\rm r}/d\theta^{2}$ and $T_{\rm r}'=dT_{\rm r}/d\theta$. The left-hand side of the above equation is only a function of $r$ (the $r$ component), and the right-hand side is only a function of $\theta$ (the $\theta$ component). Both equal the same constant. Let us set this constant as $\left(1+\lambda\right)\nu^{2}+\nu$, a choice making the solution of the $r$ component equation  concise:
\begin{equation}
\begin{aligned}
H_{\rm r}\left(r\right)=r^{\nu}.
\label{Hrrnu}
\end{aligned}
\end{equation}
We introduce a new variable,
\begin{equation}
\begin{aligned}
y=\sin^{2}\theta,
\label{y}
\end{aligned}
\end{equation}
which makes the $\theta$-component equation become
\begin{equation}
\begin{small}
\begin{aligned}
4\left(1-y\right)y^{2}T_{\rm r}T_{\rm r}''+4\lambda\left(1-y\right)y^{2}T_{\rm r}'^{2}-
2\left(3y-2\right)yT_{\rm r}T_{\rm r}'-\beta^{2}\left(\lambda+1\right)T_{\rm r}^{2}+\nu\left(\nu+1+\lambda\nu\right)yT_{\rm r}^{2}=0,
\label{DE_T_y_F2}
\end{aligned}
\end{small}
\end{equation}
where $T_{\rm r}'=dT_{\rm r}/dy$ and $T_{\rm r}''=d^{2}T_{\rm r}/dy^{2}$. Let us first consider its asymptotic properties. In the case of $\theta\ll1$ (keeping in mind $T\rightarrow0$ when $\theta\rightarrow0$), the leading-order terms give
\begin{equation}
\begin{aligned}
\lambda\left(\frac{dT_{\rm r}}{d\theta}\right)^{2}+T_{\rm r}\frac{d^{2}T_{\rm r}}{d\theta^{2}}+\cot\theta T_{\rm r}\frac{dT_{\rm r}}{d\theta}-\frac{\beta^{2}\left(\lambda+1\right)T_{\rm r}^{2}}{\sin^{2}\theta}=0.
\label{DE_T_1}
\end{aligned}
\end{equation}
It is clear that this equation has a solution of the form
\begin{equation}
\begin{aligned}
T_{\rm r}\left(\theta\right)\propto\theta^{\beta}.
\label{Tbeta}
\end{aligned}
\end{equation}
Therefore, a magnetic field line forms a ``general parabolic" configuration\footnote{In the case of $\nu=0$, $\Psi$ is only a function of $\theta$, which presents a monopole solution. The case of $\nu=\beta$ leads to $\Psi\propto R$, which gives a cylindrical solution.} at $\theta\ll1$, that is, $\Psi\propto r^{\nu}\theta^{\beta}$. Now, let us consider what value $\beta$ might take. In order to do this, we have to consider the higher-order terms of Equation \ref{DE_T_y_F2}, which may be comparable to the nonrotation term. Therefore, we have to consider the original Equation \ref{DE_Psi}. Let us substitute a general form of $\Psi\propto r^{\nu}\theta^{\beta}\left(1+a_{1}\theta+a_{2}\theta^{2}+a_{3}\theta^{3}+...\right)$ (with coefficients $a_{1},a_{2},a_{3}...$ to be determined) into the original Equation \ref{DE_Psi}. One gets the first two leading-order terms (note that $a_{1}=0$):
\begin{eqnarray}
& & \beta\left(\beta-2\right)
-\left[\nu\left(1+\nu+\lambda\nu\right)
-\frac{\beta^{2}}{3}\left(1+\lambda\right)
-\frac{\beta}{3}+4a_{2}\left(1+\beta+\beta\lambda\right)\right]\Omega^{2}\Psi^{2/\nu}\theta^{4-2\beta/\nu}  \nonumber \\
& = & \beta\left(\beta-2\right)
-\left[\nu\left(1+\nu+\lambda\nu\right)
-\frac{\beta^{2}}{3}\left(1+\lambda\right)
-\frac{\beta}{3}+4a_{2}\left(1+\beta+\beta\lambda\right)\right]\Omega^{2}r^{2}\theta^{4}
\approx0.
\label{DE_T_2}
\end{eqnarray}
It can be seen that, for any values of $\beta, \lambda, and \nu$, one always has an $a_{2}$ to make the second term vanish, whereas the first term yields $\beta\approx2$ ($\beta=0$ corresponds to the nonrotation case). We note that this choice cannot guarantee that the higher-order terms vanish, so this solution is only an approximation. In the case of $\beta<2\nu$ (i.e. $\nu\gtrsim1$), the first term dominates over the second term in Equation \ref{DE_T_2}, whereas in the case of $\beta>2\nu$ (i.e. $\nu\lesssim1$), the second term dominates over the first one. We therefore expect that the approximation $\beta\approx2$ would be more efficient in the former case (i.e. $\nu\gtrsim1$). In another aspect, from Equation \ref{Bmag_Psi}, one has $B_{r}\propto\theta^{\beta-2}$, which implies that the choice of $\beta=2$ can avoid singularity or vanishing magnetic fields on the magnetic polar axis (either $B_{\theta}$ or $B_{\phi}$  vanishes). This choice is also supported by some numerical simulations, which showed that this choice corresponds to a minimum torque (i.e. the least amount of toroidal magnetic fields) and is the one picked by a ``real" system \citep[see, e.g.,][]{1969ApJ...158..727M, 1995ApJ...446...67C, 2007MNRAS.375..548N, 2008MNRAS.388..551T}. Although the coefficient $\beta\approx2$ is derived asymptotically ($\theta\ll1$), which must hold throughout the jet region because $\Phi$, $\Omega$ and $\Psi$ are each conserved along magnetic field lines. In Appendix \ref{app:relation_Phi}, we come back to this question and present another proof of the relation $\Phi\approx-2\Omega\Psi$ based on a more physical consideration, showing that this relation is only valid in the limit of a highly magnetized jet flow \citep[e.g.,][]{2009ApJ...698.1570L}.

Equation \ref{DE_T_y_F2} can be solved analytically, with a general solution
\begin{equation}
\begin{aligned}
T_{\rm r}\left(y\right)=A_{2}e^{\frac{\nu}{s+\nu}\int_{1}^{y}\frac{G_{1}(t)+A_{1}G_{2}(t)}{A_{1}G_{3}(t)+G_{4}(t)}dt},
\label{T2_general_beta}
\end{aligned}
\end{equation}
where
\begin{eqnarray}
a_{1} & = & \frac{b}{2}-\frac{s}{2}+\frac{\beta s}{2\nu}-\frac{\nu}{2};\
b_{1} = \frac{1}{2}+\frac{b}{2}+\frac{s}{2}+\frac{\beta s}{2\nu}+\frac{\nu}{2};\
c_{1} =  1+\beta+\frac{s\beta}{\nu}, \nonumber \\
a_{2} & = & -\frac{\beta}{2}-\frac{s}{2}-\frac{\beta s}{2\nu}-\frac{\nu}{2};\
b_{2}=\frac{1}{2}-\frac{b}{2}+\frac{s}{2}-\frac{\beta s}{2\nu}+\frac{\nu}{2};\
c_{2}=1-b-\frac{\beta s}{\nu}, \nonumber \\
G_{1} & = & \frac{\beta\left(s+\nu\right)}{2\nu}\ _{2}F_{1}\left(a_{1},b_{1},c_{1},t\right)t^{-1+\frac{\beta\left(s+\nu\right)}{2\nu}}+
\frac{a_{1}b_{1}}{c_{1}}\ _{2}F_{1}\left(a_{1}+1,b_{1}+1,c_{1}+1,t\right)t^{\frac{\beta\left(s+\nu\right)}{2\nu}}, \nonumber \\
G_{2} & = & \frac{-\beta\left(s+\nu\right)}{2\nu}\ _{2}F_{1}\left(a_{2},b_{2},c_{2},t\right)t^{-1-\frac{\beta\left(s+\nu\right)}{2\nu}}+
\frac{a_{2}b_{2}}{c_{2}}\ _{2}F_{1}\left(a_{2}+1,b_{2}+1,c_{2}+1,t\right)t^{\frac{-\beta\left(s+\nu\right)}{2\nu}}, \nonumber \\
G_{3} & = & _{2}F_{1}\left(a_{2},b_{2},c_{2},t\right)t^{\frac{-\beta\left(s+\nu\right)}{2\nu}}, \nonumber \\
G_{4} & = & _{2}F_{1}\left(a_{1},b_{1},c_{1},t\right)t^{\frac{\beta\left(s+\nu\right)}{2\nu}},
\label{T2_details_beta}
\end{eqnarray}
where $A_{1,2}$ are integration constants, $_{2}F_{1}\left(a,b,c,x\right)$ are the Hypergeometric functions, and the constant
\begin{equation}
\begin{aligned}
s\equiv\nu\lambda
\label{snulambda}
\end{aligned}
\end{equation}
measures the slope of the radial profile of the angular velocity on the AD plane. The constant $A_{2}$ measures the amplitude of $T_{\rm r}$ and therefore the strength of the magnetic field. Here, we just normalize $T_{\rm r}\left(\theta=\pi/2\right)=1$ for simplicity, but in reality $T_{\rm r}$ can be multiplied by an arbitrary constant and still satisfy the original equation (see below and Appendix \ref{app:sign}). Therefore, one can set $A_{2}=1$ to make $T_{\rm r}\left(y=1\right)=1$ and $T_{\rm r}\left(y=0\right)=0$. However, $A_{1}$ is still unconstrained, which means that there are infinite solutions with different values of $A_{1}$ that can all satisfy $\Psi\left(r,\theta=0\right)=0$. Before further discussing how to determine $A_{1}$, let us consider the case near the CO/AD, where the nonrotation term would dominate.  The nonrotation term Equation \ref{DE_F1_Psi} can be easily solved \citep[see also][]{2008MNRAS.388..551T}, in terms of variable-separated $\Psi=H_{\rm nr}\left(r\right)T_{\rm nr}\left(\theta\right)$ (where the subscript `nr' denotes `non-rotation') as follows:
\begin{eqnarray}
\Psi & = & r^{\nu}T_{\rm nr}\left(\theta\right), \\
T_{\rm nr}\left(\theta\right) & = & C_{2}y
\ _{2}F_{1}\left(1-\frac{\nu}{2},\frac{1}{2}+\frac{\nu}{2},2,y\right) =\ _{2}F_{1}\left(\frac{\nu}{2}-\frac{1}{2},-\frac{\nu}{2},\frac{1}{2},\mu^{2}\right)
-C_{1}\mu\ _{2}F_{1}\left(\frac{1}{2}-\frac{\nu}{2},\frac{\nu}{2},\frac{3}{2},\mu^{2}\right),
\label{Solution_nonr}
\end{eqnarray}
where we define\footnote{Notice that both $C_{1}$ and $C_{2}$ are symmetric functions of $\nu$ centered at $\nu=1/2$. The former increases from $C_{1,\nu=0}=1$ to the maximum value at $C_{1,\nu=1/2}=\left[\Gamma\left(1/4\right)\right]^{4}/16\pi^{2}=1.094$ and continuously decreases through $C_{1,\nu=1}=1$ to $C_{1,\nu=2}=0$. The latter decreases from $C_{2,\nu=0}=1/2$ to the minimum value $C_{2,\nu=1/2}=\left[\Gamma\left(1/4\right)\right]^{2}/16\sqrt{\pi}=0.4635$ and then continuously increases through $C_{2,\nu=1}=1/2$ to $C_{2,\nu=2}=1$.}
\begin{eqnarray}
\mu & = & \cos\theta,\\
C_{1} & = & \frac{\nu\Gamma\left(3/2-\nu/2\right)\Gamma\left(\nu/2\right)}{\Gamma\left(1-\nu/2\right)\Gamma\left(\nu/2+1/2\right)},\\ C_{2} & = & \frac{\Gamma\left(3/2-\nu/2\right)\Gamma\left(1+\nu/2\right)}{\sqrt{\pi}}.
\label{C1C2}
\end{eqnarray}
Notice that we have used the condition $\Psi\left(r,\theta=0\right)=0$ and the normalization $T_{\rm nr}\left(\theta=\pi/2\right)=1$, keeping in mind that the amplitude of $T_{\rm nr}$ (or $\Psi$) can be multiplied by an arbitrary constant (see below and Appendix \ref{app:sign}). The same applies to the rotation term equation. The radial component also follows a power-law distribution: $H_{\rm r/nr}\left(r\right)=r^{\nu}$. This is the so-called self-similar solution employed in many papers \citep[e.g.][]{2007MNRAS.375..548N}. The scaling relations in these solutions can capture key aspects of the jet problem as also found in MHD simulations \citep[see, e.g,][]{2007MNRAS.375..548N, 2008MNRAS.388..551T}.

As discussed above, there are infinite solutions with different values of $A_{1}$ that can satisfy the rotation term Equation \ref{DE_T_y_F2}. They are also expected to present different trends when approaching $\theta\rightarrow\pi/2$ even though they all satisfy $T_{\rm r}\left(\theta=\pi/2\right)=1$. One may wonder whether there is an $A_{1}$ that can make the rotation and non-rotation solutions have approximately the same trend when approaching $\theta\rightarrow\pi/2$, i.e. $T_{\rm r}\left(y\right)\approx T_{\rm nr}\left(y\right)$.

Differentiating Equation \ref{T2_general_beta} with $y$, one gets
\begin{equation}
\begin{aligned}
\frac{1}{T_{\rm r}}\frac{dT_{\rm r}}{dy}=\frac{\nu}{s+\nu}\frac{G_{1}(y)+A_{1}G_{2}(y)}{A_{1}G_{3}(y)+G_{4}(y)},
\label{T2_ratio_beta}
\end{aligned}
\end{equation}
which can be set to equal $\frac{1}{T_{\rm nr}}\frac{dT_{\rm nr}}{dy}$ at $y=1$ to guarantee that $T_{\rm nr}$ and $T_{\rm r}$ follow the same trend approaching $y\rightarrow1$. Therefore, one can get a constraint on $A_{1}$:
\begin{equation}
\begin{aligned}
A_{1}=D\frac{D_{1}+D_{2}}{D_{3}+D_{4}}.
\label{T2_A1_beta}
\end{aligned}
\end{equation}
where
\begin{small}
\begin{eqnarray}
D & = & \left[\beta\left(s+\nu\right)-\nu\right]
\frac{\Gamma\left(1+\beta+\beta s/\nu\right)\Gamma\left(-\nu/2-\beta s/2\nu-\left(\beta+s+1\right)/2\right)\Gamma\left(\nu/2-\beta s/2\nu+\left(2-\beta+s\right)/2\right)}{\Gamma\left(-\nu/2+\left(1+\beta-s\right)/2+\beta s/2\nu\right)\Gamma\left(\nu/2+\left(2+\beta+s\right)/2+\beta s/2\nu\right)}, \nonumber \\
D_{1} & = & \left(-2-\nu+\nu^{2}\right)
\frac{\Gamma\left(3/2-\nu/2\right)\Gamma\left(1+\nu/2\right)}{\Gamma\left(2-\nu/2\right)\Gamma\left(3/2+\nu/2\right)}, \nonumber \\
D_{2} & = & \left(\beta-\nu\right)\left[\beta(s+\nu)+\nu(1+s+\nu)\right]
\frac{\Gamma\left(2+\beta+\beta s/\nu\right)\Gamma\left(-\nu/2+(1+\beta-s)/2+\beta s/\nu\right)\Gamma\left(\nu/2+(2+\beta+s)/2+\beta s/\nu\right)}{\Gamma\left(-\nu/2+(2+\beta-s)/2+\beta s/2\nu\right)\Gamma\left(\nu/2+(3+\beta+s)/2+\beta s/2\nu\right)}, \nonumber \\
D_{3} & = & \left(2+\nu-\nu^{2}\right)\left(\beta s+\beta\nu-\nu\right)
\frac{\Gamma\left(3/2-\nu/2\right)\Gamma\left(1+\nu/2\right)\Gamma\left(1-\beta-\beta s/\nu\right)}{\Gamma\left(2-\nu/2\right)\Gamma\left(3/2+\nu/2\right)}, \nonumber \\
D_{4} & = & \left(\beta+\nu\right)\left[\beta(s+\nu)-\nu(1+s+\nu)\right]
\frac{\Gamma\left(2-\beta-\beta s/\nu\right)\Gamma\left(-\nu/2-(\beta+s-1)/2-\beta s/2\nu\right)\Gamma\left(\nu/2+(2-\beta+s)/2-\beta s/2\nu\right)}{\Gamma\left(-\nu/2-(\beta+s-2)/2-\beta s/2\nu\right)\Gamma\left(\nu/2+(3-b+s)/2-\beta s/2\nu\right)}.
\label{T2_A1_details_beta}
\end{eqnarray}
\end{small}
With this choice of $A_{1}$, both rotation and nonrotation terms reach $T_{\rm r}\left(y=1\right)=T_{\rm nr}\left(y=1\right)=1$. Furthermore, they follow the same trend of approaching $y\rightarrow1$ with an error on the order of\footnote{Either $T_{\rm r}$ or $T_{\rm nr}$ can be expanded to the series $a_{0}+a_{1}\mu+a_{2}\mu^{2}+a_{3}\mu^{3}+\cdot\cdot\cdot$, and $dT/d\mu=\left(dy/d\mu\right)\left(dT/dy\right)$.} $\sqrt{1-y}=\mu$. In another limit in the region of $\Omega r\sin\theta\gg1$, the rotation term dominates, so that  $\Psi=r^{\nu}T_{\rm r}$ can be considered as a good approximation for the solution of the original Equation \ref{DE_Psi}. In the case of $\beta\approx2$ in particular, at $\theta\ll1$, even though the nonrotation term can be ignored compared with the rotation term, they follow the same trend, that is, $T_{\rm r}\propto T_{\rm nr}\propto\theta^{2}$ (with an error on the order of $\sin^{2}\theta\approx\theta^{2}$). This means that, for the choice of $\beta=2$, either the form $\Psi=r^{\nu}T_{\rm nr}\left(\theta\right)$ or $\Psi=r^{\nu}T_{\rm r}\left(\theta\right)$ can be considered as an approximation for the solution of the original Equation \ref{DE_Psi}. This is why some MHD simulations found that the poloidal structure of magnetic fields changes quite mildly from the nonrotation to the rotation cases \citep[e.g.,][]{2008MNRAS.388..551T}. For the parameter $\nu$, we consider $0\leq\nu\leq2$ because we are interested in the case of a collimated jet whose (1) enclosed magnetic flux increases with increasing radius, and whose (2)  strength of magnetic field decreases with increasing radius \citep[e.g.,][]{1995ApJ...446...67C, 2003ApJ...596.1080V, 2007MNRAS.375..548N}. The $\nu$ value is a free parameter in the model, which can be determined from observations (see Section \ref{sec:magnetic_configuration}). Throughout the paper, we take $\nu=3/4$ as a typical value in the later discussion \citep[see more discussion on this choice in Section \ref{Bp_AD} and][]{1982MNRAS.199..883B, 2007MNRAS.375..548N, 2007MNRAS.375..513M, 2007MNRAS.375..531M}.

\begin{figure}
\begin{center}
{\includegraphics[width=0.8\linewidth]{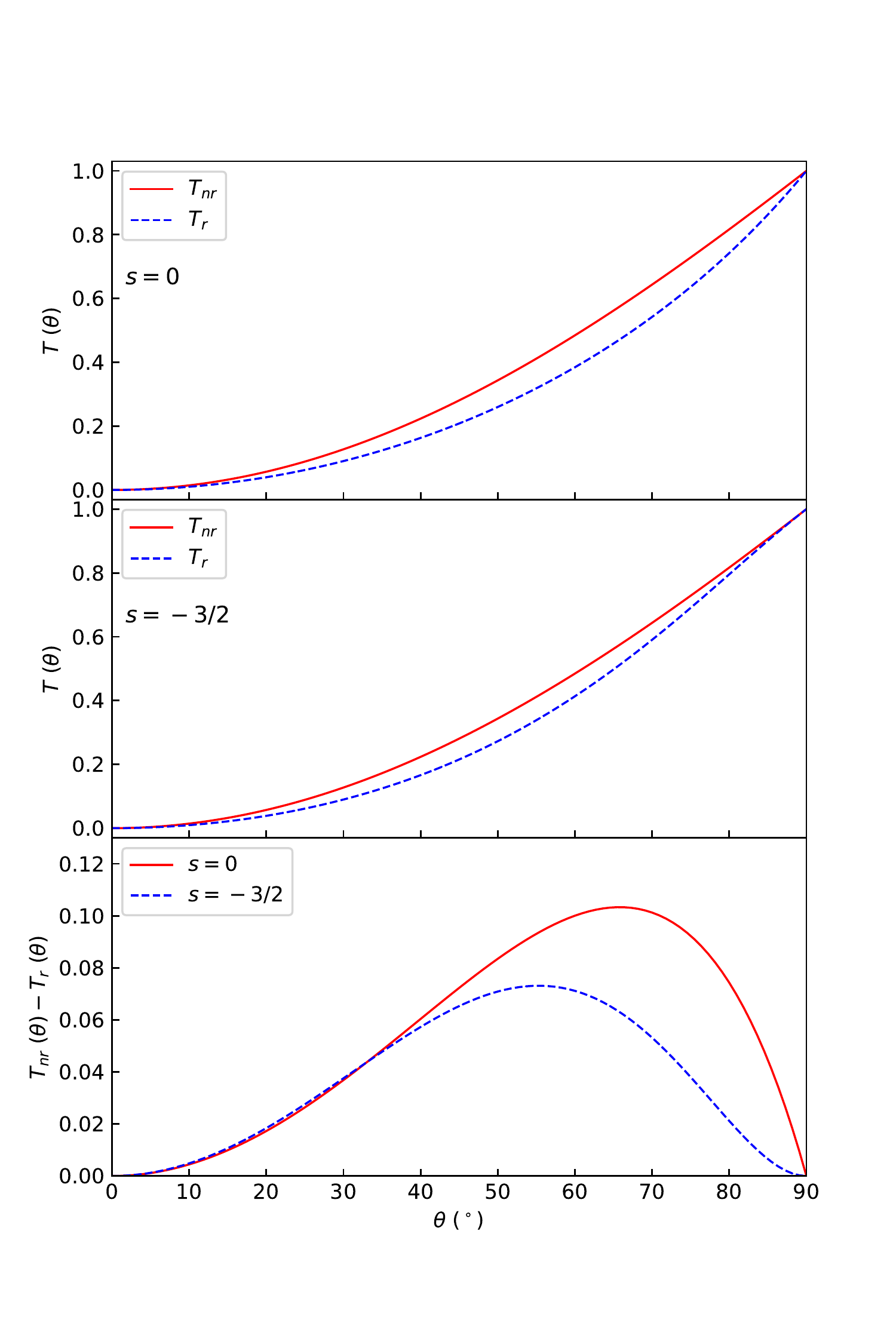}}
\end{center}
\caption{Comparison between the solutions of nonrotation and rotation terms: $T_{\rm nr}\left(\theta\right)$ vs. $T_{\rm r}\left(\theta\right)$ (assuming $\nu=3/4$). The upper panel shows the case of $s=\lambda\nu=0$ (the magnetic field lines threading the CO), and the middle panel indicates the case of $s=-3/2$ (the magnetic field lines threading a Keplerian AD with $\Omega\propto R_{0}^{-3/2}$). In both cases, they match each other at $\theta\ll1$ and $\theta\rightarrow\pi/2$.}
\label{fig:Tnr.vs.Tr}
\end{figure}

In the case of $\lambda=0$ (threading a CO), the rotation term solution (Equation \ref{T2_general_beta}) can be reduced to the simple form
\begin{eqnarray}
T_{\rm r}\left(\theta\right)&=&\frac{1}{2^{\beta}\sqrt{\pi}}\Gamma\left(\frac{\nu}{2}-\frac{\beta}{2}+1\right)\Gamma\left(\frac{1}{2}-\frac{\nu}{2}-\frac{\beta}{2}\right)
P_{\nu}^{\beta}\left(\mu\right) \nonumber \\
&=&\frac{1}{\Gamma\left(1+\beta\right)\sqrt{\pi}}\Gamma\left(-\frac{\nu}{2}+\frac{\beta}{2}+\frac{1}{2}\right)\Gamma\left(1+\frac{\nu}{2}+\frac{\beta}{2}\right)
y^{\beta/2}\ _{2}F_{1}\left(\frac{\beta}{2}-\frac{\nu}{2},\frac{\beta}{2}+\frac{1}{2}+\frac{\nu}{2},1+\beta,y\right),
\label{Solution_nonr_s_0_b}
\end{eqnarray}
where $P_{\nu}^{\beta}\left(\mu\right)$ is the Legendre function with the order $\beta$ and the degree $\nu$. This formula can also be directly derived by solving Equation \ref{DE_T_y_F2}. In the following discussion, we will focus on the case of $\beta=2$:
\begin{equation}
\begin{aligned}
\Phi=-2\Omega\Psi.
\label{Phi2OmegaPsi}
\end{aligned}
\end{equation}
In Figure \ref{fig:Tnr.vs.Tr}, we plot the comparison between $T_{\rm r}\left(\theta\right)$ and $T_{\rm nr}\left(\theta\right)$ for various cases: $s=-3/2$ corresponds to  foot-points on a Keplerian AD, and $s=0$ corresponds to foot-points on a CO. In both cases, they match each other in the regions of $\theta\ll1$ and $\theta\rightarrow\pi/2$.

The parameter $\Psi$ determines the poloidal magnetic field, while $\Phi=-2\Omega\Psi$ determines the toroidal magnetic field. Given $\Psi$ and $\Omega$ are specified, other physical qualities can be estimated (as a reminder, this approximate solution applies only in the magnetically dominated region). Because $T_{\rm nr}$ and $T_{\rm r}$ match each other with an error no larger than $\theta^{2}$ at $\theta\ll1$ and no larger than $\cos\theta$ at $\theta\rightarrow\pi/2$, one therefore expects that (1) if a quantity relates to the second-order derivative of $\Psi$ (e.g., toroidal current or charge density), its estimation will bring an error in the region $\theta\rightarrow\pi/2$ (although not very large; see Section \ref{subsec:charge} for details) but is still accurate enough in the region $\theta\ll1$; (2) if a quantity is only related to the first-order derivative of $\Psi$ (e.g. electromagnetic fields, velocity, poloidal current, and jet power), the estimation will be accurate enough for both regions $\theta\rightarrow\pi/2$ and $\theta\ll1$ (see discussion below). Therefore, even though the approximation $\Psi=r^{\nu}T\left(\theta\right)=r^{\nu}T_{\rm nr}\left(\theta\right)$ may not exactly guarantee a smooth transition through the singular point \citep[this condition is usually employed to solve the Equation \ref{DE_Psi} numerically, e.g,][]{1992ApJ...394..459L, 1999ApJ...511..351C, 2013ApJ...765..113C}, it still presents enough precision for practical purposes.

Equations \ref{DE_F2_Psi} (rotation term) and \ref{DE_F1_Psi} (nonrotation term) are originally separated from Equation \ref{DE_Psi} based on $\Omega r\sin\theta\gg1$ or $\Omega r\sin\theta\ll1$, respectively, with the conditions corresponding to the jet being relativistic or nonrelativistic (see Section \ref{sec:jet_acceleration}). The fact that $T_{\rm nr}$ matches $T_{\rm r}$ in the regions $\theta\ll1$ or $\theta\rightarrow\pi/2$ implies that our approximate solution applies to either $\theta\ll1$ or $\theta\rightarrow\pi/2$, and either nonrelativistic or relativistic. Therefore, the solution can describe the acceleration of collimated AGN jets ($\theta\ll1$) continuously from the nonrelativistic to the relativistic regimes \citep[e.g., M87 jet, ][]{2007ApJ...668L..27K, 2017PASJ...69...71H}. As a result, either $\theta\ll1$ or $\theta\rightarrow\pi/2$, and either nonrelativistic or relativistic regimes can be adapted to the approximate solution. Recall that this approximate solution does not assume $\Omega\propto R_{0}^{-1}$ \citep[a ``flat rotation" on the AD plane; see, e.g.,][]{1992ApJ...394..459L, 2007MNRAS.375..548N} and can apply to the general $\Omega\propto R_{0}^{s}$, including $s=-3/2$ for a Keplerian AD. It is the ansatz Equation \ref{Omega} that helps to solve the problem to get the approximated solution. The fact that $\Psi=r^{\nu}T_{\rm nr}\left(\theta\right)$ is not a function of $\alpha$ or $\lambda$ also indicates that the configuration of the magnetic stream surface is independent of rotation (see next Section \ref{sec:magnetic_configuration} for more discussion). In this aspect, $\Omega$ could be any piecewise-defined function of $\Psi$ (each subfunction should follow the form $\alpha\Psi^{\lambda}$, but could have different $\alpha$ and $\lambda$ values). One can always derive the approximate ($\Omega$-independent) solution $\Psi=r^{\nu}T_{\rm nr}\left(\theta\right)$.

In principle, $\Psi$ can be positive or negative, which corresponds to $\mathbf{B}_{p}$ pointing in two opposite directions. The scalar angular velocity is defined as $\Omega=\mathbf{\Omega}\cdot\hat{z}$, which implies that $\Omega$ can also be positive or negative, corresponding to the $\mathbf{\Omega}$ direction being along or opposite to the polar axis. Whether $\mathbf{B}_{\phi}$ rotates forward or backward is determined by the sign of $\Phi\approx-2\Omega\Psi$. In the following sections, we just discuss the case where both $\Psi$ and $\Omega$ are positive; that is, both the $\mathbf{\Omega}$ rotating vector and $\mathbf{B}_{p}$ projection vector are along the polar axis (one of the pair of  opposite jets). For other choices of $\mathbf{\Omega}$ and $\mathbf{B}_{p}$, some formulae in the following sections may need to change signs. The sign convention for different cases is presented in Appendix \ref{app:sign}. One can also multiply an arbitrary constant to $\Psi$, which still satisfies the original Equation \ref{DE_Psi}. This constant determines the absolute value of the magnetic field strength, which further affects other quantities. The effect is discussed in Appendix \ref{app:sign} in detail.

\section{Electromagnetic Field Configuration}
\label{sec:magnetic_configuration}

For the specific cases of $\nu=0$, 1, and 2, the magnetic stream function $\Psi=r^{\nu}T$ expresses very simple functions: $\Psi=1-\cos\theta$ for $\nu=0$ (monopole), $\Psi=r\left(1-\cos\theta\right)$ for $\nu=1$ (parabola), and $\Psi=r^{2}\sin^{2}\theta$ for $\nu=2$ (cylinder), which are  exact solutions of Equation \ref{DE_Psi} and have been studied in previous works under some special assumptions \citep[see Appendix \ref{sec:exactsolutions} and][]{1973ApJ...180L.133M, 1994MNRAS.267..629I, 1976MNRAS.176..465B, 2007MNRAS.375..548N}. General asymptotic properties of $T\left(\theta\right)$ at $\theta\ll1$ and $\theta\rightarrow\pi/2$ can be found in Appendix \ref{app:T}. Note that $\Psi=$ const defines a magnetic stream surface, in which the magnetic field lines lie, plasma fluids stream, and the currents flow. Therefore, a magnetic stream surface can measure the configuration of magnetic fields and the jet flow. In the region of $\theta\ll1$, given a magnetic stream surface specified, the jet half-opening angle and half-width are defined by
\begin{eqnarray}
\theta & = & C_{2}^{-1/2}\Psi^{1/2}r^{-\nu/2}, \\
R & = & C_{2}^{-1/2}\Psi^{1/2}z^{1-\nu/2},
\label{half_openangle_R}
\end{eqnarray}
which show a ``general parabolic" configuration \citep[see][]{2007MNRAS.375..548N}. Generally, if the magnetic field line is rooted at the foot-point $r_{0}$ and $\theta_{0}$, the conservation of $\Psi$ implies $r_{0}^{\nu}T\left(\theta_{0}\right)=C_{2}r^{\nu}\theta^{2}$, which indicates that the foot-point location and the parameter $\nu$ can be derived by measuring the jet configuration (i.e. the $r$-$\theta$ or $R$-$z$ relations, for the special case of magnetic field lines threading a BH\footnote{In principle, one needs to consider GR effects to deal with magnetic field lines threading a BH. As shown in \citet{2007MNRAS.375..513M, 2007MNRAS.375..531M} and \citet{2019PhRvL.122c5101P}, GR effects do not qualitatively change the magnetic field configuration even close to the BH. See Section \ref{BHjet} for more information.}; see Section \ref{BHjet}.).

As presented in the MHD simulations by, for example, \citet[][]{2008MNRAS.388..551T}, the poloidal magnetic field configuration in the final rotating state is nearly the same as in the initial nonrotating state, despite the fact that the final steady solution has a strong axisymmetric toroidal field $B_{\phi}$. The same result is also shown in our solutions: that is, either $B_{r}$ or $B_{\theta}$ is independent of angular velocity (see Equations \ref{Bmag_Psi} and \ref{Solution_nonr}), which indicates that the poloidal magnetic structure is independent of rotation and collimation seems to be achieved by the poloidal field itself (see Equation \ref{half_openangle_R}). Rotation twists the poloidal magnetic field to make a toroidal component and induce a poloidal current. The force that this poloidal current received in the magnetic field is  balanced by the force exerted by the appearing poloidal electric field to guarantee a force-free condition (see Section \ref{sec:Lorentz_Force} for a detailed analysis). Therefore, the conservation of magnetic field flux makes the poloidal field configuration almost change-free. Roughly, the collimating hoop stress associated with the toroidal field is canceled by the decollimating effect of the pressure gradient associated with the same field \citep[balanced with centrifugal forces for plasma rotation; see discussion in Section \ref{sec:Lorentz_Force} and, e.g.,][]{1997ApJ...486..291O, 2007MNRAS.375..531M, 2007MNRAS.375..548N, 2008MNRAS.388..551T}. In fact, each field line is collimated by the pressure associated with the field line itself farther out. This result applies for the case of $\Omega\propto R_{0}^{s}$ (on the AD plane), compared with the case of a ``flat rotation" $\Omega\propto R_{0}^{-1}$ studied by \citet[][]{2007MNRAS.375..548N}. This type of ``general parabolic" collimating jet has been observed in many AGNs; for example, the continuous collimation of the M87 jet indicates $R\propto z^{0.56}$ \citep[e.g.,][]{2012ApJ...745L..28A, 2013ApJ...775...70H}, which is consistent with Equation \ref{half_openangle_R} with $\nu=0.88$.

Now let us consider the magnetic topology, which can be described by the ratio of toroidal to poloidal magnetic field strength:
\begin{equation}
\begin{aligned}
\frac{-B_{\phi}}{B_{p}}=\frac{-\Phi}{\left|\nabla\Psi\right|}
=\frac{2\Omega\Psi}{\left|\nabla\Psi\right|}=\Omega r\frac{2}{\sqrt{\nu^{2}+\left(\frac{1}{T}\frac{dT}{d\theta}\right)^{2}}}.
\label{B_t_p}
\end{aligned}
\end{equation}
It can be easily proved that, in the region of $\theta\ll1$, one has $-B_{\phi}/B_{p}=\Omega r\sin\theta$, which implies that the toroidal magnetic field vanishes on the polar axis (so do the velocity and Poynting flux; see below). In the region of $\theta\rightarrow\pi/2$, one can also approximate\footnote{The maximum error of the last equality/approximation is $50\%$ when $\nu\rightarrow0$. A more accurate approximation is $-B_{\phi}/B_{p}\approx\Omega r\sqrt{2\left(1-\cos\theta\right)}$.} $-B_{\phi}/B_{p}=2\Omega r/\sqrt{\nu^{2}+C_{1}^{2}}\approx\Omega r\sin\theta$. Therefore, the approximation $-B_{\phi}/B_{p}\approx\Omega R$ roughly applies throughout the entire jet region \citep[ignoring the GR effects, see GRMHD simulations by, e.g.,][]{2006MNRAS.368.1561M, 2020ApJ...892...37P}, which implies that the helical magnetic field is approximately shaped as an ``Archimedean spiral" (arithmetic spiral) on the magnetic stream surface. Such a configuration also describes the structure of interplanetary magnetic fields being twisted by the Sun's rotation \citep[often called Parker's spiral,][]{1958ApJ...128..664P}. See Appendix \ref{app:magnetic_field_morphology} for the details of calculating the 3D morphology of helical magnetic field lines.

\begin{figure}
\begin{center}
{\includegraphics[width=0.7\linewidth]{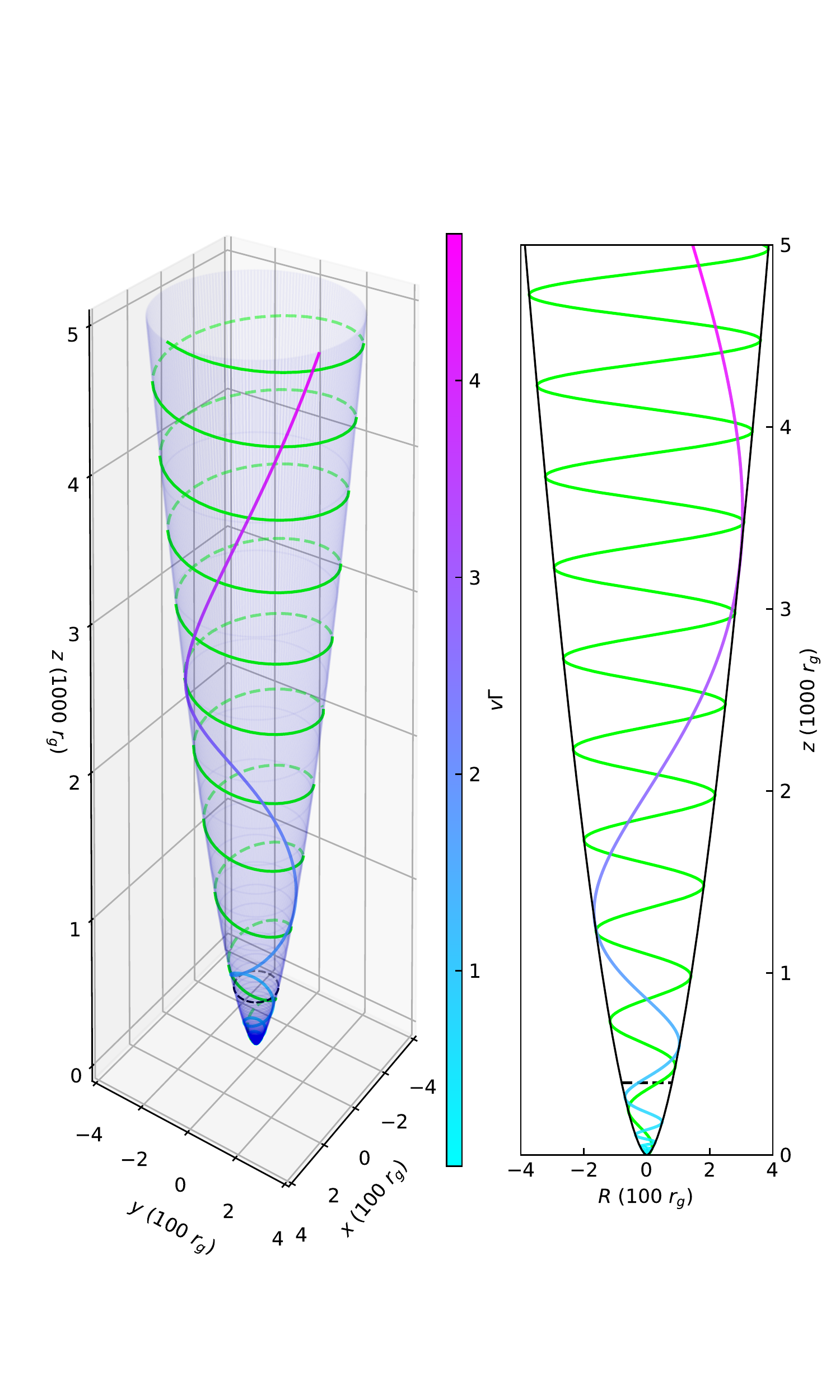}}
\end{center}
\caption{Configuration of the magnetic field line and velocity on a magnetic stream surface (illustrated by the light blue semitransparent surface in the left panel) threading the equator of a BH with a spin parameter $a=0.1$: the green line refers to the structure of a magnetic field line, and the gradient colored line represents the velocity profile with the color measuring the value of the four-velocity $v\Gamma$. The right panel is a 2D projection as seen edge-on of the left 3D configuration panel. The black dashed line shows the ACS ($z=396$ $r_{\rm g}$).}
\label{fig:3Djet}
\end{figure}

At the foot-point, the rotation velocity of magnetic fields may not be very relativistic, which indicates that the magnetic field lines are initially predominantly poloidal up to the ACS\footnote{Notice that in the region of $\theta\ll1$, the ACS follows $\alpha C_{2}^{s/\nu}z^{s-2s/\nu}R^{1+2s/\nu}=1$. For $s=0$ (i.e. $\Omega=$ const), the ACS $R=1/\Omega$ forms a light cylinder (similar to the pulsar case). For $s=-3/2$ and $\nu=3/4$, the ACS is $z=\alpha^{-2/5}C_{2}^{4/5}R^{6/5}$ (as a comparison, the magnetic stream surface reads as $z=C_{2}^{4/5}\Psi^{-4/5}R^{8/5}$).} ($\Omega R=1$), beyond which the toroidal magnetic field dominates. Figure \ref{fig:3Djet} clearly presents the scheme that the poloidal magnetic field component dominates initially but later the toroidal magnetic field component takes over as the jet propagates outward (see Section \ref{BHjet} for parameter setting), which was also presented in many MHD simulations \citep[e.g.,][]{2009A&A...507.1203M}. To maintain a toroidal magnetic field, one expects an accompanying poloidal current along the outflow \citep[see discussion in Section \ref{subsec:current} and some simulation papers, e.g.,][]{2007MNRAS.375..513M}.

In the region near the AD plane (in the case of $\theta\rightarrow\pi/2$), one has
\begin{eqnarray}
B_{r} & = & \frac{r^{\nu-2}}{\sin\theta}\frac{\partial T}{\partial\theta}=\left[C_{1}+\nu\left(\nu-1\right)\cos\theta\right]r^{\nu-2}, \nonumber \\
B_{\theta} & = & -\frac{\nu r^{\nu-2}}{\sin\theta}T=-\nu\left(1-C_{1}\cos\theta\right)r^{\nu-2}, \nonumber \\
B_{\phi} & = & \frac{\Phi}{r\sin\theta}=-\left(1-C_{1}\cos\theta\right)2\Omega rr^{\nu-2}.
\label{B_disk}
\end{eqnarray}
It can be seen that the poloidal magnetic field makes an angle with respect to the midplane of the disk\footnote{The approximation $\tan^{-1}\left(\nu/C_{1}\right)\approx\nu\pi/4$ gives an error $\lesssim2^{\circ}$, which is down to $\lesssim0.4^{\circ}$ for a more accurate approximation, $\tan^{-1}\left(\nu/C_{1}\right)\approx\nu\pi/4+0.04\sin\left(\nu\pi\right)$.}, $\theta_{\rm B,AD}^{\rm AD}=\tan^{-1}\left(\left|B_{\theta}\right|/B_{r}\right)=\tan^{-1}\left(\nu/C_{1}\right)\approx\nu\pi/4=35^{\circ}$ for the case of $\nu=3/4$, which meets the requirements of the BP model \citep[the angle has to be less than roughly $60^{\circ}$ to guarantee a ``centrifugally" driven outflow from the disk; see][]{1982MNRAS.199..883B, 2012MNRAS.426.2813C}. For comparison, with the aid of Equations \ref{Bmag_Psi} and \ref{Solution_nonr}, one can derive an angle between the poloidal magnetic field and the polar axis in the case of magnetic field lines threading a CO ``surface" ($r=$ const)\footnote{The approximation $\tan^{-1}\left(B_{R}/B_{z}\right) \approx\left(2-\nu\right)\theta/2$ gives an error $\lesssim0.3^{\circ}$ for $\nu\lesssim1$, which continuously increases to $\lesssim2^{\circ}$ for $\nu\lesssim1.7$ and reaches the maximum $\lesssim6^{\circ}$ at $\nu\lesssim2$.}: $\theta_{\rm B,CO}^{z}=\tan^{-1}\left(B_{R}/B_{z}\right) \approx\left(2-\nu\right)\theta/2$.

In the region of $\theta\ll1$, one has
\begin{eqnarray}
B_{r} & = & 2C_{2}r^{\nu-2}
=2C_{2}^{2/\nu}\Psi^{1-2/\nu}\theta^{-2+4/\nu}, \nonumber \\
B_{\theta} & = & -\nu C_{2}^{1/2}\Psi^{1/2}r^{\nu/2-2}
=-\nu C_{2}^{2/\nu}\Psi^{1-2/\nu}\theta^{-1+4/\nu}
=-\nu C_{2}z^{\nu-3}R, \nonumber \\
B_{\phi} & = & -2C_{2}^{1/2}\Omega\Psi^{1/2}r^{-1+\nu/2}
=-2C_{2}^{1/\nu}\Omega\Psi^{1-1/\nu}\theta^{-1+2/\nu}
=-2\alpha C_{2}^{\lambda+1}z^{\left(\lambda+1\right)\left(\nu-2\right)}R^{2\lambda+1}, \nonumber \\
B_{R} & = & \frac{2-\nu}{2}\theta B_{r}, \nonumber \\
B_{p} & = & B_{z}=B_{r}.
\label{B_tll1}
\end{eqnarray}
It can be seen that $B_{\phi}$ decreases the slowest while $B_{\theta}$ decreases the fastest as $r$ increases. Therefore, relative to $B_{\theta}$, $B_{r}$ always dominates the poloidal component $B_{p}$, which is exceeded by the toroidal component $B_{\phi}$ beyond the ACS ($B_{p}=\left|B_{\phi}\right|\Rightarrow\Omega R\approx1$), as discussed above. Because $B_{r}$ is independent of $R$ at a given height $z$, the poloidal magnetic field does not present a stratified structure\footnote{As one can see in the GRMHD simulations \citep[e.g.,][]{2018ApJ...868..146N}, the electromagnetic pressure (magnetic pressure measured in the fluid frame),
\begin{equation}
\begin{aligned}
P_{\rm B}=\frac{B'^{2}}{8\pi}=\frac{B^{2}-E^{2}}{8\pi}
\approx\frac{1}{\Gamma_{\perp}^{2}}\frac{B^{2}}{8\pi}
\approx\frac{B_{p}^{2}}{8\pi},
\label{EB_pressure}
\end{aligned}
\end{equation}
is independent of the radial distance $R$ at a given height from the CO/AD, where $\Gamma_{\perp}=1/\sqrt{1-v_{\perp}^{2}}$, and $v_{\perp}=E/B$ is the velocity perpendicular to the magnetic field, which approximately equals the drift velocity (see Section \ref{sec:jet_acceleration} below).}, while the toroidal magnetic field shows a stratified structure. The case $s=0$ (the magnetic field line threading a CO) implies $B_{\phi}\propto R$, while $s=-3/2$ (the magnetic field line threading a Keplerian AD, with $\nu=3/4$) indicates $B_{\phi}\propto R^{-3}$. The helical magnetic field implies that jet synchrotron emission would be polarized and its RMs\footnote{When polarized emission passes through a magnetized medium, the electric vector position angle would rotate with a wavelength ($\lambda_{\rm w}$)-dependent law $\Delta\chi=RM\lambda_{\rm w}^{2}$. One roughly has the rotation measure $RM\propto\int n_{\rm e}B_{\|}dl$, where $n_{\rm e}$ is the electron density, $B_{\|}$ the line of sight component of the magnetic field and the integral should be taken over the entire path \citep[][]{1966ARA&A...4..245G}.} would present a systematic gradient with respect to the jet axis, a phenomenon that is exactly observed in AGN jets \citep[see, e.g.,][]{2002PASJ...54L..39A, 2004MNRAS.351L..89G, 2012AJ....144..105H}. Furthermore, the RM is also expected to decrease along the jet axis, which was also observed in the M87 jet recently \citep[][]{2019ApJ...871..257P}. When projected on the sky plane with a viewing angle to the jet axis, one roughly expects a transverse magnetic field at the spine and an orthogonal magnetic field at the edge of the jet. Therefore, the polarization angle (measured by the polarization $\mathbf{E}$ vector) would be aligned with the jet along the spine and the orthogonal at the edge, which is the spine-sheath polarization angle pattern shown in some AGN jets \citep[e.g.,][]{1999ApJ...518L..87A, 2005MNRAS.356..859P, 2017MNRAS.467...83K}.

From Equation \ref{E_general}, one always has $E=\Omega r\sin\theta B_{p}\approx-B_{\phi}$. Near the disk plane,
\begin{equation}
\mathbf{E}=-\Omega r^{\nu-1}\left\{\nu\left(1-C_{1}\cos\theta\right)\hat{r}+\left[C_{1}+\nu\left(\nu-1\right)\cos\theta\right]\hat{\theta}\right\}.
\label{E_disk}
\end{equation}
Since $\mathbf{E}$ is always perpendicular to $\mathbf{B}_{p}$, the inclined angle of $\mathbf{E}$ with respect to the polar axis would be equal to the angle of $\mathbf{B}_{p}$ with respect to the mid-plane of the AD, $\theta_{\rm B,AD}^{\rm AD}$. In the region of $\theta\ll1$, one has
\begin{eqnarray}
E_{r} & = & -\nu\Omega\Psi r^{-1}=-\nu\Omega
C_{2}^{1/\nu}\Psi^{1-1/\nu}\theta^{2/\nu}=-\alpha\nu C_{2}^{\lambda+1}z^{\lambda\nu+\nu-2\lambda-3}R^{2\lambda+2}, \nonumber \\
E_{\theta} & = & -2C_{2}^{1/2}\Omega\Psi^{1/2}r^{\nu/2-1}=-2 C_{2}^{1/\nu}\Omega\Psi^{1-1/\nu}\theta^{-1+2/\nu}=-2\alpha C_{2}^{\lambda+1}z^{\lambda\nu+\nu-2\lambda-2}R^{2\lambda+1}, \nonumber \\
E_{p} & = & E_{R}=E_{\theta}=B_{\phi}, \nonumber \\
E_{z} & = & -\frac{2-\nu}{2}\theta E_{\theta}.
\label{E_tll1}
\end{eqnarray}
It can be seen that $E_{r}$ decreases faster than $E_{\theta}$ along the magnetic field line, and therefore $E_{\theta}$ would dominate eventually. Both $E_{r}$ and $E_{\theta}$ present stratified configurations.

\section{Jet Velocity and Acceleration}
\label{sec:jet_acceleration}

By assumption, the ideal MHD condition implies that the electric field vanishes in the fluid rest frame. Magnetic field lines lie and the fluid streams in the magnetic stream surfaces. Combining Equations \ref{ideal_MHD} and \ref{E_general} yields
\begin{equation}
\begin{aligned}
\mathbf{v}=\Omega r\sin\theta\hat{\phi}+\kappa\mathbf{B},
\label{vgv}
\end{aligned}
\end{equation}
where $\kappa$ is a function of position. Therefore, one has the poloidal velocity being parallel to the poloidal magnetic field and $v_{\phi}B_{p}-B_{\phi}v_{p}=\Omega r\sin\theta B_{p}$. This implies that the fluid elements behave like beads on a rigid wire (magnetic field line), which rotates with angular velocity $\Omega$; that is, the plasma fluid element slides along the rotating magnetic field lines (the plasma is actually frozen on the magnetic field line as seen in the ``corotating frame"). A simple analysis can present a trend of velocity profile in the region far from the ACS. For example, in the region far below the ACS where $B_{p}$ dominates over $B_{\phi}$ while the toroidal velocity may be larger than the poloidal velocity, one immediately has $v_{\phi}\approx\Omega R$ (i.e., the plasma fluid almost corotates with the magnetic field). On the other hand, in the region far above the ACS, conservation of angular momentum implies $v_{\phi}\approx1/\Omega R$ \citep[see Equation \ref{v_flow_pphi} below in Section \ref{sec:dynamics_velocity} and also][]{1982MNRAS.199..883B}. In principle, one has to consider the inertia to calculate the plasma velocity, which will be discussed in Section \ref{sec:dynamics_velocity}. Here we consider a magnetically dominated jet with a Poynting flux $\mathbf{S}=\mathbf{E}\times\mathbf{B}/4\pi=\mathbf{v}_{\perp}B^{2}/4\pi$, with $\mathbf{v}_{\perp}=\mathbf{E}\times\mathbf{B}/B^{2}$ being the velocity component perpendicular to $\mathbf{B}$. This relation can be interpreted as that the magnetic energy is advected with the plasma fluid along a direction perpendicular to $\mathbf{B}$, and therefore the Poynting flux is converted to the kinetic energy of the plasma, just like the conversion of enthalpy into kinetic energy in a hydrodynamic flow \citep[][]{2010LNP...794..233S}. One therefore expects that, for a highly magnetized jet, the plasma fluid may move roughly perpendicular to $\mathbf{B}$ with a velocity $\mathbf{v}=\mathbf{v}_{\perp}$. Generally, the velocity can also be decomposed into two mutually perpendicular components $\mathbf{v}=\mathbf{E}\times\mathbf{B}/B^{2}+\zeta\mathbf{B}$, with $\zeta=\kappa+\Omega r\sin\theta B_{\phi}/B^{2}$. In the force-free limit, $\kappa$ or $\zeta$ cannot be constrained because of the omission of the fluid inertia. In the following discussion, we choose $\zeta=0$ \citep[following, e.g.,][]{1975ctf..book.....L, 1982MNRAS.198..345M, 2007MNRAS.375..548N}, which corresponds to a net velocity of the plasma being at the minimum, that is, the so called ``drift velocity,"
\begin{equation}
\begin{aligned}
\mathbf{v}=\mathbf{v}_{\rm d}=\frac{\mathbf{E}\times\mathbf{B}}{B^{2}}=\Omega r\sin\theta\left(\hat{\phi}-\frac{B_{\phi}}{B^{2}}\mathbf{B}\right)
=\Omega r\sin\theta\left(\frac{B_{p}^{2}}{B^{2}}\hat{\phi}-\frac{B_{\phi}}{B^{2}}\mathbf{B}_{p}\right),
\label{v_drift}
\end{aligned}
\end{equation}
where $\mathbf{B}_{p}$ is the poloidal magnetic field vector. In fact, any cold plasma fluid that is carried along with a highly magnetized flow only has a slightly modified velocity relative to the drift velocity, which will be discussed in Section \ref{sec:dynamics_velocity}. In terms of charged particles inside a plasma fluid, the drift motion of the center of particle gyration in the magnetic field  induces a magnetic force, which almost balances the force exerted by the electric field. Therefore, the drift velocity is independent of the particle properties but is determined by the electromagnetic field configuration, which implies that every particle follows the same drift velocity and forms an overall motion of the plasma fluid. For practical purposes, we assume that the fluid has the same velocity as the drift velocity throughout the paper. The fact of $v_{p}/v_{\phi}=-B_{\phi}/B_{p}$ implies that velocity also forms a helical structure, which is always perpendicular to the magnetic field direction. Therefore, contrary to the case of the magnetic field, one expects that the toroidal velocity dominates in the region of $\Omega R\lesssim1$ (near the CO/AD), while the poloidal velocity dominates in the region $\Omega R\gtrsim1$ (see Equation \ref{v_drift_total_s}). This is clearly seen in Figure \ref{fig:3Djet}. In this case, the velocity field can be roughly expressed as
\begin{eqnarray}
v_{\phi}&=&\Omega r\sin\theta\frac{B_{p}^{2}}{B^{2}}\approx\frac{\Omega R}{1+\left(\Omega R\right)^{2}}, \nonumber \\
v_{p}&=&-\Omega r\sin\theta\frac{B_{\phi}B_{p}}{B^{2}}\approx\frac{\left(\Omega R\right)^{2}}{1+\left(\Omega R\right)^{2}}, \nonumber \\
v&=&\Omega r\sin\theta\frac{B_{p}}{B}=\frac{\Omega r\sin\theta}{\sqrt{1+\frac{4\Omega^{2}r^{2}}{\nu^{2}+\left(\frac{1}{T}\frac{\partial T}{\partial\theta}\right)^{2}}}}\approx\frac{\Omega R}{\sqrt{1+\left(\Omega R\right)^{2}}}.
\label{v_drift_total_s}
\end{eqnarray}
In the limit of $\Omega R\rightarrow+\infty$, the velocity reaches the speed of light\footnote{In a real jet system, one has to consider the loaded gas to calculate the fluid velocity. During jet propagation outward, the magnetically dominated condition would finally be broken. As a result, the jet cannot be accelerated further to reach the speed of light. We will discuss this in Section \ref{sec:dynamics_velocity}.} \citep[for the radiation conditions at infinity, see, e.g.,][]{2016ApJ...816...77P}. In this case, the fast critical surface is located at infinity, so the magnetic field lines are entirely inside the fast critical surface \citep[see, e.g.,][]{2003ApJ...596.1080V, 2007MNRAS.375..548N}. Roughly speaking, at the ACS, the jet reaches about a fraction $\approx1/\sqrt{2}$ of the final speed. This expression also presents an appropriate asymptotic trend of the toroidal velocity, that is, $v_{\phi}\approx\Omega R$ in the region $\Omega R\ll1$ and $v_{\phi}\approx1/\Omega R$, in the region region $\Omega R\gg1$ as discussed above. In the following subsections, we will present detailed velocity distributions during jet propagation outward.

\subsection{Total Velocity}
\label{sec:total_velocity}

From Equations \ref{v_drift}, one has a four-velocity
\begin{equation}
\begin{aligned}
\frac{1}{\left(v\Gamma\right)^{2}}
=\frac{1}{\left(\Omega r\sin\theta\right)^{2}}+\frac{B_{\phi}^{2}}{\left(\Omega r\sin\theta B_{p}\right)^{2}}-1
=\frac{1}{\left(v\Gamma\right)_{1}^{2}}+\frac{1}{\left(v\Gamma\right)_{2}^{2}},
\label{v_total}
\end{aligned}
\end{equation}
where one can define mathematical functions $V_{1}$ and $V_{2}$ (with no physical meaning but will be frequently used in the following discussion):
\begin{eqnarray}
V_{1}&=&\left(v\Gamma\right)_{1}=\frac{E}{B_{p}}=\Omega r\sin\theta=\Omega R_{0}\sin\theta T^{-1/\nu}, \nonumber \\
V_{2}&=&\left(v\Gamma\right)_{2}=\frac{E}{\sqrt{B_{\phi}^{2}-E^{2}}}=\frac{1}{\sqrt{\frac{B_{\phi}^{2}}{\left(\Omega r\sin\theta B_{p}\right)^{2}}-1}},
\label{v_total_12}
\end{eqnarray}
where $R_{0}$ is the cylindrical radius of the foot-point of the magnetic field line. Notice that $\left(v\Gamma\right)$ is dominated by $\left(v\Gamma\right)_{1}$ or $\left(v\Gamma\right)_{2}$ whichever the value is smaller. Near the AD plane ($\theta\rightarrow\pi/2$, $\cos\theta\ll1$), one has
\begin{eqnarray}
V_{1,\theta\rightarrow\pi/2}&=&\Omega r\left(1-\frac{1}{2}\cos^{2}\theta\right), \nonumber \\
V_{2,\theta\rightarrow\pi/2}&=&\sqrt{\frac{\nu^{2}+C_{1}^{2}}{4-\nu^{2}-C_{1}^{2}}}
\left[1-\frac{4C_{1}\left(C_{1}^{2}+\nu^{2}-\nu\right)}
{\left(C_{1}^{2}+\nu^{2}-4\right)\left(C_{1}^{2}+\nu^{2}\right)}\cos\theta\right].
\label{v_total_12_disk}
\end{eqnarray}
Under the assumption that the rotation velocity of the electromagnetic field is not very relativistic on the AD plane, the drift velocity is dominated by the first term\footnote{On the AD plane ($\theta=\pi/2$), $V_{2}$ is an increasing function of $\nu$, which has a minimum value at $\nu=0$, $V_{2,\nu=0,\rm min}=1/\sqrt{3}$, increasing through $V_{2,\nu=1}=1$, and reaching $V_{2,\nu\rightarrow2}\rightarrow+\infty$.} $V_{1}$. In the region $\theta\ll1$, one has
\begin{eqnarray}
V_{1,\theta\ll1}&=& C_{2}^{-1/\nu}\Omega\Psi^{1/\nu}\theta^{-\left(2-\nu\right)/\nu}
= C_{2}^{-1/2}\Omega\Psi^{1/2}r^{1-\nu/2}
=\alpha C_{2}^{\lambda}z^{-\lambda\left(2-\nu\right)}R^{2\lambda+1}, \nonumber \\
V_{2,\theta\ll1}&=&\frac{2}{\sqrt{2-\nu}}\theta^{-1}
=\frac{2C_{2}^{1/2}}{\sqrt{2-\nu}}\Psi^{-1/2}r^{\nu/2}=\frac{2}{\sqrt{2-\nu}}zR^{-1}=
\sqrt{\frac{\nu\mathcal{R}}{R}},
\label{v_total_12_tll1}
\end{eqnarray}
where $\mathcal{R}=-1/\left(\partial^{2}R/\partial z^{2}\right)=\frac{4}{\nu\left(2-\nu\right)}r/\theta$ is the curvature radius of the poloidal magnetic field line\footnote{This definition of $\mathcal{R}$ is valid in the limit of $\theta\ll1$, which guarantees a positive value for a concave surface. One always has the form $V_{2,\theta\ll1}\propto\theta^{-1}$, although its coefficient ($2/\sqrt{2-\nu}$ here) may  depend on the fourth order coefficient $a_{4}$ of the series $T\left(\theta\right)=C_{2}\theta^{2}+a_{4}\theta^{4}+\cdot\cdot\cdot$.}.

As discussed above, the condition $\theta\ll1$ does not mean that the plasma fluid has to be relativistic. Actually, the above velocity profiles (Equation \ref{v_total_12_tll1}) apply from the nonrelativistic to the relativistic regimes. In the case of $\nu\geq1$, the first term always dominates since $V_{2}$ increases faster than $V_{1}$ with increasing $r$ \citep[see][for GRMHD simulation of this case]{2019MNRAS.490.2200C}. In the case of $\nu<1$, $V_{2}$ increases slower than $V_{1}$, which implies that the second term dominates the first one beyond a certain distance where the four-velocity reaches $\left(v\Gamma\right)_{1}=\left(v\Gamma\right)_{2}=\left(2/\sqrt{2-\nu}\right)\theta^{-1}$ and remains valid during jet propagation. A larger velocity with $\left(v\Gamma\right)\theta>2/\sqrt{2-\nu}$ cannot guarantee the existence of a global solution, since different parts of the jet are not causally connected with each other \citep[see, e.g.,][]{2008ApJ...679..990Z, 2009ApJ...698.1570L, 2009MNRAS.394.1182K, 2010MNRAS.402..353L}. Such a feature also meets the constraints from the VLBA observations of AGN jets \citep[with $\Gamma\theta\sim0.1-0.3$; see, e.g.,][]{2005AJ....130.1418J, 2013A&A...558A.144C}. It can be seen that $V_{2}$ does not explicitly depend on the magnetic field angular velocity but is determined purely by the local curvature of the poloidal field line \citep{2004MNRAS.347..587B, 2009ApJ...698.1570L, 2009MNRAS.394.1182K}. The transition points from efficient acceleration (dominated by $V_{1}$) to inefficient acceleration (dominated by $V_{2}$) form a (causal) critical surface \citep[CCS, e.g.,][]{2006MNRAS.367..375B}, where
\begin{eqnarray}
\Omega R^{2}z^{-1}&=&C_{2}^{-1}\Omega\Psi z^{1-\nu}=\frac{1}{2}\left|B_{\phi}\right|Rz^{1-\nu}C_{2}^{-1}=\alpha C_{2}^{\lambda}z^{-\lambda\left(2-\nu\right)-1}R^{2\lambda+2}= \frac{2}{\sqrt{2-\nu}}, \nonumber \\
\left(v\Gamma\right)_{\rm CCS}&=&\frac{1}{\sqrt{2}}\left[\left(\frac{2}{\sqrt{2-\nu}}\right) C_{2}^{1/2} \Psi^{-1/2} \Omega^{-\nu/2}\right]^{1/\left(1-\nu\right)}= \left[\frac{\left(2C_{2}\right)^{\lambda}\alpha z^{1+\lambda\nu}} {\left(2-\nu\right)^{\left(1+2\lambda\right)/2}} \right]^{1/\left(2+2\lambda\right)}.
\label{V1V2trans}
\end{eqnarray}

\begin{figure}
\begin{center}
{\includegraphics[width=0.85\linewidth]{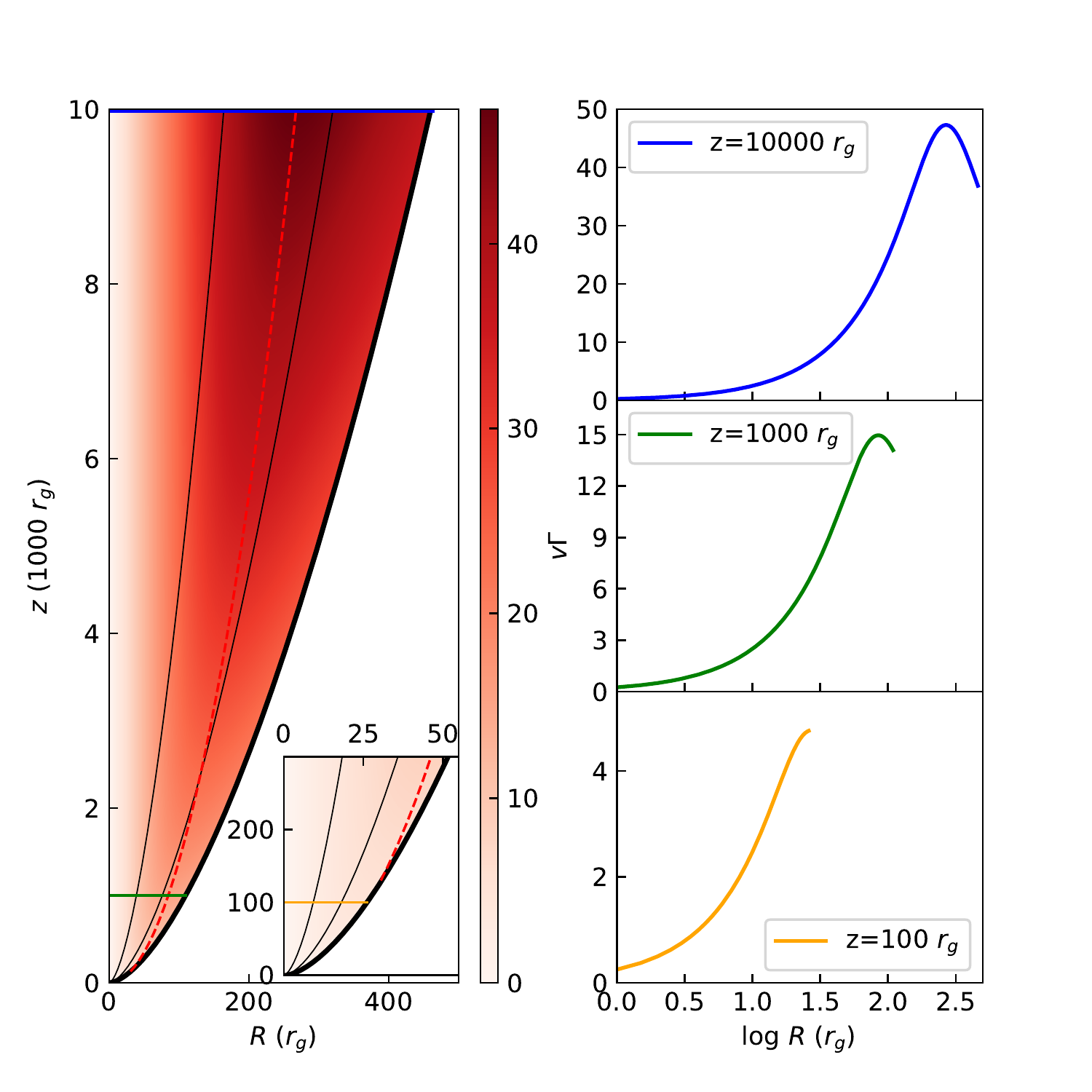}}
\end{center}
\caption{$Left$ panel: The gradient colored map of velocity distribution on the poloidal plane for the case of magnetic field lines threading a BH with a spin parameter $a=1$, where the color measures the value of four velocity $v\Gamma$. The thick black line shows the outermost magnetic field line threading the equator of the BH and the two thin black lines represent these with polar angles $\theta=60^{\circ}$ and $30^{\circ}$, respectively. The four velocity increases as the jet propagating outwards along a magnetic field line. At a fixed height of $z\lesssim130$ $r_{\rm g}$ (depending on $a$), the velocity always increases with increasing radial distance from the polar axis (e.g., at $z=100$ $r_{\rm g}$, the right/bottom panel). At a fixed height of $z\gtrsim130$ $r_{\rm g}$, the velocity increases firstly and then decreases as increasing radial distance from the polar axis (e.g., at $z=1000$ $r_{\rm g}$ and $10000$ $r_{\rm g}$ in the right/middle and right/upper panels, respectively). The red dashed line shows where the maximum velocity is reached.}
\label{fig:velocity_map}
\end{figure}

It can be seen that both $V_{1}$ and $V_{2}$ depend on $R$ at a given height $z$, so the jet should be structured. In the case of $s=-3/2$ (magnetic field lines threading an AD and $\nu=3/4$), one has $V_{1,\theta\ll1}\propto R^{-3}$ and $V_{2,\theta\ll1}\propto R^{-1}$, which always presents a faster spine surrounded by a slower layer. In the case of $s=0$ (magnetic field lines threading a CO), $V_{1,\theta\ll1}\propto R$ \citep[][]{1977MNRAS.179..433B} and $V_{2,\theta\ll1}\propto R^{-1}$ indicate a slower spine surrounded by a faster interlayer, which is further surrounded by a slower outer layer. The ``outermost" magnetic field lines threading the equator of CO form a stream surface $R=C_{2}^{-1/2}\Psi_{\rm eqt}^{1/2}z^{1-\nu/2}$, which intersects with the CCS at $\Omega C_{2}^{-1}\Psi_{\rm eqt}z_{\rm crs}^{1-\nu}=2/\sqrt{2-\nu}$. Therefore, in the region with a height less than $z_{\rm crs}$, the jet is always dominated by $V_{1,\theta\ll1}$, that is, a slower spine surrounded by a faster layer, with a maximum velocity located at the ``outermost" surface $\left(v\Gamma\right)_{\rm max,1}=\Omega R=\Omega C_{2}^{-1/2}\Psi_{\rm eqt}^{1/2}z^{1-\nu/2}$. In the region with a height larger than $z_{\rm crs}$, the jet would be separated into an inner part (dominated by $V_{1,\theta\ll1}$) and an outer part (dominated by $V_{2,\theta\ll1}$), that is, a slower spine surrounded by a faster interlayer, which is further surrounded by a slower outer layer. In this case, the maximum velocity happens at the CCS $\left(v\Gamma\right)_{\rm CCS}=\left(\Omega z/\sqrt{2-\nu}\right)^{1/2}$ at a given $z$. This ``anomalous" configuration (slower spine + faster interlayer + slower outer layer) is also seen in some simulations \citep[e.g.,][]{2007MNRAS.380...51K, 2008MNRAS.388..551T, 2018ApJ...868..146N}. The spine/layer jet structure formed plays an important role in the unified model of radio-loud (RL) AGNs \citep[see][]{2000A&A...358..104C}, explaining the $\gamma$-ray emission from misaligned radio galaxies \citep[see][]{2005A&A...432..401G, 2017ApJ...842..129C} and possibly also $\gamma$-ray prompt and afterglow emission of GRBs \citep[see, e.g.][]{2003Natur.426..154B}. This velocity map on the $R-z$ plane is plotted in Figure \ref{fig:velocity_map} (the case $s=0$; see Section \ref{BHjet} for the parameter setting).

Generally speaking, one expects a global, continuous acceleration and collimation of the jet during its propagation outward until the magnetic domination condition is violated (i.e. $\sigma$ drops below 1). This exactly matches the observations in M87 \citep[][]{2007ApJ...668L..27K, 2014ApJ...781L...2A, 2017PASJ...69...71H, 2018ApJ...855..128W, 2019ApJ...887..147P}, 1H 0323+342 \citep{2018ApJ...860..141H}, and other AGNs \citep[e.g.,][]{2015ApJ...798..134H}. From Equations \ref{half_openangle_R} and \ref{v_total_12_tll1}, we see that the collimation and acceleration of the jet follow the same law as a function of increasing distance, $v\Gamma\propto R\propto z^{1-\nu/2}$ \citep[in the case of the $V_{1}$ term dominating, see also][]{2006MNRAS.367..375B}, which is also observed in M87 as presented by, for example, \citet[][]{2016A&A...595A..54M} with $\nu\approx0.88$. As discussed above, a jet should be structured and have helical motion. This may explain why there is a slightly complicated velocity profile in M87, having various values of proper-motion velocities even at a given distance from the AGN jet core \citep[e.g., apparent speeds range from $\lesssim0.5c$ to $\gtrsim2c$ in the inner $\sim2$ mas during various epochs of observations; see, e.g.,][]{2007ApJ...668L..27K, 2014ApJ...781L...2A, 2017PASJ...69...71H, 2018ApJ...855..128W, 2019ApJ...887..147P}. In the $V_{1}$-dominated regime ($v\Gamma=\Omega R$), the angular velocity can be derived by measuring the jet width and velocity, which, in turn, may be employed to constrain the spin of the BH (see Section \ref{BHjet} for the discussion of a special case of magnetic field lines threading a BH).

In a force-free treatment and for an axisymmetric system, both the Poynting flux and the Lorentz forces (either magnetic or electric ones) vanish on the axis, where the motion of (cold) plasma flow is only controlled by gravity. The plasma near the polar axis therefore inevitably falls toward the CO, provided a not-very-large initial velocity. If one ignores the gravity pull, the plasma velocity there would vanish as shown in some simulations \citep[see, e.g.,][]{2013MNRAS.436.3741P, 2020MNRAS.498.2428W}, which is consistent with the expected drift velocity discussed here \citep[see also][]{2007MNRAS.375..548N}. It is worth noting again that the force-free treatment only applies to a high-magnetization region, while a cold plasma MHD approach can be employed to study a low-magnetization region \citep[see, e.g.,][for an interesting case study on a low-magnetization axis surrounded by a high-magnetization region]{1992SvA....36..642B, 1998ApJ...497..563H, 2000AstL...26..208B, 2007MNRAS.375..548N, 2009ApJ...698.1570L}.

\subsection{Toroidal Velocity}
\label{sec:toroidal_velocity}
From Equation \ref{v_drift}, the toroidal velocity can be expressed as
\begin{equation}
\begin{aligned}
\frac{1}{v_{\phi}}=\frac{1}{\Omega r\sin\theta}+\frac{1}{\Omega r\sin\theta}\frac{B_{\phi}^{2}}{B_{p}^{2}}=\frac{1}{\left(v_{\phi}\right)_{1}}
+\frac{1}{\left(v_{\phi}\right)_{2}},
\label{v_phi}
\end{aligned}
\end{equation}
with
\begin{eqnarray}
\left(v_{\phi}\right)_{1}&=&\Omega r\sin\theta=V_{1}, \nonumber \\
\left(v_{\phi}\right)_{2}&=&\Omega r\sin\theta\frac{B_{p}^{2}}{B_{\phi}^{2}}\approx\frac{1}{\Omega r\sin\theta}=\frac{1}{V_{1}}.
\label{v_phi_component}
\end{eqnarray}
On the AD plane, one has $\left(v_{\phi}\right)_{2,\theta=\pi/2}=\left(\nu^{2}+C_{1}^{2}\right)/4\Omega r_{0}$, which is usually larger than $\left(v_{\phi}\right)_{1,\theta=\pi/2}$. Therefore, the toroidal velocity is dominated by $\left(v_{\phi}\right)_{1}$ initially and increases as the jet propagates outward. This is followed by a decreasing phase dominated by $\left(v_{\phi}\right)_{2}$ roughly beyond the ACS ($\Omega r\sin\theta\approx1$), where the toroidal velocity reaches the maximum value $v_{\phi}\approx1/2$. Within the ACS, the plasma fluid roughly corotates with the magnetic field lines since the magnetic fields are roughly poloidal dominant. After the jet propagates outside of the ACS, the toroidal magnetic field becomes dominated. The rotation of the plasma lags behind magnetic fields since the toroidal component of the plasma (the ``slide motion'') along the magnetic field line would dominate.

\subsection{Poloidal Velocity}
\label{sec:poloidal_velocity}
The poloidal velocity can be expressed as
\begin{equation}
\begin{aligned}
\frac{1}{\left(v_{p}\Gamma_{p}\right)^{2}}=\frac{B_{p}^{2}}
{\left(\Omega r\sin\theta B_{\phi}\right)^{2}}+\frac{2}{\left(\Omega r\sin\theta\right)^{2}}+\frac{B_{\phi}^{2}}{\left(\Omega r\sin\theta B_{p}\right)^{2}}-1
=\frac{1}{\left(v_{p}\Gamma_{p}\right)_{1}^{2}}+\frac{1}{\left(v_{p}\Gamma_{p}\right)_{2}^{2}}
+\frac{1}{\left(v_{p}\Gamma_{p}\right)_{3}^{2}}
\label{v_p}
\end{aligned}
\end{equation}
where
\begin{eqnarray}
\left(v_{p}\Gamma_{p}\right)_{1}&=&\frac{\Omega r\sin\theta\left|B_{\phi}\right|}{B_{p}}
\approx\left(\Omega r\sin\theta\right)^{2}=V_{1}^{2}, \nonumber \\
\left(v_{p}\Gamma_{p}\right)_{2}&=&\Omega r\sin\theta/\sqrt{2}=V_{1}/\sqrt{2}, \nonumber \\
\left(v_{p}\Gamma_{p}\right)_{3}&=&\frac{1}{\sqrt{\frac{B_{\phi}^{2}}{\left(\Omega r\sin\theta B_{p}\right)^{2}}-1}}=V_{2}.
\label{v_p_components}
\end{eqnarray}
In the region of $\theta\rightarrow\pi/2$ (near the AD plane), one has $\left(v_{p}\Gamma_{p}\right)_{1,\theta\rightarrow\pi/2} = 2\left(\Omega r\right)^{2} /\sqrt{\nu^{2}+C_{1}^{2}}$
$\left[1-C_{1}\cos\theta\left(\nu^{2}+C_{1}^{2}-\nu\right) / \left(\nu^{2}+C_{1}^{2}\right)\right]\approx V_{1,\theta\rightarrow\pi/2}^{2}$, which dominates the initial poloidal velocity near the disk plane \citep[the square dependence on $R$ is roughly similar to that of][]{1992ApJ...394..459L}. Beyond roughly the ACS ($\Omega R=1$), $\left(v_{p}\Gamma_{p}\right)_{2}$ becomes dominant. As the jet continues propagating outward, the poloidal velocity would be always  dominated by $\left(v_{p}\Gamma_{p}\right)_{2}$ for $\nu\geq1$, and it would be dominated by $\left(v_{p}\Gamma_{p}\right)_{3}$ beyond the CCS for $\nu<1$.

\begin{figure}
\begin{center}
{\includegraphics[width=0.7\linewidth]{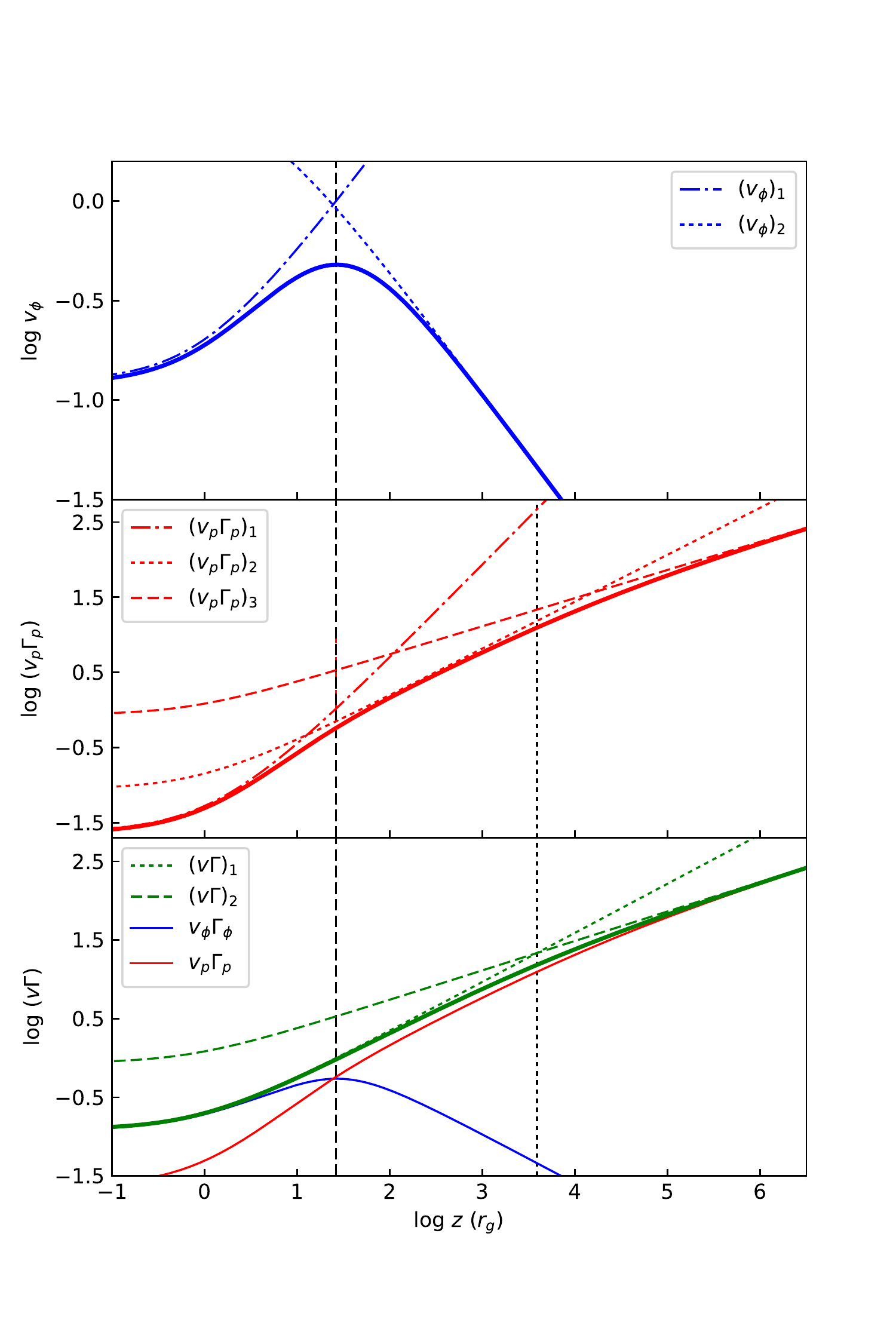}}
\end{center}
\caption{Four velocity ($v\Gamma$) profile of a flow along a magnetic field line threading the equator of a BH with a spin parameter $a=0.5$. The toroidal velocity presents two stages (within or outside of the ACS; the upper panel), while the poloidal velocity has three stages: the first turning point also happens at the ACS, and the second one is determined by the causality (the CCS, the middle panel). The complementarity of the toroidal and poloidal velocities in the first two stages makes the total velocity follow a uniform stage before the CCS, although the dominant term changes from the toroidal to the poloidal one when crossing the ACS (bottom panel; see Section \ref{sec:jet_acceleration} for details). The vertical dashed black line shows the location of the ACS ($z=26$ $r_{\rm g}$), while the vertical dotted black line refers to that of the CCS ($z=3882$ $r_{\rm g}$).}
\label{fig:je_acceleration}
\end{figure}

In summary, one has a 3-stage acceleration diagram, as presented in Figure \ref{fig:je_acceleration} (see Section \ref{BHjet} for parameter setting):
\begin{itemize}
\item \textbf{First acceleration regime:} Initially from the AD/CO, the poloidal magnetic field dominates and the fluid plasma almost corotates with the magnetic field lines. The total velocity is dominated by the toroidal component. The toroidal velocity follows $\left(v_{\phi}\Gamma_{\phi}\right)=V_{1}$, while the poloidal component follows $\left(v_{p}\Gamma_{p}\right)=V_{1}^{2}$. This stage ends at the radius where the toroidal magnetic field starts to dominate and the fluid fails to corotate, which is roughly located at the ACS ($V_{1}=\Omega R\approx1$).
\item \textbf{Second acceleration regime:} The total velocity is dominated by the poloidal component, which follows $\left(v_{p}\Gamma_{p}\right)=V_{1}/\sqrt{2}$; and the toroidal component decreases as $\left(v_{\phi}\Gamma_{\phi}\right)=1/V_{1}$ (the total velocity still follows $v\Gamma=V_{1}$). In the case of $\nu>1$, this acceleration regime continues all the way to ``infinity" or when the magnetic dominance condition is broken.
\item \textbf{Third acceleration regime:} This regime applies in the case of $\nu<1$, when the velocity reaches $\left(v\Gamma\right)_{1}\theta=2/\sqrt{2-\nu}$. Causality limits efficient acceleration in the 2nd acceleration regime. In this regime, one still has $\left(v_{\phi}\Gamma_{\phi}\right)=1/V_{1}$ and the total velocity is dominated by the poloidal velocity, which follows $\left(v_{p}\Gamma_{p}\right)=V_{2}$.
\end{itemize}

As shown by Equation \ref{v_total}, the total velocity, in terms of $v\Gamma$, has only two acceleration regimes. The physical meanings of these two regimes can also be clearly presented by  separately plotting the velocity and Lorentz factor acceleration profiles, as shown in \citet[][]{2007MNRAS.375..548N}. In fact, the bulk Lorentz factor can also be separated into three parts to match the above three stages:
\begin{equation}
\begin{aligned}
\Gamma^{2}=1+\frac{1}{\frac{1}{V_{1}^{2}}+\frac{1}{V_{2}^{2}}}.
\label{Gamma_3terms}
\end{aligned}
\end{equation}
The first stage is nonrelativistic (within the ACS), that is, $\Gamma\sim1$. In the second stage, the jet becomes relativistic, that is, $\Gamma\sim v\Gamma\approx V_{1}$, which is followed by the third stage with $\Gamma\approx v\Gamma\approx V_{2}$. In the limit of $\Gamma\gg1$, Equation \ref{Gamma_3terms} is reduced to $1/\Gamma^{2}\approx1/V_{1}^{2}+1/V_{2}^{2}$, which is the two-stage acceleration scenario that has been discussed by many authors \citep[e.g.,][]{2003ApJ...596.1080V, 2004MNRAS.347..587B, 2006MNRAS.367..375B, 2007MNRAS.375..548N, 2008MNRAS.388..551T, 2009MNRAS.394.1182K, 2020ApJ...892...37P}. In the case of $\nu>1$, $V_{1}$ always dominates, which yields $\Gamma\approx\Omega R$ for the relativistic case, as discussed by, for example, \citet[][through GRMHD simulations]{2019MNRAS.490.2200C} and \citet[][through a semianalytical study]{2007MNRAS.375..548N}.

\subsection{Helical Jet}
\label{sec:helical_jet}

One can define a ``cycle period" of motion for the helical structure of velocity
\begin{equation}
\begin{aligned}
P=\frac{2\pi r\sin\theta}{v_{\phi}}=\frac{2\pi r\sin\theta}{\left(v_{\phi}\right)_{1}}
+\frac{2\pi r\sin\theta}{\left(v_{\phi}\right)_{2}}=P_{1}+P_{2},
\label{period}
\end{aligned}
\end{equation}
where
\begin{eqnarray}
P_{1}&=&\frac{2\pi r\sin\theta}{\left(v_{\phi}\right)_{1}}=\frac{2\pi}{\Omega}, \nonumber \\
P_{2}&=&\frac{2\pi}{\Omega}\frac{B_{\phi}^{2}}{B_{p}^{2}}\approx2\pi\Omega\left(r\sin\theta\right)^{2}
=\frac{2\pi}{\Omega}V_{1}^{2}.
\label{period_P1}
\end{eqnarray}
It can be seen that $P_{1}$ is the rotation period of the electromagnetic field (see Section \ref{BHjet} for a special case of magnetic field lines threading a BH), which dominates over $P_{2}$ within the ACS. In the region outside the ACS, the second term $P_{2}$ dominates. At $\theta\ll1$, one has
\begin{equation}
\begin{aligned}
P_{2}=P_{1}\left(\Omega R\right)^{2}=2\pi C_{2}^{-2/\nu}\Omega\Psi^{2/\nu}\theta^{2-4\nu}=2\pi C_{2}^{-1}\Omega\Psi r^{2-\nu}=2\pi\alpha C_{2}^{\lambda}z^{-\lambda\left(2-\nu\right)}R^{2\lambda+2}.
\label{period_P2}
\end{aligned}
\end{equation}
The case $s=-3/2$ (magnetic field line threading the AD and $\nu=3/4$) implies $P_{2}\propto R^{-2}$; that is, the spine needs a longer time to finish a cycle motion compared with the layer. The case $s=0$ (magnetic field line threading the CO) gives $P_{2}\propto R^{2}$; that is, the cycle period increases from spine to layer.

One can also define an inclination angle between the poloidal and total velocities:
\begin{equation}
\begin{aligned}
\tan\theta_{\rm inc}=\frac{v_{\phi}}{v_{p}}=\frac{1}{2\Omega r} \sqrt{\nu^{2}+\left(\frac{1}{T}\frac{dT}{d\theta}\right)^{2}}
=\frac{B_{p}}{\left|B_{\phi}\right|}\approx\frac{1}{V_{1}},
\label{inc_ang}
\end{aligned}
\end{equation}
which measures the orderliness of the velocity. At $\theta\rightarrow\pi/2$, one has
\begin{equation}
\begin{aligned}
\tan\theta_{\rm inc,\theta\rightarrow\pi/2}=\frac{\sqrt{\nu^{2}+C_{1}^{2}}}{2\Omega r}\left[1+\frac{C_{1}\left(\nu^{2}+C_{1}^{2}-\nu\right)}{\nu^{2}+C_{1}^{2}}\cos\theta\right].
\label{inc_ang_disk}
\end{aligned}
\end{equation}
At $\theta\ll1$, this angle roughly represents the angle between the velocity direction and the polar axis, i.e.
\begin{equation}
\begin{aligned}
\tan\theta_{\rm inc,\theta\ll1}=C_{2}^{1/2}\Omega^{-1}\Psi^{-1/2}r^{-1+\nu/2}
=C_{2}^{1/\nu}\Omega^{-1}\Psi^{-1/\nu}\theta^{2\nu-1}
=\alpha^{-1}C_{2}^{-\lambda}z^{\lambda\left(2-\nu\right)}R^{-2\lambda-1},
\label{inc_ang_tll1}
\end{aligned}
\end{equation}
which is equal to the reciprocal of $V_{1}$ (see Equation \ref{v_total_12_tll1}).

By multiplying $v_{p}$ by the cycle period, one gets a cycle distance that the fluid moves along in the poloidal direction within each cycle period:
\begin{equation}
\begin{aligned}
D=v_{p}P=\frac{2\pi r\sin\theta}{v_{\phi}/v_{p}}\approx P_{2}.
\label{D_cycle}
\end{aligned}
\end{equation}
At $\theta\rightarrow\pi/2$ (near the AD), one has
\begin{equation}
\begin{aligned}
D=\frac{4\pi\Omega r^{2}}{\sqrt{\nu^{2}+C_{1}^{2}}}
\left[1-\frac{C_{1}\left(\nu^{2}+C_{1}^{2}-\nu\right)}{\nu^{2}+C_{1}^{2}}\cos\theta\right];
\label{D_cycle_disk}
\end{aligned}
\end{equation}
and at $\theta\ll1$, one has
\begin{equation}
\begin{aligned}
D=P_{2}=2\pi C_{2}^{-2/\nu}\Omega\Psi^{2/\nu}\theta^{2-4\nu}=2\pi C_{2}^{-1}\Omega\Psi r^{2-\nu}=2\pi\alpha C_{2}^{\lambda}z^{-\lambda\left(2-\nu\right)}R^{2\lambda+2},
\label{D_cycle_tll1}
\end{aligned}
\end{equation}
which shows the same formula as the cycle period $P_{2}$ outside the ACS (see Equation \ref{period_P2}).

Both velocity and magnetic field lines fall in the magnetic stream surface and show helical structures, but they are perpendicular to each other ($v_{\phi}/v_{p}=-B_{p}/B_{\phi}$). Similar qualities such as the inclination angle, cycle distance, and orderliness can also be defined for magnetic field lines, which would be a reciprocal of that of velocity. During the propagation of a helical motion of a plasma fluid element in a helical magnetic field configuration, its synchrotron emission would present a continuous rotation of the polarization angle, which can be traced in some AGN jets despite of difficulties \citep[e.g.,][]{2008Natur.452..966M, 2010ApJ...710L.126M, 2010Natur.463..919A}. The spin of the jet may also drive a rotational motion of the surrounding molecular outflow, which seems to be detected by polarization observations of AGN PG 1700+518 \citep{2007Natur.450...74Y, 2012MNRAS.419L..74Y}. Figure \ref{fig:3Djet} presents the topology of magnetic field lines (green line) and velocities (color gradient line). The black dashed line represents the location of the ACS, and the colors of the velocity lines represent $v\Gamma$. For parameter setting, see Section \ref{BHjet}.

\section{Current, Charge, and Jet Power}
\label{sec:charge_current_Power}

The plasma supplies currents and charges as needed to support the electromagnetic field. In the force-free limit, the output jet power is dominated by the Poynting flux. Therefore, the current, charge, and power of the jet can be self-consistently derived. Electromagnetic fields, velocity, and poloidal current depend on the first-order derivative of $\Psi$, so one can use $T_{\rm nr}$ to approximately measure their structures over the entire jet region. On the other hand, the toroidal current and charge densities are related to the second-order derivatives of $\Psi$, while $T_{\rm nr}$ and $T_{\rm r}$ match each other only on the order of $\cos\theta$ in the region of $\theta\rightarrow\pi/2$. This would lead to some errors in estimating the toroidal current and charge densities near the AD plane (see below).

\subsection{Current}
\label{subsec:current}

From Equation \ref{J_current}, the poloidal current density reads as
\begin{eqnarray}
\mathbf{j}_{p} = \frac{\Phi'\mathbf{B}_{p}}{4\pi}
=-\frac{\left(\lambda+1\right)\Omega \mathbf{B}_{p}}{2\pi}
= \left(\lambda+1\right)\frac{B_{\phi}}{4\pi r}\left(\frac{1}{T}\frac{dT}{d\theta}\hat{r}-\nu\hat{\theta}\right)
=\left(\lambda+1\right)\frac{\Omega}{2\pi}\frac{r^{\nu-2}}{\sin\theta}\left(-\frac{dT}{d\theta}\hat{r}+\nu T\hat{\theta}\right),    %%%%%%%%%% j\propto\Omega B
\label{j_e_poroidalj}
\end{eqnarray}
which is always parallel to the poloidal magnetic field lines. Therefore, the current streams on the magnetic stream surface. For the toroidal current, the situation seems slightly more complicated. In the regions of either outside of the ACS ($\Omega R>1$) or where the jet is collimated ($\theta\ll1$), $T_{\rm nr}$ matches $T_{\rm r}$ with an error no larger than $\sim\theta^{2}$. The toroidal current approximately vanishes. In the region inside the ACS ($\Omega R<1$) and where the jet is not collimated ($\theta\rightarrow\pi/2$), $T_{\rm nr}$ matches $T_{\rm r}$ with an error no larger than $\sim\cos\theta$. The toroidal current cannot vanish, which is a small quantity compared with the poloidal one:
\begin{equation}
j_{\phi}=\frac{-1}{4\pi R}\left(\frac{\partial^{2}\Psi}{\partial z^{2}}+\frac{\partial^{2}\Psi}{\partial R^{2}}-\frac{1}{R}\frac{\partial\Psi}{\partial R}\right)
=\left\{
  \begin{array}{ll}
    \approx0 \ \ \ \ \ \ \Omega R>1\ \rm or\ \theta\ll1,  \\
    \ll j_{p} \ \ \ \ \Omega R<1\ \&\ \theta\rightarrow\pi/2.
  \end{array}
\right.
\label{j_e_toroidalj}
\end{equation}
For the cases with exact solutions of the pulsar Equation \ref{DE_Psi} (monopole, cylinder, and parabola with $\Omega\propto R_{0}^{-1}$, see Appendix \ref{sec:exactsolutions}), the toroidal currents exactly vanish over the entire jet \citep[see also][for the slow rotation case]{1992SvAL...18..337B}. This is consistent with the above analysis for the general cases. Physically, rotation induces a toroidal magnetic field and a poloidal current. The magnetic force made by the so-called $z$-pinch effect is approximately balanced by the induced electric force. This makes the toroidal current a small quantity (i.e. it almost vanishes in the region of $\Omega R>1$ or $\theta\ll1$), as indicated by Equation \ref{FL_E} (see below), which is also shown in some GRMHD simulations \citep[i.e. the toroidal current in the ``funnel" region almost vanishes, e.g.,][]{2007MNRAS.375..513M}. Similar to velocity (Equation \ref{vgv}), the force-free condition (Equation \ref{force_free}) makes the current to be generally separate into two parts: the moving of charges and a current sliding along magnetic field lines, that is, $\mathbf{j}=\rho_{\rm e}\mathbf{v}_{\rm d}+j_{\parallel}\hat{B}$, where the parallel current is $j_{\parallel}\approx-\rho_{\rm e}\Omega r\sin\theta B_{p}^{2}/\left(BB_{\phi}\right)$ when considering the smallness of the toroidal current (to note the condition Equation \ref{j_e_toroidalj}). One therefore has a poloidal current $j_{p}\approx-\rho_{\rm e}\Omega r\sin\theta B_{p}/B_{\phi}\approx\rho_{\rm e}$ (see also the discussion on charge density in Section \ref{subsec:charge}). In a nonrotating system, the toroidal magnetic field, the electric current, and the electric field would vanish. What might exist is only the poloidal magnetic field (which may be supported by the CO/AD; see Section \ref{Bp_AD}). It is the rotation that generates a toroidal magnetic field, which induces a poloidal current. The force that this poloidal current receives in the magnetic field is balanced by the force exerted by the appearing poloidal electric field to guarantee the force-free condition (see Section \ref{sec:Lorentz_Force} for a  detailed analysis).

\begin{figure}
\begin{center}
{\includegraphics[width=0.85\linewidth]{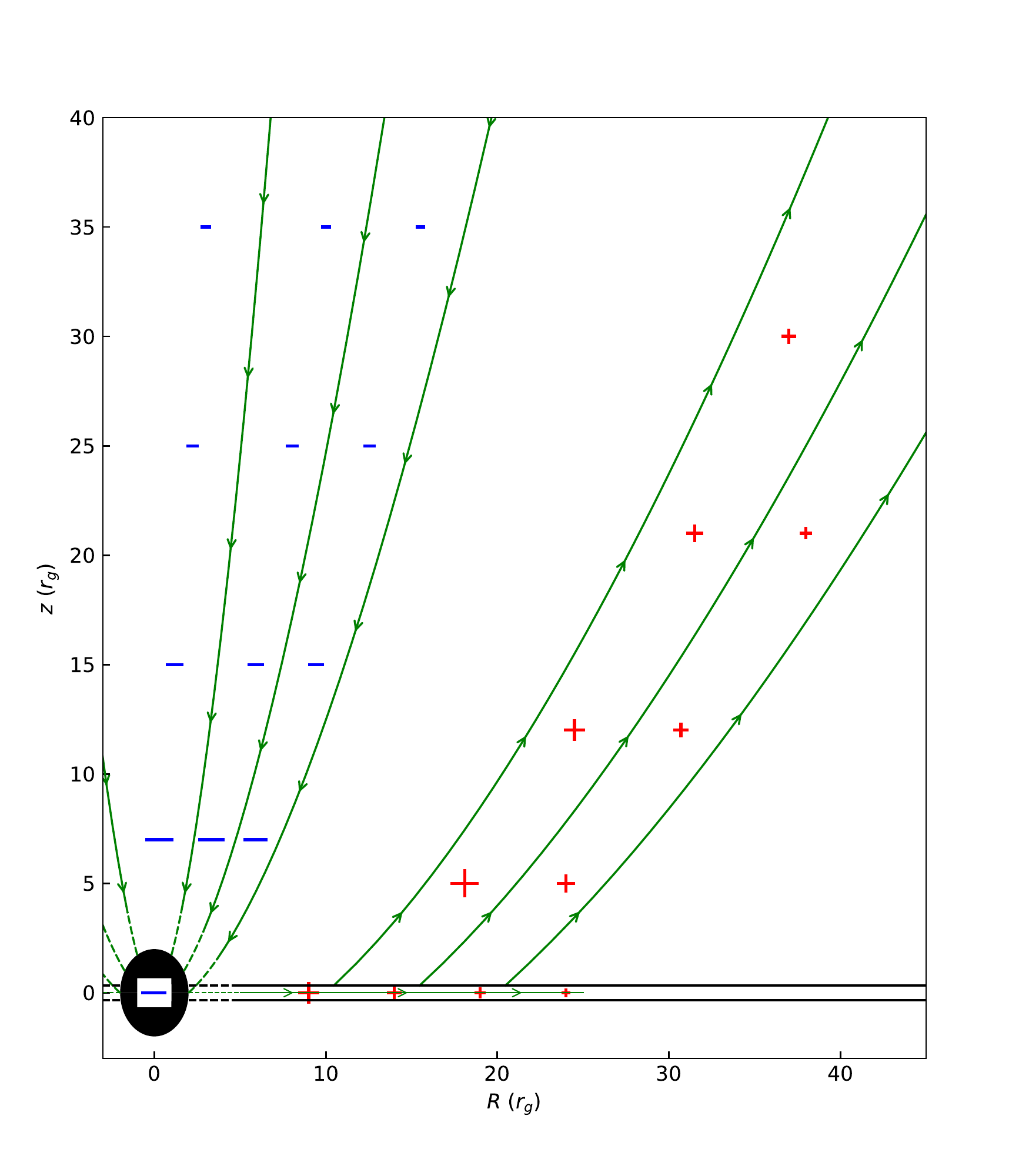}}
\end{center}
\caption{Current and charge in a BH accretion/jet system (the AD rotating along the same direction as the BH with a spin parameter $a=0.1$ and the case $\mathbf{\Omega}\cdot\mathbf{B}>0$). The green lines show the current flows and their directions. The red ``+" symbols on the flow from the AD refer to positive charges, with their sizes indicating the relative charge densities, while the blue ``--" symbols refer to that of negative charges of flow from the BH. The red ``+" symbols on the AD show positive charges, with their sizes indicating the relative charge surface densities. The radii of the foot-points of the three current lines from the AD are $R_{0}=$10 $r_{g}$, 15 $r_{g}$ and 20 $r_{g}$, respectively, while those of the current lines originating from the BH have polar angles $\theta=30^{\circ}$, $60^{\circ}$, and $90^{\circ}$, respectively.  Notice that the BH has a negative charge in this case.}.
\label{fig:charge_current}
\end{figure}

At $\theta\rightarrow\pi/2$ (near the AD), one has
\begin{equation}
\begin{aligned}
\mathbf{j}_{p}=\frac{\Omega}{2\pi}\left(\lambda+1\right)r^{\nu-2}
\left\{\left[-C_{1}-\left(\nu-1\right)\nu\cos\theta\right]\hat{r}
+\left[\nu-C_{1}\nu\cos\theta\right]\hat{\theta}\right\}.
\label{j_e_disk}
\end{aligned}
\end{equation}
At $\theta\ll1$, one has
\begin{eqnarray}
j_{r}&=&\left(\lambda+1\right)\frac{B_{\phi}}{2\pi R}=-\frac{ C_{2}}{\pi}\left(\lambda+1\right)\Omega r^{\nu-2}
=-\frac{ C_{2}^{2/\nu}}{\pi}\left(\lambda+1\right)\Omega\Psi^{1-2/\nu}\theta^{-2+4/\nu}
=-\frac{\alpha C_{2}^{\lambda+1}}{\pi}\left(\lambda+1\right)z^{\lambda\nu-2\lambda+\nu-2}R^{2\lambda}, \nonumber \\
j_{\theta}&=&\frac{\nu C_{2}^{1/2}}{2\pi}\left(\lambda+1\right)\Omega\Psi^{1/2}r^{\nu/2-2}
=\frac{\nu C_{2}^{2/\nu}}{2\pi}\left(\lambda+1\right)\Omega\Psi^{1-2/\nu}\theta^{-1+4/\nu}
=\frac{\alpha\nu C_{2}^{\lambda+1}}{2\pi}\left(\lambda+1\right)z^{\lambda\nu-2\lambda+\nu-3}R^{2\lambda+1}, \nonumber \\
j_{R}&=&\frac{2-\nu}{2}\theta j_{r}, \nonumber \\
j_{z}&=&j_{p}=j_{r}.
\label{j_e_tll1}
\end{eqnarray}
Similar to $\mathbf{B}_{p}$, $j_{r}$ dominates over $j_{\theta}$ in the region far away from the CO/AD. In the case of $s=0$ (magnetic field lines threading a CO), $j_{r}$ carries a negative value and is almost constant at a given height, while in the case of $s=-3/2$ (magnetic field lines threading an AD), $j_{r}$ carries a positive value, which decreases with increasing $R$ at a given height $z$. This current distribution is clearly presented in Figure \ref{fig:charge_current}.

\subsection{Charge}
\label{subsec:charge}

As discussed above, it is the rotation of the system that produces both the toroidal magnetic field and the poloidal electric field, which provide a balance between magnetic and electric forces and, in turn, also provide the condition for the presence of charges \citep[][]{2010ApJ...709.1100P}. Generally, one has two methods to calculate the charge density. The first method is from the force-free condition,
\begin{equation}
\begin{aligned}
\rho_{\rm e,ff}=-\frac{\left(\mathbf{j}\times\mathbf{B}\right)\cdot\mathbf{E}}{E^{2}}
\xlongequal[]{j_{\phi}\approx0}\frac{\left(\lambda+1\right)B_{\phi}}{2\pi r\sin\theta}=
\frac{-B_{\phi}}{B_{p}\Omega r\sin\theta}j_{p}\approx j_{p},
\label{rho_eff}
\end{aligned}
\end{equation}
which implies that the charge density is always equal to the poloidal electric current density, as discussed above. Similar to the enclosed current (Equation \ref{JzJr}), one can define a linear charge density along the jet $\rho_{\rm e}^{\rm l}=\int_{0}^{R}\rho_{\rm e}2\pi R'dR'\approx J=\Phi/2$, which is also a roughly conserved quality. The second method to calculate the charge density is from the divergence of the electric field:
\begin{eqnarray}
\rho_{\rm e,em}&=&\frac{\nabla\cdot\mathbf{E}}{4\pi}=-\frac{\mathbf{\Omega}\cdot\mathbf{B}}{2\pi}+\Omega r\sin\theta j_{\phi}-\frac{\Omega'}{4\pi}\left|\nabla\Psi\right|^{2}
=-\frac{\Omega}{4\pi}\left(\nabla^{2}\Psi+\frac{\Omega'}{\Omega}\left|\nabla\Psi\right|^{2}\right) \nonumber \\
&\xlongequal[]{j_{\phi}\approx0}&-\frac{\Omega r^{\nu-2}}{4\pi}\left[\frac{\lambda}{T}\left(\frac{dT}{d\theta}\right)^{2}+2\cot\theta \frac{dT}{d\theta}+\left(2+\lambda\nu\right)\nu T\right].
%%%%%%%%%%%% \rho_{\rm e}\propto\frac{\Omega B}{c}
\label{rho_eem}
\end{eqnarray}
In the case of $s=0$ (magnetic field threading a CO, $\Omega'=0$), one has $\rho_{\rm e}=-\mathbf{\Omega}\cdot\mathbf{B}/2\pi$, which is similar to that in a pulsar magnetosphere \citep[the so-called Goldreich-Julian charge density $\rho_{\rm GJ}$, see][]{1969ApJ...157..869G}. In principle, these two methods should self-consistently present a unique charge density. However, as discussed in Section \ref{sec:approximate_solution}, the charge density is related to the second-order derivative of $\Psi$, and therefore our approximate solution would only be accurate at $\theta\ll1$, and would have an error at $\theta\rightarrow\pi/2$. This is clearly shown when comparing Equations \ref{rho_eff} and \ref{rho_eem}. At $\theta=\pi/2$ (with $\nu=3/4$), the case of $s=0$ yields $\rho_{\rm e,em}/\rho_{\rm e,ff}=0.375$, and the case of $s=-3/2$ gives $\rho_{\rm e,em}/\rho_{\rm e,ff}=0.480$.

It is the global rotation of the magnetic fields that guarantees the existence of a charge density in the entire plasma region \citep[][]{2007MNRAS.375..548N}, including near the CO surface or on the AD plane (see Section \ref{Bp_AD}). On the other hand, in the case of a nonrelativistic plasma flow, the electric force is relatively unimportant compared with the magnetic force. Some MHD simulations often ignore the electric force, of which such an approximation cannot give a self-consistent result $\rho_{\rm e}=\nabla\cdot\mathbf{E}/4\pi=0$, but nevertheless a good-enough approximation to solve the magnetic field configuration \citep[e.g.,][and references therein; see Appendix \ref{app:ne0} for a detailed discussion]{2019ApJ...881...34Y}.

\subsection{Potential Difference}
\label{subsec:potential_difference}

The magnetic stream surface is equipotential, which implies that there is a potential difference between two magnetic stream surfaces \citep[layers $l_{1}$ and $l_{2}$ in the case of $\lambda\neq-1$, see Equation \ref{E_general} and also][]{2019ARA&A..57..467B}, i.e.
\begin{small}
\begin{equation}
\begin{aligned}
\Delta V=\int_{l_{1}}^{l_{2}}\mathbf{E}\cdot d\mathbf{l}_{\rm E}=
-\int_{l_{1}}^{l_{2}}\Omega\nabla\Psi\cdot d\mathbf{l}_{\rm E}
=-\frac{\Omega\Psi}{\left(\lambda+1\right)}\bigg|_{1}^{2}
=-\frac{\Omega F_{B}}{2\pi\left(\lambda+1\right)}\bigg|_{1}^{2}
=\frac{RB_{\phi}}{2\left(\lambda+1\right)}\bigg|_{1}^{2}
=\frac{J}{\left(\lambda+1\right)}\bigg|_{1}^{2}.
%%%%%%%%%% \Delta V\propto \frac{J}{c}
\label{DeltaV}
\end{aligned}
\end{equation}
\end{small}
In the case of $\lambda=-1$, the coefficient $1/\left(\lambda+1\right)$ in the above equation should be replaced by $\ln\left(\Psi_{2}/\Psi_{1}\right)$. The fluid plasma supports the charge to cancel the electric field induced by the motion of the fluid plasma, which makes the jet self-balanced to maintain the electromagnetic configuration (see Section \ref{sec:Lorentz_Force}).  Therefore, one anticipates a shorting out of the parallel electric field $E_{\parallel}$ (to the magnetic field): $\mathbf{E}\cdot\mathbf{B}=0$. However, there may be cases where in some regions the charges stream out along the magnetic field line and no new charges replenish the region. In extreme situations, the charges could be totally depleted, forming a ``gap'' with $\mathbf{E}\cdot\mathbf{B}\neq0$ \citep[see, e.g.,][]{1998ApJ...497..563H, 2018ApJ...863L..31C, 2019PhRvL.122c5101P}. The parallel electric field $E_{\parallel}$ would bring a potential difference along the magnetic field line, the maximum of which is limited by Equation \ref{DeltaV} \citep[see, e.g.,][for a gap in the pulsar magnetosphere]{1975ApJ...196...51R}. For the case of magnetic field lines threading a BH, see Section \ref{BHjet}.

\subsection{Jet Power}
\label{subsec:power}

In the force-free limit, the output jet energy flow is almost equal to the Poynting flux, $\mathbf{S}=\mathbf{E}\times\mathbf{B}/4\pi$, which follows the direction of the particle drifting velocity, transported within the plane of the magnetic stream surface. The $r$ and $z$ directions of the Poynting fluxes can be written as
\begin{eqnarray}
S_{r}&=&-\frac{\Omega R}{4\pi}B_{\phi}B_{r}=\frac{\Omega^{2}}{2\pi}\frac{\Psi}{r^{2}\sin\theta}\frac{\partial\Psi}{\partial\theta}
=\frac{B_{\phi}^{2}}{8\pi}\frac{\sin\theta}{T}\frac{dT}{d\theta}, \nonumber \\
S_{z}&=&-\frac{\Omega R}{4\pi}B_{\phi}B_{z}=-\frac{\Omega\Phi}{4\pi R}\frac{\partial\Psi}{\partial R}
=\frac{B_{\phi}^{2}}{8\pi}\sin^{2}\theta\left(\frac{1}{T}\frac{dT}{d\theta}\cot\theta+\nu\right).
\label{Poynting_flux_rz}
\end{eqnarray}
The Poynting flux enclosed between two magnetic stream surfaces $\Psi_{1}$ and $\Psi_{2}$ can be then easily calculated (two-side jet power, in the case of $\lambda\neq-1$):
\begin{small}
\begin{equation}
\begin{aligned}
P_{\rm jet}=2\int_{\theta_{1}}^{\theta_{2}}S_{r}2\pi r^{2}\sin\theta d\theta
=\frac{\Omega^{2}\Psi^{2}}{\left|\lambda+1\right|}\bigg|_{1}^{2}
=\frac{\Omega^{2}F_{B}^{2}}{4\pi^{2}\left|\lambda+1\right|}\bigg|_{1}^{2}
=\frac{R^{2}B_{\phi}^{2}}{4\left|\lambda+1\right|}\bigg|_{1}^{2}
=\frac{J^{2}}{\left|\lambda+1\right|}\bigg|_{1}^{2}.
%%%%%%%%%%  P_{\rm jet}\propto\frac{J^{2}}{c}\propto c^{3}\left(\Delta V\right)^{2}
\label{P_jetpower}
\end{aligned}
\end{equation}
\end{small}
In the case of $\lambda=-1$, the coefficient $1/\left|\lambda+1\right|$ in the above equation should be replaced by $2\ln\left(\Psi_{2}/\Psi_{1}\right)$. The last equality shows a very simple relation between the jet power and the current carried by the plasma. Radio galaxy 3C 303 is hitherto the only source that has the electric current of its jet measured, at $\sim3.9\times10^{18}$ A \citep[through mapping polarization and Faraday rotation,][]{2011ApJ...741L..15K}. Employing Equation \ref{P_jetpower}, assuming a magnetic field threading a BH ($\lambda=0$) or a Keplerian AD ($\lambda=-2$ with $\nu=3/4$), one can easily estimate a jet power $\sim4.6\times10^{45}$ erg s$^{-1}$, which is only a factor of $\sim2$ smaller than that estimated through modeling the broadband spectral energy distribution \citep[$\sim1.0\times10^{46}$ erg s$^{-1}$,][]{2018ApJ...858...27Z}.

Combining with Equation \ref{DeltaV}, one has a maximum potential difference of $\Delta V\approx1.7\times10^{15}\sqrt{P_{\rm jet}/10^{36}\rm erg\ s^{-1}}$ volts in the case of magnetic field lines threading a CO. Taking the radiative luminosity of the Crab Nebula \citep[$\gtrsim10^{37}$ erg s$^{-1}$,][]{2008ARA&A..46..127H} as a lower limit of jet/wind power, the potential difference can reach up to $\gtrsim5\times10^{15}$ volts, which is large enough to (theoretically) accelerate charged particles to emit the highest energy photons ever detected in the Crab Nebula \citep[$\approx0.45$ PeV,][]{2019PhRvL.123e1101A}. For a special case of magnetic field threading a BH, see Section \ref{BHjet}.

At $\theta\rightarrow\pi/2$ (near the AD), the $z$-direction Poynting flux reads as
\begin{equation}
\begin{aligned}
S_{z}=\frac{B_{\phi}^{2}}{8\pi}\left(\nu+C_{1}\cos\theta\right).
\label{Poynting_flux_z_disk}
\end{aligned}
\end{equation}
At $\theta\ll1$, one has
\begin{equation}
\begin{aligned}
S_{z}=\frac{B_{\phi}^{2}}{4\pi}=\frac{C_{2}}{\pi}\Omega^{2}\Psi r^{\nu-2}
=\frac{C_{2}^{2/\nu}}{\pi}\Omega^{2}\Psi^{2-2/\nu}\theta^{-2+4/\nu}
=\frac{\alpha^{2}}{\pi}C_{2}^{2\lambda+2}z^{2\lambda\nu+2\nu-4\lambda-4}R^{4\lambda+2}.
\label{Poynting_flux_z_tll1}
\end{aligned}
\end{equation}
In the case of $\lambda=0$ (a magnetic field threading a CO), one has $S_{z}\propto R^{2}$ at a given height, implying a hollow jet (the Poynting flux vanishes on the polar axis), which may account for a limb-brightening phenomenon in some AGN jets \citep[for example, M87; see, e.g.,][]{2007ApJ...668L..27K, 2007ApJ...660..200L, 2018ApJ...855..128W} and the apparent darkness of the inner pulsar wind nebulae of the Crab and other pulsars \citep[e.g.,][]{2000ApJ...536L..81W, 2002ApJ...577L..49H, 2009ASSL..357..421K, 2009ApJ...698.1570L}. The hollow jet structure is also seen in MHD simulations \citep[e.g.,][]{2006ApJ...641..103H, 2008MNRAS.388..551T}.

\section{Jet Dynamics}
\label{sec:jet_dynamics}

Since we consider a magnetohydrodynamic or
force-free jet in this paper, we are unable to study the effects of mass loading of the jet (see \citealt{2001ApJ...553..158O, 2004ApJ...601...90C} for mass-loading discussion in AGN jets and \citealt{2013ApJ...765..125L} for GRB jets). However, it is expected that the main properties of a highly magnetized jet carry over to a mass-loaded jet, provided that the latter is electromagnetically dominated.

\subsection{Force Balance}
\label{sec:Lorentz_Force}

In assumption of force-free, the Lorentz force vanishes. In a real plasma flow system, when plasma inertia is considered, the Lorentz force cannot be neglected and would be responsible for plasma acceleration, $\mathbf{F}_{\rm L}=\rho\left(\mathbf{u}\cdot\bigtriangledown\right)\mathbf{u}
=\rho_{\rm e}\mathbf{E}+\mathbf{j}\times\mathbf{B}$. It can be seen that in an axisymmetric system, the Lorentz force along the magnetic field line always vanishes: $\mathbf{F}_{\rm L,\hat{B}}=0$. Similar to the analysis in Section \ref{sec:basic_equation}, it can be proved that $\Psi$ and $\Omega$ are still conserved along magnetic field lines. Since the poloidal magnetic field $\mathbf{B}_{p}$ and the current density $\mathbf{j}_{p}$ may not be parallel to each other, $\Phi$ may not be conserved. Because the electric field is always perpendicular to the magnetic field, one can naturally separate the Lorentz force into two parts: one along the direction of $\mathbf{E}$ (i.e. $\mathbf{F}_{\rm L,\hat{E}}$) and another along the direction of $\mathbf{E}\times\mathbf{B}$ (i.e. the direction of the drift velocity, $\mathbf{F}_{\rm L,\hat{v}_{\rm d}}$), i.e.
\begin{equation}
\begin{aligned}
\mathbf{F}_{\rm L}=\overbrace{\underbrace{\mathbf{j}_{p}\times\mathbf{B}_{p}}_{1,\hat{\phi}}
+\underbrace{\left(\mathbf{j}_{p}\times\mathbf{B}_{\phi}\right)_{B_{p}}\hat{B}_{p}}_{2,\hat{B}_{p}}}^{\mathbf{F}_{\rm L,\hat{v}_{\rm d}}}
+\overbrace{\underbrace{\left(\mathbf{j}_{p}\times\mathbf{B}_{\phi}\right)_{E}\hat{E}}_{3,\hat{E}}
+\underbrace{\mathbf{j}_{\phi}\times\mathbf{B}_{p}}_{4,\hat{E}}
+\underbrace{\rho_{\rm e}\mathbf{E}}_{5,\hat{E}}}^{\mathbf{F}_{\rm L,\hat{E}}}.
\label{FL_jB}
\end{aligned}
\end{equation}

The first two terms of Equation \ref{FL_jB} correspond to the Lorentz force along the direction of the drift velocity:
\begin{equation}
\begin{aligned}
\mathbf{F}_{\rm L,\hat{v}_{\rm d}}=\frac{1}{4\pi R^{2}}
\left(\frac{\partial\Phi}{\partial z}\frac{\partial\Psi}{\partial R}-\frac{\partial\Phi}{\partial R}\frac{\partial\Psi}{\partial z}\right) \left[\hat{\phi}+\left(\frac{B_{\phi}}{B_{p}}\right)\hat{B}_{p}\right].
\label{FL_EtB}
\end{aligned}
\end{equation}
This force has two effects: one is to accelerate the plasma, and another is to make the plasma velocity direction tend to align with the drift velocity direction. Furthermore, we have the following conditions equivalent: (1) the Lorentz force vanishes in the magnetic stream surface (i.e. the force-free condition applies in the surface, $\mathbf{F}_{\rm L,\hat{v}_{\rm d}}=0$); (2) the poloidal current density, velocity, and magnetic field are parallel to each other, $\mathbf{j}_{p}=\Phi'\mathbf{B}_{p}/4\pi$; (3) the current flows in the magnetic stream surface; (4) $\Phi$ is conserved along a magnetic field line, and thus is a function of $\Psi$ only. Releasing the force-free assumption (considering the inertia of the plasma) requires that the above conditions are broken in order to accelerate the plasma fluid. As a result, these conditions apply approximately in the limit of highly magnetized jet \citep[see below and, e.g.,][]{1992ApJ...394..459L}.

The last three terms of Equation \ref{FL_jB} refer to the Lorentz force along the direction of the electric field $\hat{E}$,
\begin{small}
\begin{equation}
\begin{aligned}
F_{\rm L,\hat{E}}=\frac{B_{p}}{R}
\left\{\underbrace{\frac{\partial^{2}\Psi}{\partial R^{2}}+\frac{\partial^{2}\Psi}{\partial z^{2}}-\frac{1}{R}\frac{\partial\Psi}{\partial R}}_{4}
+\underbrace{\Phi\frac{4\pi\mathbf{j}_{p}\cdot\mathbf{B}_{p}}{B_{p}^{2}}}_{3}
-\underbrace{\Omega^{2}R^{2}\left[\frac{\partial^{2}\Psi}{\partial R^{2}}+\frac{\partial^{2}\Psi}{\partial z^{2}}+\frac{1}{R}\frac{\partial\Psi}{\partial R}+\frac{\Omega'}{\Omega}\left(\left(\frac{\partial\Psi}{\partial R}\right)^{2}+\left(\frac{\partial\Psi}{\partial z}\right)^{2}\right)\right]}_{5}\right\},
\label{FL_E}
\end{aligned}
\end{equation}
\end{small}
which partially offers the ``centripetal force" for plasma rotation and self-collimation for the outflows. Term 3 is well known as the $z$-pinch in the plasma physics \citep[see][]{2001Sci...291...84M}. In the case of poloidal current density $\mathbf{j}_{p}$ parallel to that of the magnetic field $\mathbf{B}_{p}$ (i.e. $F_{\rm L,\hat{v}_{\rm d}}=0$), the term in the braces is reduced to the left-hand side term in Equation \ref{DE_Psi}. Therefore, Equation \ref{DE_Psi} refers to the force-free condition along $\mathbf{E}$ \citep[$F_{\rm L,\hat{E}}=0$; see, e.g.,][]{2009ApJ...698.1570L}\footnote{The magnetic force is almost balanced by the electric force, which is why both acceleration and collimation proceed very slowly.}. Making term 4 equal to zero leads to Equation \ref{DE_F1_Psi}, while making combined terms 3 and 5 equal to zero leads to Equation \ref{DE_F2_Psi}. Because $\Omega R$ almost measures the four-velocity of plasma fluid (see Section \ref{sec:jet_acceleration}), it can be seen that the electric force part is only important in the relativistic case. As discussed in Section \ref{sec:approximate_solution}, Equations \ref{DE_F1_Psi} and \ref{DE_F2_Psi} have approximately the same solutions, which implies that (1) the $\hat{\phi}$ component current density approximately vanishes (term 4 in Equation \ref{FL_jB}), and that (2) the force the poloidal current receives from the magnetic field is balanced by the force exerted by the poloidal electric field to guarantee the force-free condition \citep[in both relativistic and nonrelativistic cases, see terms 3 and 5 in Equation \ref{FL_jB}, for more discussion on force balance, see, e.g.,][]{2010ApJ...709.1100P}. As discussed above, in a nonrotation system, the toroidal magnetic field, the electric current, and the electric field would vanish. What might exist is the poloidal magnetic field. It is the rotation that generates the toroidal magnetic field and induces the poloidal current. The force that this poloidal current receives from the magnetic field will be (almost exactly) balanced by the force exerted by the appearing poloidal electric field to guarantee the force-free condition (Equation \ref{DE_F2_Psi}). This leaves the toroidal current almost vanishing and the magnetic stream function being almost change-free with rotation (Equation \ref{DE_F1_Psi}). This may explain why some MHD simulations found that the poloidal configuration of magnetic fields had little change from the nonrotation to the rotation cases \citep[e.g.,][]{2008MNRAS.388..551T}.

This rotating poloidal magnetic field coil then drives the plasma fluid trapped in it outward along the magnetic field lines as they try to uncoil. As this twist propagates outward, the induced toroidal field pinches the plasma fluid toward the rotation axis. Therefore, a magnetically dominated jet would be self-collimated and accelerated \citep[self-balanced, see, e.g.,][]{1982MNRAS.199..883B, 1989ApJ...347.1055H, 1996MNRAS.279..389L, 1997ApJ...486..291O, 2001Sci...291...84M, 2006MNRAS.368.1561M, 2014ApJ...789..144C}. There may be a very special case where a bunch of magnetic field lines that anchor on and rotate with the CO/AD can load a bunch of magnetically dominated plasma moving along the field line. This bunch of magnetically dominated plasma may also be self-balanced. Therefore one may also employ the magnetic stream function to describe the system. In this viewpoint, the CO/AD and its rotation can be considered as merely a boundary providing a special electromagnetic field configuration, which further determines the outflow. In pulsars, the magnetic axis is usually misaligned from the rotation axis, so that a bunch of open magnetic field lines threading the magnetic polar region may launch a magnetically dominated collimated outflow, while the magnetic field lines near the magnetic dipolar equator would span through a torus-like region \citep[e.g.,][]{2005ApJ...630.1020R, 2012MNRAS.424..605P}. This offers an alternative explanation to the ``jet-torus" feature frequently observed in pulsar wind nebulae \citep[][]{1995ApJ...448..240H, 2000ApJ...536L..81W, 2001ApJ...554L.189P, 2001ApJ...556L.107G, 2009ASSL..357..421K}. This model also predicts a high Lorentz factor for Poynting-flux-dominated pulsar winds (e.g. $\Gamma\sim10^{3-6}$, \citealt{1974MNRAS.167....1R,1984ApJ...283..694K}). This can be more easily understood for a special pulsar case where the rotation and magnetic axes align with each other. The magnetic field configuration outside the light cylinder would roughly follow a monopole structure \citep[e.g.,][]{1999ApJ...511..351C, 2006MNRAS.367...19K, 2006MNRAS.368L..30M}, which is an exact solution of Equation \ref{DE_Psi} \citep[see Appendix \ref{sec:monopole} and][]{1973ApJ...180L.133M}. In this case, the velocity of the plasma flow follows $v\Gamma=\Omega R$ for all polar angles (see Appendix \ref{sec:monopole}), which gives rise to a very high Lorentz factor for the pulsar winds.

There is another way to understand the effect of the Lorentz force. Generally speaking, the Lorentz force can be written as
\begin{equation}
\begin{aligned}
\mathbf{F}_{\rm L}=\frac{1}{4\pi}\left(\mathbf{B}\cdot\nabla\right)\mathbf{B}-
\nabla\left(\frac{B^{2}}{8\pi}\right)
+\frac{1}{4\pi}\left(\nabla\cdot\mathbf{E}\right)\mathbf{E}
=-\frac{B^{2}}{4\pi}\frac{\hat{R}_{B}}{R_{B}}-
\nabla_{\perp}\left(\frac{B^{2}}{8\pi}\right)
+\frac{1}{4\pi}\left(\nabla\cdot\mathbf{E}\right)\mathbf{E},
\label{FL_tension_pressure}
\end{aligned}
\end{equation}
where $\hat{R}_{\rm B}$ denotes the radial vector of a curved magnetic field line, and $R_{\rm B}$ denotes the radius of curvature of the magnetic field line. Therefore, we have $\hat{R}_{\rm B}/R_{\rm B}=-d\hat{\mathbf{B}}/ds=-\left(\hat{\mathbf{B}}\cdot\nabla\right)\hat{\mathbf{B}}$, with $\hat{\mathbf{B}}=\mathbf{B}/B$ being the unit vector along the magnetic field line. The gradient operator can be decomposed into two parts parallel and perpendicular to the magnetic field line, that is, $\nabla=\nabla_{\parallel}+\nabla_{\perp}$. The first term refers to a magnetic tension force of the magnetic field line, which appears whenever the magnetic field lines are curved and can be represented as a restoring force\footnote{The magnetic tension force provides a restoring force for the Alfv\'{e}n waves \citep[][]{1942Natur.150..405A}.}. The second term represents the magnetic pressure ($p_{\rm B}=B^{2}/8\pi$) gradient force, which occurs when the field strength, $B$, varies from position to position. In principle, the magnetic pressure is isotropic. However, the component parallel to the magnetic field line is exactly canceled out by the component of the tension force in the same direction. Concerning the balance between various magnetic stream surfaces in a jet flow, the magnetic tension force points toward the polar axis, while the magnetic pressure gradient force points toward the polar axis in the case of a magnetic field threading a CO ($s=0$) and points away from the polar axis in the case of a magnetic field threading a AD ($s=-3/2$ and $\nu=3/4$). The third term is the electric field force, which points toward an opposite direction of that of the magnetic pressure gradient force. The magnetic pressure gradient and electric forces are two large numbers in the region outside the ACS\footnote{According to Equation \ref{EB_pressure}, the electromagnetic pressure is measured by the field in the rest frame of the plasma, $B'=\sqrt{B^{2}-E^{2}}=B/\Gamma$, which is a small value in the region far outside the ACS, even though both $B$ and $E$ are large values.} \citep[e.g.,][]{2010mfca.book.....B}. The difference between the two almost balances the magnetic tension force.

\subsection{Jet Flow Velocity}
\label{sec:dynamics_velocity}

In this subsection, we discuss the maximum velocity a jet can be accelerated to, and the relationship between the jet fluid velocity and the drift velocity. Hereafter (only) in this subsection, $v$ (and $\Gamma$) refers to plasma fluid velocity (and Lorentz factor), while $v_{\rm d}$ (and $\Gamma_{\rm d}$) indicates the drift velocity (and Lorentz factor).

Combining the continuity equation
\begin{equation}
\nabla\cdot\left(\rho\mathbf{u}\right)=0,
\label{continuity_Eq}
\end{equation}
and the general velocity Equation \ref{vgv}, one obtains a conserved quality along a magnetic field line
\begin{equation}
\begin{aligned}
4\pi\rho\Gamma\kappa=\eta\left(\Psi\right),
\label{eta_cons}
\end{aligned}
\end{equation}
which represents the mass flux per unit magnetic flux.

Equation \ref{motion_Eq} can be divided into the poloidal and toroidal components,
\begin{equation}
\begin{aligned}
u_{p}\frac{\partial u_{p}}{\partial l}=\frac{\sin\vartheta}{R}u_{\phi}^{2}-\frac{B_{\phi}}{4\pi\rho R}\frac{\partial\left(RB_{\phi}\right)}{\partial l},
\label{motion_Eq_p}
\end{aligned}
\end{equation}
\begin{equation}
\begin{aligned}
u_{p}\frac{\partial\left(Ru_{\phi}\right)}{\partial l}=\frac{B_{p}}{4\pi\rho}\frac{\partial\left(RB_{\phi}\right)}{\partial l},
\label{motion_Eq_phi}
\end{aligned}
\end{equation}
where $l$ is the coordinate of the poloidal direction, and $\vartheta$ is the angle between the $l$ and $z$ directions, that is, $\sin\vartheta = \partial R/\partial l$. The toroidal component (Equation \ref{motion_Eq_phi}) gives another conserved quality
\begin{equation}
\begin{aligned}
R\Gamma v_{\phi}-\frac{RB_{\phi}}{\eta}=\mathcal{L}\left(\Psi\right),
\label{L_cons}
\end{aligned}
\end{equation}
where the first term (of the left-hand side) measures the matter angular momentum, the second term gives the torque associated with the magnetic stresses, and $\mathcal{L}$ represents the total angular momentum flux per unit mass flux.

Differentiating Equation \ref{L_cons} with $l$ and substituting it into Equation \ref{motion_Eq_p}, together with Equation \ref{vgv}, one obtains another conserved quality
\begin{equation}
\begin{aligned}
\Gamma-\frac{B_{\phi}\Omega R}{\eta}=\mathcal{E}\left(\Psi\right),
\label{E_cons}
\end{aligned}
\end{equation}
where $\mathcal{E}$ represents the total energy flux per unit rest energy flux. The first term of the left-hand side measures the matter energy flux, and the second term is for the Poynting flux. The ratio between the two terms defines a magnetization parameter \citep[e.g.,][]{2009ApJ...698.1570L}
\begin{equation}
\begin{aligned}
\sigma=\frac{-B_{\phi}\Omega R}{\eta\Gamma}=\frac{\mathcal{E}-\Gamma}{\Gamma}.
\label{sigma_pol}
\end{aligned}
\end{equation}
It should be noted that, only in the case of high magnetization ($\sigma\gg1$), the properties of plasma flow almost would not affect the details of the acceleration process, which is instead controlled by the magnetic field topology \citep[see Equation \ref{v_total} below and also, e.g., the GRMHD simulation in][]{2020ApJ...892...37P}.

Therefore,
\begin{equation}
\Gamma\left(\sigma+1\right)=\mathcal{E}
\end{equation}
is conserved along a magnetic field line, which implies that $\sigma$ decreases during jet acceleration, and the maximum Lorentz factor of the plasma cannot exceed $\mathcal{E}$ even when all of the Poynting flux energy is converted to kinetic energy \citep[e.g.,][]{2006MNRAS.367..375B, 2013PTEP.2013h3E02T}. In principle, from Equation \ref{E_cons}, the conservation of $\Phi=B_{\phi}R$ in the force-free condition has to be broken in a real jet system to prevent Lorentz factor $\Gamma$ from being conserved along a magnetic field line (see also the Equation \ref{FL_EtB}). However, in the case of high magnetization, one has $B_{\phi}R=\Phi$ approximately conserved and $\Gamma\sigma=-\Phi\Omega/\eta\approx\mathcal{E}$ ($\sigma\gg1$; see Equation \ref{sigma_pol}). To determine $\mathcal{E}$ at a foot-point, one needs to consider a detailed mass-loading mechanism, which is not well understood yet \citep[see, e.g.,][and references therein]{1982MNRAS.199..883B, 2001ApJ...553..158O, 2004ApJ...601...90C, 2016ApJ...821...80B,
2013ApJ...765..125L}. In principle, the value of $\mathcal{E}$ cannot be arbitrarily large, because the plasma energy density cannot be arbitrarily small because the fact that the plasma has to carry a proper charge and current to support the electromagnetic field (see Appendix \ref{app:maximum_Lorentz} for more discussion on the maximum Lorentz factor). One expects that MHD mass-loaded outflows starting with a high value of the initial magnetization parameter $\sigma$ would roughly follow the force-free solution out to a certain distance, beyond which the MHD flows likely feel the effect of the inertia. This simply means that the force-free solutions can only apply (physically reasonable) up to a modest distance from the central engine. Furthermore, for a highly magnetized jet flow, the plasma fluid velocity is very close to the drift velocity, which will be shown below.

From Equations \ref{L_cons} and \ref{E_cons}, one has another conservation law
\begin{equation}
\begin{aligned}
\Gamma\left(1-\Omega Rv_{\phi}\right)=\mathcal{E}-\Omega\mathcal{L}=\varepsilon.
\label{varepsilon_cons}
\end{aligned}
\end{equation}
Near the jet base, let us assume that the flow is sub-Alfv\'{e}nic. One expects that the fluid motion is dominated by the toroidal velocity, which is not very relativistic, that is, $\Omega R_{0} v_{\phi}\ll1$ and $\Gamma\approx1$. This yields\footnote{Even if the velocity reaches $v_{\phi}\approx\Omega R_{0}\approx1/2$ and $\Gamma\approx1/\sqrt{1-v_{\phi}^{2}}$, one still has the $\varepsilon$ value close to 1: $\varepsilon\approx0.87$.} $\varepsilon\approx1$. Expressing $v_{\phi}$ and $v_{p}$ in terms of $\Gamma$, one has
\begin{eqnarray}
%\begin{equation}
%\begin{aligned}
v_{\phi}&=&\frac{1}{\Omega R}\left(1-\frac{\varepsilon}{\Gamma}\right), \nonumber \\
v_{p}&=&\frac{B_{p}}{B_{\phi}}\left[\frac{1}{\Omega R}\left(1-\frac{\varepsilon}{\Gamma}\right)-\Omega R\right],
\label{v_flow_pphi}
%\end{aligned}
%\end{equation}
\end{eqnarray}
which implies
\begin{equation}
\begin{aligned}
\frac{\Gamma^{2}-1}{\Gamma^{2}}=\left[\frac{1}{\Omega R}\left(1-\frac{\varepsilon}{\Gamma}\right)\right]^{2}+
\left\{\frac{B_{p}}{B_{\phi}}\left[\frac{1}{\Omega R}\left(1-\frac{\varepsilon}{\Gamma}\right)-\Omega R\right]\right\}^{2}.
\label{Gamma_flow_Eq}
\end{aligned}
\end{equation}
Let us define $r_{B}=B_{\phi}/B_{p}$ and $\varpi=\Omega R$ for simplicity of writing. The above equation has a physical solution:
\begin{equation}
\begin{aligned}
\Gamma=\frac{\sqrt{r_{B}^{2}\varpi^{2}\left(1+r_{B}^{2}-\varpi^{2}\right)\left(-1+\varepsilon^{2}+\varpi^{2}\right)}
-\varepsilon\left(1+r_{B}^{2}-\varpi^{2}\right)}
{\left(1+r_{B}^{2}-\varpi^{2}\right)\left(\varpi^{2}-1\right)}.
\label{Gamma_flow}
\end{aligned}
\end{equation}
It can be seen that once the magnetic field configuration is specified, the cold plasma velocity would be completely determined \citep[see below and, e.g.,][]{2006MNRAS.367.1797M}.

\begin{figure}
\begin{center}
{\includegraphics[width=0.75\linewidth]{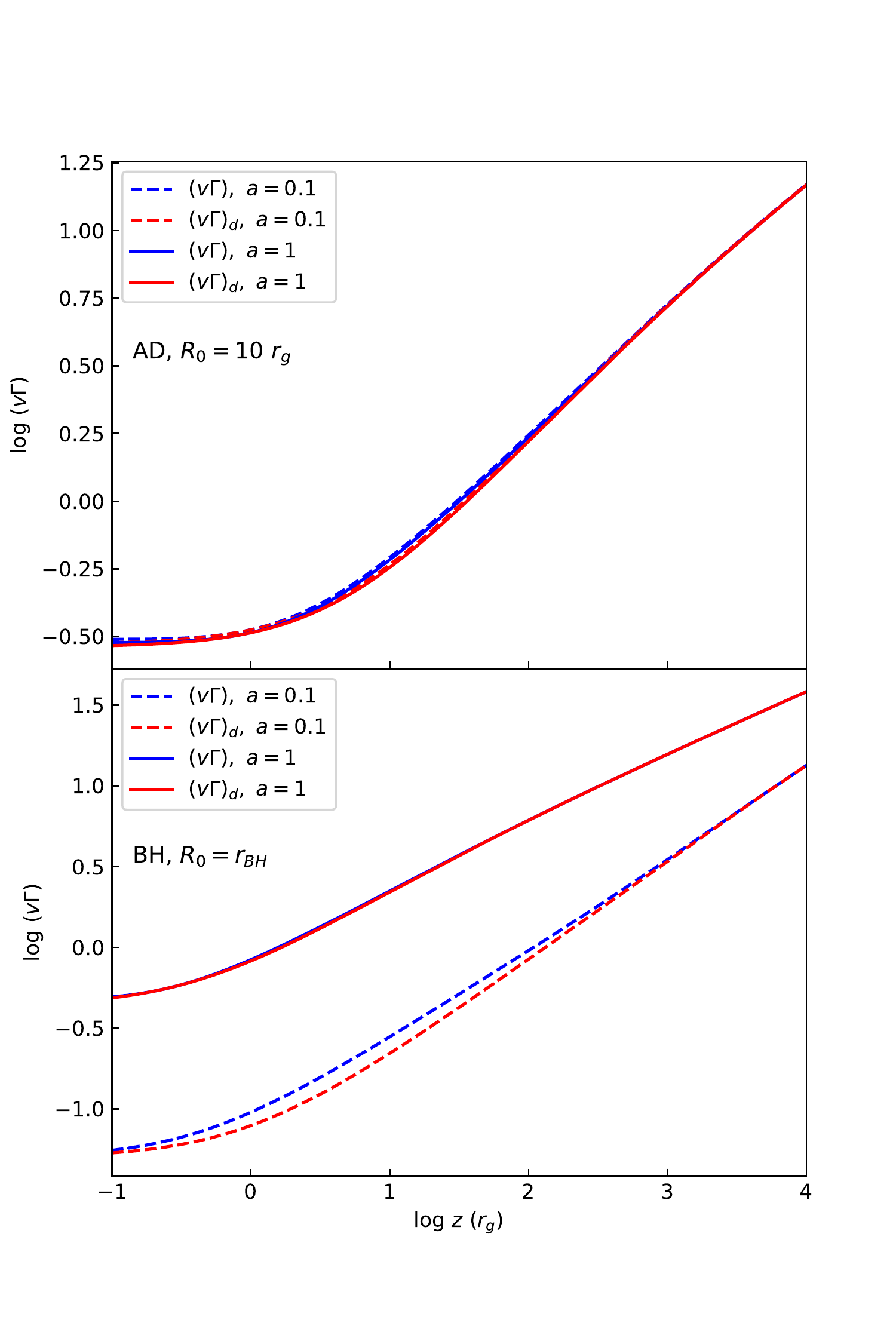}}
\end{center}
\caption{Cold plasma velocity (blue) vs. the drift velocity (red) of a highly magnetized jet. The bottom panel is for the magnetic field line threading the equator of a BH with the spin parameters $a=1$ and $a=0.1$ as presented by solid and dashed lines, respectively. The upper panel is for the magnetic field line threading the AD at radius $R_{0}=10$ $r_{\rm g}$ rotating with a Keplerian velocity, where the solid and dashed lines are the same as that in the bottom panel. It can be seen that the drift velocity matches the cold plasma velocity very well. This result does not significantly depend on the value of the conserved quality in Equation \ref{varepsilon_cons}, which is set to $\varepsilon=1-\left(\Omega R_{0}\right)^{2}/2$ here (see Equation \ref{varepsilon_approx}). See Section \ref{sec:dynamics_velocity} for details.}.
\label{fig:vG_v.vs.vd}
\end{figure}

From Equation \ref{v_drift_total_s}, the drift velocity can be expressed as
\begin{equation}
\begin{aligned}
\Gamma_{\rm d}=\sqrt{\frac{1+r_{B}^{2}}{1+r_{B}^{2}-\varpi^{2}}}.
\label{Gamma_drift}
\end{aligned}
\end{equation}
We can then define a deviation of plasma fluid velocity from drift velocity:
\begin{small}
\begin{equation}
\begin{aligned}
D_{\rm fd}\equiv\frac{\left(v\Gamma\right)^{2}-\left(v_{\rm d}\Gamma_{\rm d}\right)^{2}}{\left(v_{\rm d}\Gamma_{\rm d}\right)^{2}} =
\frac{\left(1+r_{B}^{2}-\varpi^{2}\right)\left(\varpi^{2}-1\right)
-2\varepsilon\sqrt{r_{B}^{2}\varpi^{2}\left(1+r_{B}^{2}-\varpi^{2}\right)\left(-1+\varepsilon^{2}+\varpi^{2}\right)}
+\varepsilon^{2}\left[1-\varpi^{2}+r_{B}^{2}\left(1+\varpi^{2}\right)\right]}
{\varpi^{2}\left(\varpi^{2}-1\right)^{2}}.
\label{vG_diff_general}
\end{aligned}
\end{equation}
\end{small}
As argued above, one has $\varepsilon\approx1$ and $\left|B_{\phi}\right|/B_{p}\approx\Omega R=\varpi$ (see Equation \ref{B_t_p}). Therefore, the above equation can be approximated as
\begin{equation}
\begin{aligned}
D_{\rm fd}\approx
\frac{\varpi^{4}-2\varpi^{3}+\varpi^{2}}{\varpi^{2}\left(\varpi^{2}-1\right)^{2}},
\label{vG_diff_app}
\end{aligned}
\end{equation}
which is a decreasing function of $\varpi$. In the limit of $\varpi=\Omega R\gg1$, one has $D_{\rm fd}\approx1/\varpi^{2}\ll1$, implying that the relativistic jet has almost the same fluid and drift Lorentz factors. In the limit of $\varpi=\Omega R\ll1$, one has $v\Gamma\approx\sqrt{2}v_{\rm d}\Gamma_{\rm d}\ll1$, indicating that the fluid and drift velocities are still roughly equal to each other. More exactly, in the limit of $\varpi=\Omega R\ll1$, the deviation of $\varepsilon$ from  unity could not be safely ignored (see Equations \ref{varepsilon_cons} and \ref{vG_diff_general}). In this case, let us set $v\Gamma=x v_{\rm d}\Gamma_{\rm d}$, which also implies $v_{\phi}\approx xv_{\phi,\rm d}$ since the $\phi$ component velocity dominates. In this case, $\left|B_{\phi}\right|/B_{p}$ may also slightly differ from $\Omega R$. We may simply set $\left|B_{\phi}\right|/B_{p}=c_{\rm B}\Omega R$ (see Equation \ref{B_t_p}). Combining these relations and $v_{\phi,\rm d}=\varpi/\left(1+\varpi^{2}\right)$ (see Equation \ref{v_phi}), one has
\begin{equation}
\begin{aligned}
\varepsilon=1-\left(x-\frac{x^{2}}{2}\right)\varpi^{2}+O\left(\varpi^{4}\right).
\label{varepsilon_approx}
\end{aligned}
\end{equation}
Substituting this relation into  Equation \ref{vG_diff_general}, one can derive a solution $x=1+c_{\rm B}^{2}\varpi^{2}/2+O\left(\varpi^{3}\right)$, which indicates that the fluid has a velocity very close to the drift velocity. We plot the comparison between this cold plasma velocity (Equation \ref{Gamma_flow}) and drift velocity (Equation \ref{Gamma_drift}) in Figure \ref{fig:vG_v.vs.vd}. It can be seen that the drift velocity matches the cold plasma velocity very well. This result is not significantly affected by the conserved quality in Equation \ref{varepsilon_cons}, which is set to $\varepsilon=1-\left(\Omega R_{0}\right)^{2}/2$ here, with $R_{0}$ being the radial distance of the foot-point of the magnetic field (see Equation \ref{varepsilon_approx}). We note that \citet{2007MNRAS.375..548N} and \citet{2008MNRAS.388..551T} reached a similar result in the limit of the relativistic case. As for why the plasma fluid velocity can be represented by a pure electromagnetic field quality - drift velocity, one may understand it as follows: (1) The plasma flow is highly magnetized, so it is dynamically unimportant and almost cannot affect the electromagnetic field configuration. (2) The ideal MHD condition indicates that the motion of the plasma fluid is governed by the electromagnetic field configuration (freezing effect). (3) The electromagnetic field configuration is self-consistently determined because the plasma can supply currents and charges as needed to support the electromagnetic field.

\subsection{Jet Flow Density}
\label{sec:JetProperdensity}
Within the force-free approach, the inertia is ignored, which prevents a calculation of the flow density. In another aspect, given an available mass flux per magnetic flux $\eta$ (see Equation \ref{eta_cons}) that still satisfies a highly magnetically dominated condition, an approximation of the jet flow density along the magnetic field line can be derived, with the aid of Equations \ref{vgv}, \ref{eta_cons}, \ref{magneticflux} and \ref{v_drift_total_s}, i.e.
\begin{eqnarray}
\rho & = & \frac{\eta}{4\pi}\frac{B_{p}}{u_{p}}\simeq \frac{\eta\Psi}{2\pi\Omega^{2}}\frac{\sqrt{1+\left(\Omega R\right)^{2}}}{R^{4}}\simeq \frac{\eta}{4\pi\Omega^{2}}\frac{B}{R^{2}}, \nonumber \\
\rho_{\rm l} & = & \Gamma\rho\simeq \frac{\eta}{8\pi\Psi\Omega^{2}}B^{2},
\label{JetProperdensity}
\end{eqnarray}
where $\rho_{\rm l}=\Gamma\rho$ refers the density in the lab frame \citep[see e.g.,][for $\rho_{\rm l}\propto B^{2}$]{2009ApJ...697.1164B}.  Note that the derivation of the approximations makes use of Equation \ref{v_drift_total_s}, which can be further improved by using the velocity formula from Sections \ref{sec:total_velocity} and \ref{sec:poloidal_velocity}.

\section{Black Hole Jets}
\label{BHjet}

As discussed in Section \ref{sec:magnetic_configuration}, the foot-point location can be constrained by measuring the large-scale jet configuration $r_{0}^{\nu}T\left(\theta_{0}\right)=C_{2}z^{\nu-2}R^{2}$ (see Equations \ref{Solution_nonr} and \ref{half_openangle_R}). Furthermore, the angular velocity can also be constrained by measuring the jet width and its velocity (in the $V_{1}$-dominated regime, $v\Gamma=\Omega R$), which presents a potential method of estimating the BH spin for a jet launched from a rotating BH\footnote{In some AGNs, the jet power may be even larger than the accretion power, which implies that the jet may be produced from a rotating BH \citep[see, e.g.,][]{2014Natur.515..376G, 2018ApJS..235...39C}.}.

In principle, one should consider  GR effects to study jets launching from a BH, which was not taken into account in this paper. In the following, we discuss how GR effects may play a significant role near the BH. The full GR effects have been studied in a growing number of GRMHD simulations recently, which are required to capture the complicity of outflows near the BH \citep[e.g.,][]{2006MNRAS.368.1561M, 2006MNRAS.367.1797M, 2007MNRAS.375..513M, 2012MNRAS.423.3083M, 2013MNRAS.436.3741P, 2015ApJ...801...56P, 2019MNRAS.490.2200C}. Theoretical analyses show that, because of the GR frame dragging effect in the vicinity of the BH, a magnetic field line threading a BH does not corotate with the BH but lags behind with an angular velocity approximately one-half of that of the BH \citep[e.g.,][]{1977MNRAS.179..433B, 1982MNRAS.198..345M}. Numerical simulations show that the angular velocity is not constant and has $f_{\Omega}\approx0.3-0.5$ for $\Omega=f_{\Omega}\Omega_{\rm BH}$ depending on the polar angles \citep[e.g.,][]{2007MNRAS.375..531M, 2010ApJ...711...50T, 2015ApJ...812...57P}. For our analytical treatment, we assume a constant $f_{\Omega}=0.5$ throughout the following analysis. As presented by \citet{2007MNRAS.375..531M}, the impact of this polar-angle-dependent behavior on the jet properties is almost negligible. The electromagnetic energy would be always transported out from a BH so that the jet can be launched. This implies that as long as the jet is not very close to the BH, the Lorentz force would dominate gravity, so that the gravitational effect would not qualitatively change the magnetic field configuration, as shown in simulations \citep[see, e.g.,][]{2007MNRAS.375..513M, 2007MNRAS.375..531M}.

We employ a pseudo-Newtonian potential to measure GR effects around a rotating BH. The gravitational acceleration can be written as \citep{1980A&A....88...23P, 1996ApJ...461..565A}
\begin{equation}
\begin{aligned}
g_{\rm BH}=-\frac{M}{r^{2-\gamma}\left(r-r_{+}\right)^{\gamma}},
\label{BH_acceleration}
\end{aligned}
\end{equation}
with $\gamma=r_{\rm ISCO}/r_{+}-1$, where $r_{+}=(1+\sqrt{1-a^{2}})r_{\rm g}$ is the horizon radius of the BH, $r_{\rm g}=M$ is the gravitational radius, $a=J_{\rm am}/M^{2}$ is the BH spin parameter, and $J_{\rm am}$ is the angular momentum. The innermost stable circular orbit (ISCO) is $r_{\rm ISCO}=\left\{3+Z_{2}-\sqrt{\left(3-Z_{1}\right)\left(3+Z_{1}+2Z_{2}\right)}\right\}r_{\rm g}$ with $Z_{1}=1+\left(1-a\right)^{1/3}\left[\left(1+a\right)^{1/3}+\left(1-a\right)^{1/3}\right]$ and $Z_{2}=\sqrt{3a^{2}+Z_{1}^{2}}$. As presented by \citet{1996ApJ...461..565A}, this potential captures the essentials of the GR effects from the exact relativistic Kerr metric because the freefall acceleration tends to be infinite when $r$ approaches  the event horizon of the BH, and the position of the extremum of the boundary condition function is the same as that of the last stable circular orbit in the exact relativistic Kerr metric. Notice that the gravitational acceleration is not significantly dependent on the polar angle $\theta$ \citep[e.g.,][]{1982MNRAS.198..345M, 2018PhRvD..98h3004D}.

With gravity, the motion equation now reads $\rho\left(\mathbf{u}\cdot\bigtriangledown\right)\mathbf{u}
=\rho_{\rm e}\mathbf{E}+\mathbf{j}\times\mathbf{B}+\rho g_{\rm BH}\mathbf{\hat{r}}$. In the case of the magnetic force dominating over gravity (the inertia is negligible), the force-free condition (Equation \ref{force_free}) applies, so that the electromagnetic configuration would evolve self-consistently and our approximate solution applies. This implies that there is a critical surface where gravity balances the magnetic force, outside of which our approximate solution would roughly apply. The magnetic force can be expressed in terms of acceleration:
\begin{equation}
\begin{aligned}
\left|g_{\rm B}\right|=\sigma\left(\frac{B^{2}}{8\pi}\right)^{-1}\left|\left(\mathbf{j}\times\mathbf{B}\right)_{r}\right|
=8\nu\sigma\Omega^{2}r\frac{1}{\left(1+B_{\phi}^{2}/B_{p}^{2}\right)\left(\nu^{2}+T'^{2}/T^{2}\right)} \approx4\nu\sigma\Omega^{2}r\frac{1-\cos\theta}{\left(1+B_{\phi}^{2}/B_{p}^{2}\right)},
\label{Lorentz_acceleration_Bmag}
\end{aligned}
\end{equation}
where $\sigma=U_{\rm B}/\rho$ is the magnetization parameter for a nonrelativistic flow. Note that the magnetic field would be poloidally dominated near the BH, that is, $B_{\phi}^{2}/B_{p}^{2}<1$. It can be seen that the magnetic force acceleration becomes smaller with a smaller $\theta$, which implies that this critical surface will extend to a larger distance from the BH at a smaller $\theta$. For $a=0.998$, $\sigma=10$, $\theta=\pi/4$, and $\nu=3/4$, the combination of Equations \ref{BH_acceleration} and \ref{Lorentz_acceleration_Bmag} yields the critical balance point $r_{\rm cb}\approx1.4$ $r_{\rm g}$.

To magnetically launch a jet from a BH, the rotation of the BH forces the magnetic field lines to rotate,  forming a helical structure. This  generates a Lorentz force that is exerted on the charged particles to force them to move outward. If the jet is very close to the BH, strong gravity would force the particles to move inward. One would therefore expect the existence of the so-called stagnation surface, within which  gravity dominates over the Lorentz force and particles move inward, and beyond which the Lorentz force dominates over gravity and  particles move outward. This stagnation surface can be derived through the condition of gravity balancing the Lorentz force. Near the BH, the Lorentz force is dominated by the magnetic force, which implies that the stagnation surface is very close to the critical balance surface discussed above, but is a little bit more outside since the electric force would partially cancel the magnetic force. In principle, the force-free condition implies that the Lorentz force would vanish and, therefore, would prevent a further constraint on the stagnation surface. Our approximate solution can be used to make a very rough estimate by expressing the Lorentz force in terms of acceleration:
\begin{equation}
\begin{aligned}
\left|g_{\rm L}\right|\approx\nu\sigma\Omega^{2}r\frac{\left(1-\cos\theta\right)^{2}\left(\nu^{2}-5\nu+6\right)}{\left(1+B_{\phi}^{2}/B_{p}^{2}\right)}.
\label{Lorentz_acceleration}
\end{aligned}
\end{equation}
Similar to Equation \ref{Lorentz_acceleration_Bmag}, the stagnation surface is expected to extend to a larger distance from the BH at a smaller $\theta$. For  $\nu=3/4$, $a=0.998$, $\sigma=10$ and $\theta=\pi/4$, the combination of Equations \ref{BH_acceleration} and \ref{Lorentz_acceleration} gives the stagnation point $r_{\rm st}\approx2.5$ $r_{\rm g}$. This picture is roughly consistent with GRMHD simulations \citep[see, e.g.,][]{1990ApJ...363..206T, 2006MNRAS.368.1561M, 2015ApJ...801...56P, 2018ApJ...868..146N}.

We now discuss how to constrain the BH spin and jet foot-points by measuring the large-scale jet properties. The rotation angular velocity of a BH (the angular
velocity of the zero-angular-momentum observer at the BH's horizon) is $\Omega_{\rm BH}=a/2r_{+}$. The magnetic field line penetrating the BH rotates with
\begin{equation}
\begin{aligned}
\Omega=f_{\Omega}\Omega_{\rm BH}=f_{\Omega}\frac{a}{2\left(1+\sqrt{1-a^{2}}\right)r_{\rm g}}.
\label{BH_rotation_Omega}
\end{aligned}
\end{equation}
The magnetic field line with the foot-point $r_{0}$ on the AD cannot rotate faster than the Keplerian velocity (in Boyer-Lindquist coordinates):
\begin{equation}
\begin{aligned}
\Omega\lesssim\Omega_{\rm K}=\frac{1}{r_{\rm g}\left[\left(r_{0}/r_{\rm g}\right)^{3/2}+a\right]}\leq\frac{1}{r_{\rm g}\left(r_{0}/r_{\rm g}\right)^{3/2}},
\label{Keplerian_rotation_Omega}
\end{aligned}
\end{equation}
which reaches the maximum value $\Omega_{\rm K,max}$ at the ISCO. Based on the measured angular velocity and given a BH mass, one can estimate the value (Equation \ref{BH_rotation_Omega}) or the lower-limit value  ($\Omega\leq\Omega_{\rm K,max}$, Equation \ref{Keplerian_rotation_Omega}) of $a$. In another aspect, for any choices of $a$, the angular velocity of a magnetic field line cannot be larger than $1/r_{\rm g}$ (corresponding to $r_{0}=r_{\rm g}$ at ISCO for case $a=1$ in Equation \ref{Keplerian_rotation_Omega}), which gives a constraint on the BH mass\footnote{The magnetic field threading the AD may rotate more slowly than the Keplerian velocity. However, it is still unclear how much slower it is. The magnetic field threading the BH has a maximum rotation angular velocity of $\Omega=f_{\Omega}/2r_{\rm g}$ (Equation \ref{BH_rotation_Omega} for $a=1$). If this value is also assumed to be the maximum rotation velocity of the magnetic field threading the AD \citep[][]{2007MNRAS.375..513M}, one has a more severe constraint on the BH mass $M\leq f_{\Omega}R/\left(2v\Gamma\right)$.} $M\leq1/\Omega=R/\left(v\Gamma\right)$.

Assuming that the jet configuration and velocity profile represent the same magnetic stream surface, one can combine these relations with Equations \ref{BH_rotation_Omega} and \ref{Keplerian_rotation_Omega} to obtain two equations of three quantities: BH mass, BH spin, and foot-point location (measured by $\theta_{0}$ and $r_{0}$). In principle, given one of these three quantities, one can derive the remaining two. Generally, one can get a constraint on these three quantities based on the fact that $0\leq a\leq1$ and $0\leq T\left(\theta\right)\leq1$ (noticing that the velocity profile does not depend on the absolute amplitude of $\Psi$; see Section \ref{sec:jet_acceleration} and Appendix \ref{app:sign}).

\textbf{Case I}: Magnetic field line threading a BH:
\begin{eqnarray}
T\left(\theta_{0}\right) & \geq & C_{2}\theta^{2-\nu}\left(\frac{2v\Gamma}{f_{\Omega}}\right)^{\nu}, \nonumber \\
a & \geq & a_{\rm min}=C_{2}^{1/\nu}\theta^{2/\nu-1}\left(\frac{2v\Gamma}{f_{\Omega}}\right), \nonumber \\
M & \geq & \frac{f_{\Omega}a_{\rm min}R}{2\left(1+\sqrt{1-a_{\rm min}^{2}}\right)v\Gamma}, \nonumber \\
r_{0} & = & r_{+}=(1+\sqrt{1-a^{2}})r_{\rm g}.
\label{COAD_constrain_BH}
\end{eqnarray}

\textbf{Case II}: Magnetic field line threading a AD:
\begin{eqnarray}
\theta_{0} & = & \pi/2, \nonumber \\
r_{0} & = &C_{2}^{1/\nu}r\theta^{2/\nu}, \nonumber \\
M & \geq & C_{2}^{3/\nu}\left(v\Gamma\right)^{2}r\theta^{6/\nu-2}, \nonumber \\
\frac{1}{\left(r_{\rm ISCO}/r_{\rm g}\right)^{3/2}} & \geq & C_{2}^{3/\nu}\left(v\Gamma\right)^{3}\theta^{6/\nu-3}.
\label{COAD_constrain_AD}
\end{eqnarray}

\subsection{Properties of a Black Hole Jet}
\label{BHjet_2}

In this section, we consider a jet with magnetic field lines threading a BH. Let us further assume that the outermost magnetic stream surface is connected with the equator of a spinning BH, and the innermost stream surface shrinks to the polar axis. For the outermost magnetic stream surface, one has ($\theta\ll1$)
\begin{eqnarray}
\frac{R_{\rm out}}{r_{\rm g}}&=&C_{2}^{-1/2}\left(1+\sqrt{1-a^{2}}\right)^{\nu/2}\left(\frac{z}{r_{\rm g}}\right)^{1-\nu/2}, \nonumber \\
V_{1,\theta\ll1,\rm out}&=& C_{2}^{-1/2}\left(\frac{f_{\Omega}}{1/2}\right)\frac{a}{4\left(1+\sqrt{1-a^{2}}\right)^{1-\nu/2}}
\left(\frac{z}{r_{\rm g}}\right)^{1-\nu/2}, \nonumber \\
V_{2,\theta\ll1,\rm out}&=& \frac{2}{\sqrt{2-\nu}} C_{2}^{1/2}\left(1+\sqrt{1-a^{2}}\right)^{-\nu/2}\left(\frac{z}{r_{\rm g}}\right)^{\nu/2}, \nonumber \\
P_{1}&=&3.44\frac{1+\sqrt{1-a^{2}}}{a}\left(\frac{1/2}{f_{\Omega}}\right)
\left(\frac{M}{10^{8}M_{\odot}}\right)\ \rm hr, \nonumber \\
\left(v\Gamma\right)^{2}B'&=& \left(\Omega R\right)^{2}B_{p}= \frac{2}{\nu^{2}+C_{1}^{2}}\left(\Omega R_{0}\right)^{2}B_{p,0}= \frac{2}{\nu^{2}+C_{1}^{2}}\left(\frac{f_{\Omega}}{1/2}\right)^{2} \frac{a^{2}B_{p,0}}{16},
\label{equator_BH_real}
\end{eqnarray}
where $P_{1}$ is the cycle period of the magnetic field. It can be seen that a BH with a larger spin parameter will produce a faster jet at a given height $z$ from the central BH \citep[see the GRMHD simulation by][]{2018ApJ...868..146N}, which implies that one can constrain the BH spin by measuring the jet acceleration profile. For  $f_{\Omega}=1/2$, $\nu=3/4$, and $a=1$, at distance $z=10^{3}\ r_{\rm g}$, one has $R_{\rm out}=109\ r_{\rm g}$, $V_{1,\theta\ll1,\rm out}=27.3$, $V_{2,\theta\ll1,\rm out}=16.4$, and $\left(v\Gamma\right)_{\theta\ll1,\rm out}=14.1$ (the ACS $\left(v\Gamma\right)_{\rm out}=1$ is located at $z=4.99\ r_{\rm g}$ for $a=1$, at $z=67.4\ r_{\rm g}$ for $a=0.3$, and at $z=800\ r_{\rm g}$ for $a=0.1$). For BH spin parameters varying in the range $a\sim0.1-1$ and with typical values of Lorentz factor $\Gamma\sim10-30$ in AGN jets \citep[e.g.,][]{2014Natur.515..376G, 2018ApJS..235...39C}, the cycle period in the inertial frame would be on a timescale from submonth to years ($P\approx\left(1+\Gamma^{2}\right)P_{1}$, see Equation \ref{period_P1}). These values are in the range of the observed periodic timescales of flux variability in AGN jets  \citep[e.g.,][]{2015ApJ...813L..41A, 2018NatCo...9.4599Z, 2020arXiv200200805P} or the oscillation timescales of the jet position angle \citep[e.g.,][]{2013AJ....146..120L, 2018ApJ...855..128W}. The last formula in above Equation \ref{equator_BH_real} applies to the region where\footnote{In other cases, it is an upper limit.} $\Omega R=v\Gamma$ is satisfied, where $B'=B/\Gamma=B_{p}$ is the strength of the magnetic field measured in fluid rest frame, and where $B_{p,0}$ is the strength of the poloidal magnetic field near the CO equator.

For the case of $\nu<1$, $V_{1,\theta\ll1}$ crosses $V_{2,\theta\ll1}$ at the CCS (see Equation \ref{V1V2trans}): $\Omega R^{2}z^{-1}=2/\sqrt{2-\nu}$. The inside region of CCS is dominated by $V_{1,\theta\ll1}$, while the outside region is dominated by $V_{2,\theta\ll1}$. The CCS crosses the outermost magnetic stream surface at
\begin{equation}
\begin{aligned}
\frac{z_{\rm crs}}{r_{\rm g}}=\left(1+\sqrt{1-a^{2}}\right)\left[\frac{2}{\sqrt{2-\nu}}\left(\frac{1/2}{f_{\Omega}}\right)\frac{4C_{2}}{a}\right]^{1/\left(1-\nu\right)}.
\label{z_cross}
\end{aligned}
\end{equation}
For $f_{\Omega}=1/2$, $\nu=3/4$, and $a=1$, one has $z_{\rm crs}=130 \ r_{\rm g}$. Since $V_{1,\theta\ll1}$ is an increasing function and since $V_{2,\theta\ll1}$ is a decreasing function of the cylindrical radius $R$, one therefore expects that (1) in the region of $z>z_{\rm crs}$, the jet has a slower spine surrounded by a faster interlayer, which is further surrounded by a slower outer layer; and that (2) in the region of $z<z_{\rm crs}$, the jet only has a slower spine surrounded by a faster layer. For the former case, there is a maximum velocity at the CCS (Equation \ref{V1V2trans}):
\begin{equation}
\begin{aligned}
\left(v\Gamma\right)_{\rm CCS}=V_{1,\theta\ll1,\rm crs}/\sqrt{2}=\frac{1}{\left(2-\nu\right)^{1/4}}
\left[\frac{a}{4\left(1+\sqrt{1-a^{2}}\right)}\left(\frac{f_{\Omega}}{1/2}\right)\frac{z}{r_{\rm g}}\right]^{1/2}.
\label{v_cross}
\end{aligned}
\end{equation}
For $f_{\Omega}=1/2$, $\nu=3/4$, and $a=1$, at $z=10^{3}\ r_{\rm g}$, one has $\left(v\Gamma\right)_{\rm CCS}=15.0$, which is larger than that at the outermost surface $\left(v\Gamma\right)_{\rm out}=14.1$. At $z=10^{5}\ r_{\rm g}$, one has $\left(v\Gamma\right)_{\rm CCS}=150$ and the jet open angle $\theta_{\rm op}=2\sin^{-1}\left(R_{\rm out}/z\right)=2.22^{\circ}$, which are consistent with the typical values for GRBs\footnote{Note that the maximum velocity does not occur at the outermost magnetic stream surface and thus does not conflict with $\left(v\Gamma\right)\theta\leq2/\sqrt{2-\nu}$. Also, for a stellar mass black hole, $z=10^{5}$ $r_{\rm g}\sim10^{11}$ cm is compatible with the size of the progenitor star.} \citep[e.g.,][]{2001ApJ...555..540L, 2001ApJ...562L..55F, 2018ApJ...859..160W}. At the critical point $z=z_{\rm crs}$, one has $\left(v\Gamma\right)_{\rm CCS,crs}=5.40$ for $a=1$, and this value  increases as the spin parameter $a$ decreases. This means that for the jet region with $v\Gamma\leq5.4$, $V_{1,\theta\ll1}$ always dominates (with $f_{\Omega}=1/2$ and $\nu=3/4$). Notice that this maximum velocity $\left(v\Gamma\right)_{\rm CCS}$ is over the entire jet region, as opposed to that along a magnetic field line \citep[see][for a similar discussion]{2008MNRAS.388..551T}. Such overall jet velocity properties are clearly presented in Figure \ref{fig:velocity_map} (for $\nu=3/4$).

The total jet power is dominated by the Poynting flux, which can be estimated as \citep[see Equation \ref{P_jetpower} and also][]{1977MNRAS.179..433B, 2008MNRAS.388..551T},
\begin{eqnarray}
P_{\rm jet}&\sim &2\times10^{45}\frac{2}{\nu^{2}+C_{1}^{2}} \left(1+\sqrt{1-a^{2}}\right)^{2}a^{2}\left(\frac{f_{\Omega}}{1/2}\right)^{2}
\left(\frac{M}{10^{8}M_{\odot}}\right)^{2}
\left(\frac{B_{p,0}}{10^{5}\rm Gs}\right)^{2}\ \rm erg \ \rm s^{-1} \nonumber \\
&\sim&2\times10^{51}\frac{2}{\nu^{2}+C_{1}^{2}} \left(1+\sqrt{1-a^{2}}\right)^{2}a^{2}\left(\frac{f_{\Omega}}{1/2}\right)^{2}
\left(\frac{M}{10M_{\odot}}\right)^{2}
\left(\frac{B_{p,0}}{10^{15}\rm Gs}\right)^{2}\ \rm erg \ \rm s^{-1},
%%%%%P_{\rm jet}\propto\frac{G^{2}B^{2}M^{2}}{c^{3}}
\label{P_jetpower_real}
\end{eqnarray}
which is large enough to provide the jet power in AGNs \citep[e.g.,][]{2014Natur.515..376G, 2018ApJS..235...39C, 2019ARA&A..57..467B} and GRBs \citep[e.g.,][]{2004RvMP...76.1143P, 2006RPPh...69.2259M,2015PhR...561....1K, 2018pgrb.book.....Z}. The jet power increases with an increasing $a$ and roughly follows a power-law dependence with a slope $\sim2$. This is also consistent with what is observed in X-ray binaries \citep[][]{2012MNRAS.419L..69N, 2013ApJ...762..104S}.

The potential difference between the magnetic stream surfaces at polar angle $\theta$ and at the polar axis near the BH can be estimated as (Equation \ref{DeltaV})
\begin{eqnarray}
\Delta V&\sim&8\times10^{19}\sqrt{\frac{2}{\nu^{2}+C_{1}^{2}}} \left(1+\sqrt{1-a^{2}}\right)a\left(\frac{f_{\Omega}}{1/2}\right)
\left(\frac{M}{10^{8}M_{\odot}}\right)
\left(\frac{B_{p,0}}{10^{5}\rm Gs}\right)T\left(\theta\right)\ \rm volts  \nonumber \\
&\sim&8\times10^{22}\sqrt{\frac{2}{\nu^{2}+C_{1}^{2}}} \left(1+\sqrt{1-a^{2}}\right)a\left(\frac{f_{\Omega}}{1/2}\right)
\left(\frac{M}{10M_{\odot}}\right)
\left(\frac{B_{p,0}}{10^{15}\rm Gs}\right)T\left(\theta\right)\ \rm volts.
%%%%%\Delta V\propto\frac{GBM}{c^{3}}
\label{DeltaV_real}
\end{eqnarray}
In principle, any charged particle that crosses the magnetic stream surface would be accelerated by the potential difference. In another aspect, this provides a maximum potential difference reachable for a gap (if formed) near the polar cap region, which indicates that even making use of a small potential in the gap (corresponding to $\theta\ll1$) would be able to accelerate a charged particle to a very high energy \citep[see the case of a pulsar magnetosphere,][]{1975ApJ...196...51R}. Similar to the case in pulsar inner gaps, an induced strong electric field along the magnetic field line can serve to accelerate particles and further produce $e^{\pm}$ pairs through a cascade of interaction between these high-energy particles and background photons \citep[see][for detailed simulations]{1998ApJ...497..563H, 2019PhRvL.122c5101P}. This also offers a possible source of ultra-high-energy cosmic rays and neutrinos \citep[][]{2011ARA&A..49..119K, 2013Sci...342E...1I, 2018Sci...361.1378I, 2018Sci...361..147I, 2019PhR...801....1A}.

One needs to mention the caveat that the ``pulsar" Equation \ref{DE_Psi} is written for a flat space. The application of its solution near a BH only yields approximate results.

\section{The CO/AD as a Boundary}
\label{Bp_AD}

In principle, a detailed jet-launching mechanism is needed to produce the magnetic field configuration for launching a jet. This is beyond the scope of this paper. Here we consider the basic properties of the CO/AD (such as charges and currents as the boundary conditions of the plasma outflow) required to support a magnetic jet. Section \ref{sec:magnetic_configuration} presents the electromagnetic field properties near the CO/AD, which can be used to constrain the current and charge properties of the CO/AD. When applying the Stokes's theorem to $\mathbf{j}=\nabla\times\mathbf{B}/4\pi$ and the Gauss's theorem to $\rho_{\rm e}=\nabla\cdot\mathbf{E}/4\pi$ on an infinitely thin AD with an electromagnetic field described by Equations \ref{B_disk} and \ref{E_disk}, one immediately gets the surface current and charge densities
\begin{eqnarray}
J_{\phi}^{\rm s}\left(R\right)&=&\frac{B_{R}\left(R\right)}{2\pi}= \frac{C_{1}}{2\pi}R^{\nu-2},\nonumber \\
J_{R}^{\rm s}\left(R\right)&=&-\frac{B_{\phi}\left(R\right)}{2\pi}= \frac{\Omega R}{\pi}R^{\nu-2},\nonumber \\
\rho_{\rm e}^{\rm s}\left(R\right)&=&-\frac{C_{1}B_{\phi}\left(R\right)}{4\pi}= \frac{C_{1}\Omega R}{2\pi}R^{\nu-2}.
\label{current_charge_AD}
\end{eqnarray}
The total charges carried by the CO can also be derived by a surface integration of electric fields over a sphere near the CO surface, which reads as
\begin{small}
\begin{equation}
\begin{aligned}
Q=\frac{1}{4\pi}\oint \mathbf{E}\cdot d\mathbf{S}=
r_{0}^{2}\int_{0}^{\pi/2}E_{r,\rm CO}\left(\theta\right)\sin\theta d\theta
=
\frac{-\nu C_{1}\Omega r_{0}^{3}B_{p,0}}{\left(1+\nu\right)\left(2-\nu\right)\sqrt{\nu^{2}+C_{1}^{2}}}=
-Sg\left(\mathbf{\Omega}\cdot\mathbf{B}\right)\frac{\nu C_{1}r_{0}\sqrt{P_{\rm jet}}}{\left(1+\nu\right)\left(2-\nu\right)}
=
-\frac{Sg\left(\mathbf{\Omega}\cdot\mathbf{B}\right)\nu C_{1}f_{\Omega}aF_{\rm B}}{4\pi\left(1+\nu\right)\left(2-\nu\right)},
%%%%%%%% Q=r0\sqrt{Pjet/c}
\label{CO_charge}
\end{aligned}
\end{equation}
\end{small}
where $r_{0}$ is the radius of the CO, and $E_{r,\rm CO}\left(\theta\right)$ is the $\hat{r}$ component of the electric field near the CO surface. For the typical value $\nu=3/4$, one has $\left|Q\right|=0.281\Omega r_{0}^{3}B_{p,0}$. Notice that in the case of a pulsar with a magnetosphere with an aligned dipolar magnetic field, the charge of the neutron star is $\left|Q\right|=\Omega r_{0}^{3}B_{p}/3$ \citep[see][]{1975ApJ...196...51R}. Therefore, a BH/AD has to be charged in order to launch a magnetically dominated jet, which, however, cannot be produced by a neutral BH/AD even if it rotates rapidly or accretes lowly/highly. This is consistent with observations that (1) many AGNs host fast-rotating BHs but are still radio-quiet \citep[RQ, no jet; see][]{2014SSRv..183..277R, 2016MNRAS.458.2012V, 2017ApJ...837...21X, 2018MNRAS.473.4377W, 2018MNRAS.478.1900S}; (2) RL and RQ AGNs present similar optical emission properties \citep{1994ApJS...95....1E, 2011ApJS..196....2S}, which are mainly determined by BH mass, spin, and accretion rate; and (3) jet can be produced from either a BH-advection-dominated accretion flow system \citep[geometrically thick, radiatively inefficient, and with a low accretion rate, e.g., BL Lacertae objects; see][]{1995ApJ...444..231N, 1995ApJ...452..710N, 2014ARA&A..52..529Y, 2018ApJS..235...39C} or a BH-standard Shakura-Sunyaev disk system \citep[geometrically thin, radiatively efficient, and with a high accretion rate, e.g., flat spectrum radio quasars; see][]{1973A&A....24..337S, 2018ApJS..235...39C}, with the difference between the two cases being mainly determined by the accretion rate \citep[e.g.,][]{2009ApJ...694L.107X, 2014ARA&A..52..529Y}. In this aspect, whether the BH/AD is charged may account for the RL/RQ dichotomy in AGNs \citep{1989AJ.....98.1195K, 1993ARA&A..31..473A, 1995PASP..107..803U, 2017A&ARv..25....2P, 2019ARA&A..57..467B}. This may be further related to the properties of the accreted plasma. Many theoretical works explore a magnetized plasma flow accreted by a spinning BH and find that a relativistic jet can be launched via the BZ process in the funnel region, while the region outside the funnel is not preferred to launch a relativistic outflow \citep[e.g.,][]{2004ApJ...611..977M, 2019ApJS..243...26P, 2013MNRAS.436.3856S, 2018ApJ...868..146N}. This may suggest that a charged BH is more essential than a charged AD in launching a relativistic jet. For a relativistic jet considered here, which is powered by the rotation of the BH/AD, the BH charge is much smaller than the maximum value that is achievable theoretically (see Appendix \ref{app:BHCharge}). The poloidal current of a BZ jet would exhaust the BH charge on timescale of $\Delta t\sim Q/J\sim10M_{8}~\rm min$ ($M_{8}=M/10^{8}M_{\odot}$; see Equation \ref{P_jetpower} and \ref{CO_charge}). Such a small timescale indicates that a continuous accretion is necessary to maintain a ``stable" BZ jet.

The above result suggests that the azimuthal surface current density on the AD would follow a power-law distribution, the slope of which determines the configuration of the poloidal magnetic field in the jet. This simple boundary condition is consistent with some simulations, which show that although the fluid and electromagnetic quantities in the AD are chaotic, the azimuthal surface current density (i.e. a vertically integrated toroidal current) exhibits a smooth and simple power-law behavior \citep[e.g.,][]{2007MNRAS.375..513M, 2007MNRAS.375..531M}. Let us consider an AD with an azimuthal current but no plasma above or below it. Generally speaking, the circulating current $I$ at ($\theta',r'$) contributes to a magnetic stream function at ($\theta,r$) \citep[see][]{1975clel.book.....J, 1994MNRAS.267..235L, 2019ApJ...872..149L},
\begin{equation}
\begin{aligned}
\Psi\left(\theta,r\right)=2Ir\frac{\left(1+x^{2}-2x\cos\theta'\cos\theta\right)K\left(k\right)
-\left(1+x^{2}-2x\cos\left(\theta'+\theta\right)\right)E\left(k\right)}
{\sqrt{1+x^{2}\cos^{2}\theta'-2x\cos\theta'\cos\theta}\sqrt{1+x^{2}-2x\cos\left(\theta'+\theta\right)}},
\label{ADC_B}
\end{aligned}
\end{equation}
where $x=r'/r$ and $K\left(k\right)$ and $E\left(k\right)$ are the first and second kind of complete elliptic integrals with argument $k=4x\sin\theta'\sin\theta/\left(1+x^{2}-2x\cos\left(\theta'+\theta\right)\right)$. For a current following a power-law distribution on the AD ($\theta'=\pi/2$), where $I=J_{\phi}^{\rm s}\left(R\right)dR=J_{0}R^{\nu-2}dR$ within the region ($R_{1},R_{2}$), one has
\begin{equation}
\begin{aligned}
\Psi\left(r,\theta\right)=2r^{2}J_{\phi}^{\rm s}\left(r\right)\int_{R_{1}/r}^{R_{2}/r}
\frac{\left(1+x^{2}\right)K\left(k\right)-\left(1+x^{2}+2x\sin\theta\right)E\left(k\right)}{\sqrt{1+x^{2}+2x\sin\theta}}x^{\nu-2}dx
\xlongequal[\left(R_{1}/r\right)\ll1]{\left(R_{2}/r\right)\gg1}\frac{2\pi}{C_{1}}r^{2}J_{\phi}^{\rm s}\left(r\right)T\left(\theta\right),
\label{ADC_B}
\end{aligned}
\end{equation}
with the argument reducing to $k=4x\sin\theta/\left(1+x^{2}+2x\sin\theta\right)$, which surprisingly matches what we derived in the force-free model (Equation \ref{Solution_nonr}) in the limit of $\left(R_{1}/r\right)\ll1\ll\left(R_{2}/r\right)$. In fact, the condition $\mathbf{j}=\nabla\times\mathbf{B}=0$ above the AD can be expressed in the form of Equation \ref{DE_F1_Psi}, and therefore they have the same solution. When the magnetic field starts rotation,  nature chooses a magnetic stream function that is almost conserved, which determines the magnetic configuration.

From Equation \ref{Bmag_Psi}, the components of the magnetic field can be expressed as \citep[see also][]{1987PASJ...39..821S},
\begin{eqnarray}
B_{\theta}\left(r,\theta\right)&=&2J_{\phi}^{\rm s}\left(r\right)\int_{R_{1}/r}^{R_{2}/r}
\frac{\left(1+x^{2}\cos2\theta\right)E\left(k\right)-\left(1+x^{2}-2x\sin\theta\right)K\left(k\right)}
{\sin\theta\left(1+x^{2}-2x\sin\theta\right)\sqrt{1+x^{2}+2x\sin\theta}}x^{\nu-2}dx
\xlongequal[\left(R_{1}/r\right)\ll1]{\left(R_{2}/r\right)\gg1}-2\pi J_{\phi}^{\rm s}\left(r\right)\frac{\nu}{C_{1}}\frac{T\left(\theta\right)}{\sin\theta}, \nonumber \\
B_{r}\left(r,\theta\right)&=&4J_{\phi}^{\rm s}\left(r\right)\int_{R_{1}/r}^{R_{2}/r}\frac{\cos\theta x^{\nu}E\left(k\right)}
{\left(1+x^{2}-2x\sin\theta\right)\sqrt{1+x^{2}+2x\sin\theta}}dx
\xlongequal[\left(R_{2}-r\right)\gg r\cos\theta]{\left(r-R_{1}\right)\gg r\cos\theta}2\pi J_{\phi}^{\rm s}\left(r\right)\frac{1}{C_{1}}\frac{1}{\sin\theta}\frac{dT}{d\theta}.
\label{ADC_B}
\end{eqnarray}
The latter matches the poloidal magnetic fields we derived in Section \ref{sec:magnetic_configuration}. Notice that we did not consider GR effects when deriving this result. In principle, to calculate the BH charge, one must consider a GR BH in Einstein-Maxwell theory \citep[e.g.,][]{1977MNRAS.179..457Z, 1977MNRAS.179..433B, 1982MNRAS.198..345M} for a Kerr-Newman metric \citep[][]{1963PhRvL..11..237K, 1965JMP.....6..918N}. This is left for future explorations.

Therefore, it seems that a simple disk model with a power-law scaling of azimuthal current may capture key properties of the jet such as jet configuration, collimation, and Lorentz factor. Theoretical analyses on the equipartition between magnetic and ram pressures suggest that the magnetic field or azimuthal current on an AD may follow a power-law distribution $\propto R_{0}^{-5/4}$ \citep[corresponding to $\nu=3/4$,][]{1982MNRAS.199..883B, 1995ApJ...452..710N}. This is confirmed by numerical simulations \citep[][]{2007MNRAS.375..513M, 2007MNRAS.375..531M}, where they found that despite the chaotic fluid and electromagnetic quantities in the AD, the vertically integrated azimuthal current follows the above smooth and simple power-law behavior. One always has this result once the simulations reach a quasi-steady state regardless of the different initial conditions, such as multiple magnetic loops in the initial torus, a net vertical field, or loops of alternating poloidal directions. We, therefore, take $\nu=3/4$ as the example value throughout the paper.

\section{Jet Stability}
\label{sec:stability}

Since we assume an axial symmetry in a steady state in this paper, we cannot address the question of jet stability. However, jets in this paper have $B_{\phi}\approx-\Omega RB_{p}$ (especially in the collimated region $\theta\ll1$), which marginally satisfies the stability criterion of \citet{2001PhRvD..64l3003T}, suggesting that the jets are marginally stable to the kink instability, which can be avoided or mitigated through, for example, the jet plasma being spinning faster than its internal Alfv\'{e}n velocity \citep[e.g.,][]{2004ApJ...617..123N}. Therefore, it is not expected that the jet would be (violently) kink unstable. This is also shown in some time-dependent simulations \citep[e.g.,][]{2001MNRAS.326L..41K, 2006MNRAS.368.1561M}. Also, the spontaneous development of the spine-layer structure (case $\nu<1$) may be naturally stabilized \citep[see, e.g.,][]{2007ApJ...664...26H, 2007ApJ...662..835M}. Indeed, the jet in M87 is clearly Poynting-flux-dominated, accelerating and collimated over $\sim 300$ pc \citep[at a distance $\lesssim10^{4}-10^{6}R_{\rm S}$][]{2001Sci...291...84M, 2008Natur.452..966M}, and yet does not kink until well beyond HST-1.

In the case of magnetically dominated jets for GRBs, significant dissipation of the magnetic energy is needed to power the prompt $\gamma$-ray emission. One possibility is that kink instability may be induced via internal interactions between different parts of the jet \citep{2011ApJ...726...90Z} or by the deceleration of the jet as the jet encounters a large-enough inertia from the ambient medium \citep{2019ApJ...882..184L}.

\section{Conclusions and Summary}
\label{sec:summary}

In this paper, starting from the first principles, we study an analytical solution of a magnetized jet/wind flow. The explicit expression of the solution makes it easy for further developments (e.g. adding radiative processes) and to directly compare with the observational data. Our findings can be summarized as follows:
\begin{enumerate}
    \item By separating the force-free equation of a jet/wind plasma flow into the nonrotating and rotating parts, we found that each of the two equations can be solved analytically and that the two solutions match each other very well (in the regimes either  $\theta\ll1$ or $\theta\rightarrow\pi/2$, and either nonrelativistic or relativistic). Therefore, we have a general approximate solution of a highly magnetized jet.
    \item An ordered rotating magnetic field is the indispensable ingredient to globally launch, accelerate, and collimate a relativistic jet. The magnetic stream function $\Psi=C_{2}r^{\nu}\sin^{2}\theta\ _{2}F_{1}\left(1-\frac{\nu}{2},\frac{1}{2}+\frac{\nu}{2},2,\sin^{2}\theta\right)$ (with $0\leq\nu\leq2$ a free parameter) controls the poloidal magnetic field configuration being a general parabola, while the toroidal magnetic field is determined by $\Phi=-2\Omega\Psi$ with an angular velocity $\Omega=\alpha\Psi^{\lambda}$. The resulting helical magnetic field is dominated by the poloidal component within the ACS, and by the toroidal component outside. The large-scale jet configuration can be used to constrain the foot-point locations of the magnetic fields.
    \item For a highly magnetized jet/wind flow, the drift velocity matches the cold plasma velocity very well, which is almost always perpendicular to the magnetic field and therefore also forms a helical structure. Acceleration from nonrelativistic through relativistic regimes is described, which is divided into three stages in the case of $\nu<1$: (1) within the ACS, where the toroidal velocity dominates and the four-velocity reads as $v\Gamma=\Omega R$; (2) outside the ACS but within the CCS, where the poloidal velocity dominates and  $v\Gamma=\Omega R$; and (3) outside the CCS, where the dominant poloidal velocity follows $v\Gamma\approx2/\left(\theta\sqrt{2-\nu}\right)$ due to a causality constraint. The acceleration in the case of $\nu\geq1$ only has the former two stages and therefore always follows $v\Gamma=\Omega R$. The large-scale jet velocity can be used to constrain the angular velocity and, hence, the spin of the central BH.
    \item The energy transportation of a magnetized jet/wind is dominated by the Poynting flux. The jet power is determined by the angular velocity and the magnetic flux, that is, $P_{\rm jet}=\Omega^{2}F_{\rm B}^{2}/\left(4\pi^{2}\left|\lambda+1\right|\right)\approx2\times10^{45}a^{2}M_{8}^{2}B_{5}^{2}$ erg s$^{-1}$ in the case of magnetic fields threading a BH ($\lambda=0$ and $B_{5}=B/10^{5}$Gs).
    \item A jet/wind flow has to carry charges so that the electric field force can balance the magnetic field force. The charge density reads as $\rho_{\rm e}=-\mathbf{\Omega}\cdot\mathbf{B}/2\pi$ (in the case of magnetic fields threading a CO, $\lambda=0$). The central rotating CO also has to be charged to globally launch a magnetized (BZ) jet with a total charge $\left|Q\right|\approx aF_{\rm B}/8\pi\approx r_{0}\sqrt{P_{\rm jet}} \approx2.8\times10^{20}\left(1+\sqrt{1-a^{2}}\right) M_{8}\sqrt{P_{44}}$ C (the BH case, the sign determined by $-Sg\left(\mathbf{\Omega}\cdot\mathbf{B}\right)$, $P_{44}=P_{\rm jet}/10^{44}$ erg s$^{-1}$). Whether the BH (AD) is charged may account for the RL/RQ dichotomy in AGNs, with the aid of the spin of the BH.
    \item In a jet/wind flow, the toroidal current almost vanishes, while the poloidal current is about $j_{p}\approx\rho_{\rm e}$. The total current carried by the jet reads as $J=\sqrt{\left|\lambda+1\right|P_{\rm jet}}\approx5.8\times10^{17}\sqrt{P_{44}}$ A ($\lambda=0$).
    \item The magnetic stream surface is equipotential, within which the magnetic field lines lie, the current streams, and the plasma flows. Crossing these surfaces can in principle lead to acceleration of the charged particles, which may be achieved by forming a gap in the polar region. The acceleration is limited by the potential difference between two magnetic stream surfaces $\Delta V=\sqrt{P_{\rm jet}/\left|\lambda+1\right|}\approx1.7\times10^{19}\sqrt{P_{44}}$ volts ($\lambda=0$).
    \item Given an available mass flux per magnetic flux ($\eta$) that still satisfies a highly magnetically dominated condition, one has approximations of the proper density $\rho\approx\left(\eta/4\pi\Omega^{2}\right)B/R^{2}$ and the lab frame density $\rho_{\rm l}=\Gamma\rho\approx\left(\eta/8\pi\Psi\Omega^{2}\right)B^{2}$.
    \item This approximate solution can (roughly) match known numerical simulation results and interpret most  observations of AGN and GRB jets.
\end{enumerate}

\acknowledgments

\textbf{Acknowledgments:} We are grateful to the anonymous referee for useful comments and suggestions, and we also appreciate the encouraging comments from Roger Blandford. We would like to thank Zhenhui Zhang, Defu Bu, Jiawen Li, Jinming Bai, Zhaoming Gan, Lei Huang, Hai Yang, Gang Li, Ming Chen, Hao Tong, Yunwei Yu, Henk Spruit, Xinwu Cao, Minfeng Gu, Shuangliang Li, Zhen Pan, and Yuanping Hong for helpful discussions. L.C. thanks Tongxin Chen and Tongyun Chen. The work of L.C. is supported by the National Natural Science Foundation of China (grant U1831138). L.C. also acknowledges the University of Nevada, Las Vegas, for hospitality when this work was carried out.

\newpage

\appendix
\label{append}

\section{The Relation $\Phi=-2\Omega\Psi$}
%%%%%%% $\Phi=-2\Omega\Psi/c$
\label{app:relation_Phi}

In Section \ref{sec:basic_equation}, we show that in the limit of force-free conditions, $\Phi=B_{\phi}R$ is conserved along a magnetic field line. In Section \ref{sec:approximate_solution}, the relation $\Phi=-2\Omega\Psi$ is further employed to solve Equation \ref{DE_Psi}, which is mainly based on a mathematical requirement (see Equation \ref{DE_T_2}). Here, we present a brief discussion on a more physical consideration \citep[see, e.g.,][for a similar discussion]{2009ApJ...698.1570L}.

In principle, $\Phi=B_{\phi}R=-\eta\left(\mathcal{E}-\Gamma\right)/\Omega$ cannot be conserved along a magnetic field line (the fluid Lorentz factor $\Gamma$ would change; see Equation \ref{E_cons}). In the limit of a highly magnetized jet ($\sigma\gg1$; see Equation \ref{sigma_pol}), there is a reduction $\Phi\approx-\eta\mathcal{E}/\Omega$, which is therefore approximately conserved along the magnetic field line. The fluid velocity generally follows $v_{\phi}B_{p}-B_{\phi}v_{p}=\Omega RB_{p}$ (see Section \ref{sec:jet_acceleration}). Let us consider the region $\Omega R\gg1$, where the jet is already relativistic and $v_{\phi}\ll1$ and $v_{p}\approx1$ are satisfied. One immediately gets $-B_{\phi}/B_{p}\approx\Omega R$. In this collimated jet region (polar angle $\theta\ll1$), the poloidal magnetic field can be assumed to be approximately uniform at a given height from the central CO/AD \citep[being self-consistent with Equation \ref{B_tll1}; see also][]{2009ApJ...698.1570L}, which yields an enclosed magnetic flux $F_{\rm B}=\pi R^{2}B_{p}$. Comparing this with Equation \ref{magneticflux} and with the aid of relation $-B_{\phi}/B_{p}\approx\Omega R$, one immediately gets $\Phi\approx-2\Omega\Psi$. Due to conservation of $\Psi$, $\Omega$, and $\Phi$ (in the limit of high magnetization), this relation would hold throughout the jet region, even though it was derived asymptotically (see Section \ref{sec:approximate_solution}).

\section{The Magnetic Field Direction and Amplitude}
\label{app:sign}

The parameters $\Psi$ and $\Omega=\mathbf{\Omega}\cdot\hat{z}$ can be both positive and negative, which correspond to the magnetic field lines pointing in two opposite directions. All of the formulae in the main text are  written assuming a positive $\Psi$ and $\Omega$. In the case of a negative $\Psi$ or a negative $\Omega$, one needs to multiply a sign factor to these formulae. The sign convention can be easily derived, which can be summarized as
\begin{eqnarray}
Sg\left(\mathbf{S}_{p}\right)&=&Sg\left(\mathbf{v}_{p}\right)=+1, \nonumber \\
Sg\left(S_{\phi}\right)&=&Sg\left(v_{\phi}\right)=Sg\left(\Omega\right), \nonumber \\
Sg\left(\mathbf{B}_{p}\right)&=&Sg\left(j_{\phi}\right)=Sg\left(\Psi\right), \nonumber \\
Sg\left(B_{\phi}\right)&=&Sg\left(\mathbf{E}\right)
=Sg\left(\mathbf{j}_{p}\right)=Sg\left(\rho_{\rm e}\right)
=Sg\left(\Phi\right)=Sg\left(\mathbf{\Omega}\cdot\mathbf{B}\right)=Sg\left(\Omega\right)\cdot Sg\left(\Psi\right).
\label{Sign}
\end{eqnarray}
It can be seen that $\mathbf{j}_{p}$ and $B_{\phi}$ have the same sign convention ($\mathbf{E}$ and $\rho_{\rm e}$ also have the same sign convention), and therefore the Lorentz force does not change. This implies that the sign choice is dynamically not important. One therefore expects that the acceleration and velocity always point outward from the CO/AD. This can be also illustrated by $Sg\left(\mathbf{S}_{p}\right)=Sg\left(\mathbf{v}_{p}\right)=+1$, which means that regardless of the sign of $\Psi$ or $\Omega$, the velocity and Poynting flux always point outward for either of the pairs of antiparallel jets/winds. It can be also easily inferred that, in the case of $\mathbf{\Omega}$ and $\mathbf{B}$ projected in the same directions, the currents in both sides of the jet (having negative charges) flow inward toward the central CO ($s=0$). In the case of $\mathbf{\Omega}$ and $\mathbf{B}$ projected in opposite directions (having a positive charge), on the other hand, both currents flow outwards from the central CO. The case of the AD may be different because there is a gradient of angular velocity (see Section \ref{subsec:current}).

For simplicity, the form $\Psi=r^{\nu}T\left(\theta\right)$ is taken in the main text. In fact, one can multiply $\Psi$ by an arbitrary constant $C_{\Psi}$, and the result should still be a solution of the original equations. In this case, $\Phi$, $\mathbf{B}$, $\mathbf{E}$, $\mathbf{j}$ and $\rho_{\rm e}$ should be amplified by $C_{\Psi}$, and the jet power (Poynting flux) should be multiplied by $C_{\Psi}^{2}$. The velocity does not change.

\section{Exact solutions}
\label{sec:exactsolutions}

In some special cases, the force-free Equation \ref{DE_Psi} has exact solutions. These solutions are well known in the literature \citep[e.g.,][]{2007MNRAS.375..548N}, and we summarize them below for completeness. As shown below, these solutions can also be derived from our general solutions when the special conditions are met.

\subsection{Monopole}
\label{sec:monopole}

In the case of a monopole magnetic configuration, the magnetic stream function is only a function of $\theta$. In this case, one has an exact solution for Equation \ref{DE_Psi} \citep[with $\Omega=\alpha\Psi^{\lambda}$, see][]{1969ApJ...158..727M},
\begin{eqnarray}
\Psi&=&1-\cos\theta, \nonumber \\
\Phi&=&-\Omega\sin^{2}\theta,
\label{Psi_monopole}
\end{eqnarray}
which corresponds to our approximate solution for the case of $\nu=0$, where we have the same $\Psi=1-\cos\theta$ but a different $\Phi=-2\Omega\left(1-\cos\theta\right)$. This value of $\Phi$ and the exact value of $\Phi$ match each other at  $\theta\ll1$, but differ by $50\%$ at $\theta=\pi/2$. This exact monopole solution shows a magnetic field that still follows $B_{\phi}/B_{p}=-\Omega r\sin\theta$ and
\begin{eqnarray}
B_{\theta}&=&0, \nonumber \\
B_{r}&=&\frac{1}{r^{2}}B_{0}, \nonumber \\
B_{\phi}&=&-\frac{\Omega\sin\theta}{r}B_{0}.
\label{B_monopole}
\end{eqnarray}
The drift velocity is also the same as our approximate solution (Equation \ref{v_drift_total_s}):
\begin{eqnarray}
v_{\theta}&=&0, \nonumber \\
v_{r}&=&\frac{\left(\Omega R\right)^{2}}{1+\left(\Omega R\right)^{2}}, \nonumber \\
v_{\phi}&=&\frac{\Omega R}{1+\left(\Omega R\right)^{2}}, \nonumber \\
v&=&\frac{\Omega R}{\sqrt{1+\left(\Omega R\right)^{2}}}, \nonumber \\
v\Gamma&=&\Omega R.
\label{v_monopole}
\end{eqnarray}

The Poynting flux and jet power read as
\begin{eqnarray}
S&=& \frac{\Omega R\sqrt{1+\left(\Omega R\right)^{2}}}{4\pi r^{4}}B_{0}^{2}, \nonumber \\
S_{r}&=&\frac{\Omega^{2}R^{2}}{4\pi r^{4}}B_{0}^{2},\nonumber \\
P_{\rm jet}&=&\frac{2}{3}\Omega^{2}B_{0}^{2}.
\label{S_monopole}
\end{eqnarray}

\subsection{Parabola}
\label{sec:parabola}

With an assumption of $\Omega=\alpha\Psi^{-1}$, one can derive another exact solution of Equation \ref{DE_Psi}:
\begin{eqnarray}
\Psi&=&r\left(1-\cos\theta\right), \nonumber \\
\Phi&=&-2\alpha,
\label{Psi_parabola}
\end{eqnarray}
which is the same as our approximate solution for the case $\nu=1$, where the rotation curve is ``flat" on the AD plane $\Omega\propto R_{0}^{-1}$ \citep[see][]{1976MNRAS.176..465B, 2007MNRAS.375..548N}. The magnetic field configuration forms a parabola in this case.

\subsection{Cylinder}
\label{sec:cylinder}

Another exact solution of Equation \ref{DE_Psi} reads as \citep[see][]{1994MNRAS.267..629I}
\begin{eqnarray}
\Psi&=&r^{2}\sin^{2}\theta, \nonumber \\
\Phi&=&-2\Omega\Psi,
\label{Psi_parabola}
\end{eqnarray}
which is the same as our approximate solution for the case of $\nu=2$ (applies for $\Omega=\alpha\Psi^{\lambda}$, see Section \ref{sec:approximate_solution}). The magnetic field configuration forms a cylinder in this case.

\section{Asymptotic Behavior of $T\left(\theta\right)$}
\label{app:T}
Setting $\mu=\cos\theta$ and $\xi=\sin\theta$, one obtains the asymptotic behavior of $T\left(\theta\right)$ (see Equation \ref{Solution_nonr}) at  $\theta\rightarrow\pi/2$ (i.e. $\mu\ll1$, $\xi\rightarrow1$)
\begin{eqnarray}
T&=&1-C_{1}\mu-\frac{\nu\left(\nu-1\right)}{2}\mu^{2}, \nonumber \\
\frac{dT}{d\theta}&=&\xi\left[C_{1}+\nu\left(\nu-1\right)\mu-
C_{1}\frac{1}{2}\mu^{2}\right], \nonumber \\
\frac{d^{2}T}{d\theta^{2}}&=&-\nu\left(\nu-1\right)\xi^{2}+\left(1+\xi^{2}\right)
C_{1}\mu+\nu\left(\nu-1\right)\left[1+\frac{1}{2}\left(\nu+1\right)\left(\nu-2\right)\xi^{2}\right]\mu^{2}
\label{T_approx_disk}
\end{eqnarray}
and at $\theta\ll1$ (i.e. $\xi\ll1$, $\mu\rightarrow1$)
\begin{eqnarray}
T&=& C_{2}\left[\xi^{2}+\frac{1}{2}\left(1-\frac{\nu}{2}\right)
\left(\frac{1}{2}+\frac{\nu}{2}\right)\xi^{4}+\frac{1}{12}
\left(1-\frac{\nu}{2}\right)\left(2-\frac{\nu}{2}\right)\left(\frac{1}{2}+\frac{\nu}{2}\right)
\left(\frac{3}{2}+\frac{\nu}{2}\right)\xi^{6}\right], \nonumber  \\
\frac{dT}{d\theta}&=& C_{2}\mu\left[2\xi-\frac{1}{2}\left(\nu-2\right)
\left(\nu+1\right)\xi^{3}+\frac{1}{2}
\left(1-\frac{\nu}{2}\right)\left(2-\frac{\nu}{2}\right)\left(\frac{1}{2}+\frac{\nu}{2}\right)
\left(\frac{3}{2}+\frac{\nu}{2}\right)\xi^{5}\right], \nonumber \\
\frac{d^{2}T}{d\theta^{2}}&=& C_{2}\left\{2\mu^{2}+\left[-2-\frac{3}{2}\left(\nu-2\right)\left(\nu+1\right)\mu^{2}\right]\xi^{2}
+\frac{1}{2}\left(\nu+1\right)
\left(\nu-2\right)\left[1+\frac{5}{16}\left(\nu-4\right)
\left(\nu+3\right)\mu^{2}\right]\xi^{4}\right\}.
\label{T_approx_axis}
\end{eqnarray}

\section{Magnetic Field 3D Morphology}
\label{app:magnetic_field_morphology}

With the approximate solution $\Psi=r^{\nu}T\left(\theta\right)$ and $\Phi=-2\Omega\Psi$, one can derive the 3D morphology of the magnetic field lines. It is known that the ratios of different components ($r,\theta,\phi$) of the magnetic field determine its 3D morphology.  Combining with Equation \ref{Bmag_Psi}, one can derive an equation that tracks a magnetic field line:
\begin{equation}
\begin{aligned}
dr:rd\theta:r\sin\theta d\phi =B_{r}:B_{\theta}:B_{\phi} =-\frac{1}{T}\frac{dT}{d\theta}:\nu:2\Omega r.
\label{magnetic_field_3D_morphology}
\end{aligned}
\end{equation}
Supposing a magnetic field line anchored at $\left(r_{0},\theta_{0},\phi_{0}\right)$, one immediately has ($T_{0}\equiv T\left(\theta_{0}\right)$)
\begin{eqnarray}
r&=&r_{0}T_{0}^{1/\nu}T^{-1/\nu},\nonumber\\
\phi-\phi_{0}&=&\frac{2\Omega r_{0}T_{0}^{1/\nu}}{\nu}\int_{\theta_{0}}^{\theta} \frac{T^{-1/\nu}}{\sin\theta'}d\theta'.
\label{magnetic_field_3D_morphology2}
\end{eqnarray}
In the case of $\theta\ll\theta_{0}$, the angle rotation ($\phi-\phi_{0}$) mainly happens in the region near $\theta$. The above Equation \ref{magnetic_field_3D_morphology2} is reduced to
\begin{eqnarray}
r&=&r_{0}T_{0}^{1/\nu}C_{2}^{-1/\nu}\theta^{-2/\nu},\nonumber\\
\phi-\phi_{0}&=&-\Omega r.
\label{magnetic_field_3D_morphology3}
\end{eqnarray}
Let us consider the number of rotation cycles, from foot-point to the ACS (assuming $\theta\ll1$), of a magnetic field line anchored at the equatorial plane ($\theta_{0}=\pi/2$):
\begin{eqnarray}
N=\frac{\left|\phi_{\rm ACS}-\phi_{0}\right|}{2\pi}&\approx& \frac{C_{2}^{1/\left(2-\nu\right)}}{2\pi\left(\Omega r_{0}\right)^{\nu/\left(2-\nu\right)}}
\xlongequal[]{\nu=3/4}\frac{1}{2^{9/5}\pi\left(\Omega r_{0}\right)^{3/5}}.
\label{magnetic_field_3D_morphology4}
\end{eqnarray}
This indicates that a magnetic field line rotates even less than one cycle from its foot-point to the ACS, provided that the foot-point rotation velocity is not very slow ($\Omega r_{0}\gtrsim0.02$ for $\nu=3/4$).

\section{Jet Flow Neutrality?}
\label{app:ne0}

Some MHD simulations ignore the electric force because of its relative unimportance compared with the magnetic force. Therefore, the force-free condition reduces to $\left(\nabla\times\mathbf{B}\right)\times\mathbf{B}=0$. Now, let us consider whether the solution of this equation can guarantee that charge vanishes, that is, $\rho_{\rm e}=\nabla\cdot\mathbf{E}/4\pi=0$. Ignoring the electric force part, the force-free condition yields (steady and axisymmetric; see Equation \ref{DE_Psi})
\begin{equation}
\begin{aligned}
\frac{\partial^{2}\Psi}{\partial r^{2}}+\frac{1}{r^{2}}\frac{\partial^{2}\Psi}{\partial\theta^{2}}
-\frac{\cot\theta}{r^{2}}\frac{\partial\Psi}{\partial\theta} +\Phi'\Phi=0.
\label{No_charge_DE_Psi}
\end{aligned}
\end{equation}
Instead of appealing to possible numerical methods to solve this equation, here we try to seek analytical solutions. The $\Psi$ function can be separated, $\Psi=H\left(r\right)T\left(\theta\right)$, only in the case of $\Phi=-\sqrt{\omega^{2}\Psi^{2}+\varrho^{2}}$ with $\omega$ and $\varrho$ being constants. In this case, we have a solution of $T\left(\theta\right)$ the same as $T_{\rm nr}\left(\theta\right)$ (Equation \ref{Solution_nonr}). The function $H\left(r\right)$ follows
\begin{equation}
\begin{aligned}
r^{2}\frac{d^{2}H}{dr^{2}}+\left[\omega^{2}r^{2}-\nu\left(\nu-1\right)\right]H=0,
\end{aligned}
\label{No_charge_Hequation}
\end{equation}
with $\nu\left(\nu-1\right)$ being a separate constant. With transformations $H=r^{1/2}F$ and $x=\omega r$, one then has a standard Bessel equation:
\begin{equation}
\begin{aligned}
x^{2}\frac{d^{2}F}{dx^{2}}+x\frac{dF}{dx}+\left[x^{2}-\left(\nu-\frac{1}{2}\right)^{2}\right]F=0.
\end{aligned}
\label{No_charge_Fequation_bess}
\end{equation}
Considering the boundary condition of $\Psi$ vanisheing when $\omega r\rightarrow0$, one can get a solution of Equation \ref{No_charge_Hequation} that reads as
\begin{equation}
\begin{aligned}
H\left(r\right)=C_{\rm h}r^{\frac{1}{2}}
J_{\nu-\frac{1}{2}}\left(\omega r\right),
\end{aligned}
\label{No_charge_Fequation_bess2}
\end{equation}
where $C_{\rm h}$ is the integration constant and $J_{\nu-1/2}\left(x\right)$ is the first kind of Bessel function with an order of $\nu-1/2$. Because $\Psi$ vanishes at $\theta\rightarrow0$, one expects that the constant $\varrho$ should vanish (i.e. $\varrho=0$) to guarantee $B_{\phi}=-\sqrt{\omega^{2}\Psi^{2}+\varrho^{2}}/r\sin\theta$ being not singular at $\theta\rightarrow0$. Therefore one has $\Phi=-\omega\Psi$, which implies that $\omega$ represents the angular velocity of the electromagnetic field (although they may differ by a constant; see Section \ref{sec:approximate_solution}). In the case of $\omega r\ll1$ (i.e. nonrelativistic), $H\left(r\right)$ reduces to $H\left(r\right)=C_{\rm h}\left[\Gamma\left(\nu+1/2\right)\right]^{-1}
\left(\omega/2\right)^{\nu-1/2}r^{\nu}$. One can safely choose $C_{\rm h}=\Gamma\left(\nu+1/2\right)\left(\omega/2\right)^{1/2-\nu}$ to guarantee that $H\left(r\right)=r^{\nu}$ is the same as $H_{\rm r/nr}\left(r\right)$. The electric field can be derived through Equation \ref{E_general}, and the charge density can be then derived through $\rho_{\rm e}=\nabla\cdot\mathbf{E}/4\pi$, which would be the same as Equation \ref{rho_eem}. Therefore, the above solution (ignoring the electric force part) cannot generally guarantee that the charge density vanishes. However, one still has, in the nonrelativistic limit, the magnetic field configuration being approximately the same as that in the case of considering the electric force (see Equation \ref{Solution_nonr}) and the case with the electric force neglected (see more in Section \ref{sec:Lorentz_Force}).

\section{The Maximum Lorentz Factor}
\label{app:maximum_Lorentz}

One has $\Gamma\left(\sigma+1\right)=\mathcal{E}$ conserved along a magnetic field line, which also approximates the theoretical maximum Lorentz factor that a jet can be accelerated to, if all of the Poynting flux is converted to kinetic energy. In principle, the value of $\mathcal{E}$ cannot be arbitrarily large, because the plasma energy density cannot be arbitrarily small, which is due to the fact that the plasma has to carry a proper charge and current density to support the electromagnetic field (see Section \ref{sec:charge_current_Power}). Given the plasma being completely made up of charged particles, i.e. without any other neutral particles (pairs), one has a minimum plasma density $\rho_{\rm me}=\rho_{\rm e}/r_{\rm cm}$, where $r_{\rm cm}$ is the charge-to-mass ratio of the particle, while this minimum plasma density is still too small to maintain a poloidal current $j_{p}\approx\rho_{\rm e}$ (see Section \ref{sec:charge_current_Power}) due to a small poloidal velocity $v_{p}\approx\left(\Omega R\right)^{2}\ll1$ (nonrelativistic at the foot-point; see Section \ref{sec:poloidal_velocity}). Therefore, in order to sustain this poloidal current, additional plasma particles are required with a density roughly at least $\rho_{\rm min}\sim\rho_{\rm me}/\left(\Omega R\right)^{2}$. In this case, one has a maximum magnetization parameter at the foot-point, which is also the maximum Lorentz factor the plasma flow can reach, $\gamma_{\rm max}\approx\sigma_{\rm max}\approx U_{\rm B_{0}}/\rho_{\rm min}\sim r_{\rm cm}B_{0}\Omega R^{2}/4\sim r_{\rm cm}\Delta V/4$ (see Equations \ref{DeltaV} and \ref{DeltaV_real} for the case of a BH engine).

\section{On Black Hole Charge}
\label{app:BHCharge}
In principle, a BH can be charged. Similar to the BH spin parameter, a dimensionless BH charge parameter can be defined in Kerr-Newman metric
\begin{equation}
r_{\rm Q}=\frac{Q}{M},
\label{rQ}
\end{equation}
which follows a constrain $r_{\rm Q}^{2}+a^{2}\leq1$ \citep[see][]{1963PhRvL..11..237K, 1965JMP.....6..918N}. It is not easy for an astrophysical black hole to be charged unless there is a magnetosphere induced rotating magnetic field threading its horizon. For a charged CO that can launch a jet and that we are interested in here, the charge parameter roughly follows (see Equation \ref{CO_charge})
\begin{equation}
r_{\rm Q}\approx
\frac{r_{0}}{r_{\rm +}}\sqrt{\frac{P_{\rm jet}}{P_{\rm max}}
%{3.63\times10^{59}\ \rm erg\ s^{-1}}
},
\label{rQCO}
\end{equation}
where $P_{\rm max}=c^{5}/G=3.63\times10^{59}$ erg s$^{-1}$ is the universal maximum power defined by fundamental constants. It can be seen that, for any astrophysical system, the charge parameter is a mall value, provided that the astrophysical jet power being always very small compared with $P_{\rm max}$, even though such a ``small" charge is essential to launch a relativistic powerful jet powered by BH rotation.

\section{Some Formulae in Gaussian Units}
\label{app:Gaussianformula}

The formulae in the main text are written in natural units. For the convenience of  readers, below we write some important formulae also in Gaussian units.

The ideal MHD condition Equation (\ref{ideal_MHD}):
\begin{equation}
\mathbf{E}+\frac{\mathbf{v}\times\mathbf{B}}{c}=0.
\end{equation}
%%%%%%%%%%%%%%%%%%%%%%%%%%%%%%%%%%%%%%%%%
The force-free Equation (\ref{force_free}):
\begin{equation}
\rho_{\rm e}\mathbf{E}+\frac{\mathbf{j}\times\mathbf{B}}{c}=0.
\end{equation}
%%%%%%%%%%%%%%%%%%%%%%%%%%%%%%%%%%%%%%%%%
The electric field Equation (\ref{E_general}):
\begin{equation}
\mathbf{E}=-\frac{1}{c}\Omega\nabla\Psi=-\frac{\Omega r\sin\theta}{c}\hat{\phi}\times\mathbf{B}.
\end{equation}
%%%%%%%%%%%%%%%%%%%%%%%%%%%%%%%%%%%%%%%%%
The enclosed current Equations \ref{JzJr} and (\ref{Phi2OmegaPsi}):
\begin{equation}
\begin{aligned}
\Phi=-\frac{2\Omega\Psi}{c}=\frac{2J}{c}.
\end{aligned}
\end{equation}
%%%%%%%%%%%%%%%%%%%%%%%%%%%%%%%%%%%%%%%%%
The Poynting flux Equation (\ref{Poynting_flux_rz}):
\begin{equation}
\begin{aligned}
\mathbf{S}=\frac{c}{4\pi}\mathbf{E}\times\mathbf{B}.
\end{aligned}
\end{equation}
%%%%%%%%%%%%%%%%%%%%%%%%%%%%%%%%%%%%%%%%%
The drift velocity Equation (\ref{v_drift}):
\begin{equation}
\begin{aligned}
\mathbf{v}_{\rm d}=c\frac{\mathbf{E}\times\mathbf{B}}{B^{2}}.
\end{aligned}
\end{equation}
%%%%%%%%%%%%%%%%%%%%%%%%%%%%%%%%%%%%%%%%%
The current density Equations \ref{J_current} and (\ref{j_e_poroidalj}):
\begin{equation}
\begin{aligned}
\mathbf{j}_{p}=c\frac{\left(\nabla\times\mathbf{B}\right)_{\hat{p}}}{4\pi}
=\left(\lambda+1\right)\frac{cB_{\phi}}{4\pi r}\left(\frac{1}{T}\frac{dT}{d\theta}\hat{r}-\nu\hat{\theta}\right)
=\left(\lambda+1\right)\frac{\Omega}{2\pi}\frac{r^{\nu-2}}{\sin\theta}\left(-\frac{dT}{d\theta}\hat{r}+\nu T\hat{\theta}\right).
\end{aligned}
\end{equation}
%%%%%%%%%%%%%%%%%%%%%%%%%%%%%%%%%%%%%%%%%
The charge density Equations \ref{rho_eff} and (\ref{rho_eem}):
\begin{small}
\begin{equation}
\begin{aligned}
\rho_{\rm e}=\frac{\nabla\cdot\mathbf{E}}{4\pi}=-\frac{\mathbf{\Omega}\cdot\mathbf{B}}{2\pi c}+\frac{\Omega r\sin\theta j_{\phi}}{c^{2}}-\frac{\Omega'}{4\pi c}\left|\nabla\Psi\right|^{2}
\xlongequal[]{j_{\phi}=0}-\frac{\Omega r^{\nu-2}}{4\pi c}\left[\frac{\lambda}{T}\left(\frac{dT}{d\theta}\right)^{2}+2\cot\theta \frac{dT}{d\theta}+\left(2+\lambda\nu\right)\nu T\right]=\frac{j_{p}}{c}.
\end{aligned}
\end{equation}
\end{small}
%%%%%%%%%%%%%%%%%%%%%%%%%%%%%%%%%%%%%%%%%
The potential difference Equation (\ref{DeltaV}):
\begin{equation}
\begin{aligned}
\Delta V=-\frac{\Omega\Psi}{\left(\lambda+1\right)c}\bigg|_{1}^{2}
=-\frac{\Omega F_{B}}{2\pi\left(\lambda+1\right)c}\bigg|_{1}^{2}
=\frac{RB_{\phi}}{2\left(\lambda+1\right)}\bigg|_{1}^{2}
=\frac{J}{\left(\lambda+1\right)c}\bigg|_{1}^{2}.
\end{aligned}
\end{equation}
%%%%%%%%%%%%%%%%%%%%%%%%%%%%%%%%%%%%%%%%%
The jet power Equation (\ref{P_jetpower}):
\begin{equation}
\begin{aligned}
P_{\rm jet}=\frac{\Omega^{2}\Psi^{2}}{c\left|\lambda+1\right|}\bigg|_{1}^{2}
=\frac{\Omega^{2}F_{B}^{2}}{4\pi^{2}c\left|\lambda+1\right|}\bigg|_{1}^{2}
=\frac{cR^{2}B_{\phi}^{2}}{4\left|\lambda+1\right|}\bigg|_{1}^{2}
=\frac{J^{2}}{c\left|\lambda+1\right|}\bigg|_{1}^{2}.
\end{aligned}
\end{equation}
%%%%%%%%%%%%%%%%%%%%%%%%%%%%%%%%%%%%%%%%%
The CO/BH total charge Equation (\ref{CO_charge}):
\begin{small}
\begin{equation}
\begin{aligned}
Q=-Sg\left(\mathbf{\Omega}\cdot\mathbf{B}\right)\frac{\nu C_{1}r_{0}\sqrt{P_{\rm jet}/c}}{\left(1+\nu\right)\left(2-\nu\right)}
=-\frac{Sg\left(\mathbf{\Omega}\cdot\mathbf{B}\right)\nu C_{1}f_{\Omega}aF_{\rm B}}{4\pi\left(1+\nu\right)\left(2-\nu\right)},
\end{aligned}
\end{equation}
\end{small}
%%%%%%%%%%%%%%%%%%%%%%%%%%%%%%%%%%%%%%%%%
The BH gravitational radius:
\begin{equation}
\begin{aligned}
r_{\rm g}=\frac{GM}{c^{2}}.
\end{aligned}
\end{equation}
%%%%%%%%%%%%%%%%%%%%%%%%%%%%%%%%%%%%%%%%%
The BH spin parameter:
\begin{equation}
\begin{aligned}
a=\frac{cJ_{\rm am}}{GM^{2}}.
\end{aligned}
\end{equation}
%%%%%%%%%%%%%%%%%%%%%%%%%%%%%%%%%%%%%%%%%
The BH rotating angular velocity:
\begin{equation}
\begin{aligned}
\Omega_{\rm BH}=\frac{ac}{2\left(1+\sqrt{1-a^{2}}\right)r_{g}}.
\end{aligned}
\end{equation}
%%%%%%%%%%%%%%%%%%%%%%%%%%%%%%%%%%%%%%%%%
The Keplerian angular velocity on a BH equator:
\begin{equation}
\begin{aligned}
\Omega_{\rm K}=\frac{c}{r_{\rm g}\left[\left(r_{0}/r_{\rm g}\right)^{3/2}+a\right]}.
\end{aligned}
\end{equation}
%%%%%%%%%%%%%%%%%%%%%%%%%%%%%%%%%%%%%%%%%
The flow density Equation (\ref{JetProperdensity}):
\begin{equation}
\begin{aligned}
\rho=\frac{\eta}{4\pi}\frac{B_{p}}{u_{p}}\simeq \frac{c\eta\Psi}{2\pi\Omega^{2}}\frac{\sqrt{1+\left(\Omega R/c\right)^{2}}}{R^{4}}\simeq \frac{c\eta}{4\pi\Omega^{2}}\frac{B}{R^{2}}\simeq \frac{c\eta}{8\pi\Psi\Omega^{2}}\frac{B^{2}}{\Gamma}\simeq \frac{\rho_{\rm l}}{\Gamma}.
\end{aligned}
\end{equation}
%%%%%%%%%%%%%%%%%%%%%%%%%%%%%%%%%%%%%%%%%
The BH charge parameter:
\begin{equation}
\begin{aligned}
r_{\rm Q}=\frac{Q}{\sqrt{G}M}\approx\frac{r_{0}}{r_{\rm +}}\sqrt{\frac{GP_{\rm jet}}{c^{5}}}.
\end{aligned}
\end{equation}

\bibliography{sample63}{}
\bibliographystyle{aasjournal}

\end{document}